\documentclass[aps,prd,superscriptaddress,showpacs,twocolumn,amsmath,amssymb,groupedaddress,nofootinbib]{revtex4}
\usepackage{amsmath}
\usepackage{amssymb}
\usepackage{amsfonts}
\usepackage{graphicx}
\usepackage{latexsym}
\usepackage[]{latexsym}
\usepackage{bm}
\usepackage{array}
\usepackage{dcolumn}
\usepackage{longtable}
\usepackage{slashed}
\usepackage{epsfig}
\usepackage{color}
\usepackage{natbib}
\usepackage{units}
\usepackage[unicode,psdextra,hyperfootnotes = true]{hyperref}
\usepackage{bookmark}
\newcommand{\be}{\begin{equation}}\newcommand{\ee}{\end{equation}}
\newcommand{\bea}{\begin{eqnarray}}\newcommand{\eea}{\end{eqnarray}}
\newcommand{\brr}{\begin{array}}\newcommand{\err}{\end{array}}
\newcommand{\bit}{\begin{itemize}}\newcommand{\eit}{\end{itemize}}
\newcommand{\ben}{\begin{enumerate}}\newcommand{\een}{\end{enumerate}}

\newcommand{\ide}{1\hspace{-1mm}{\rm I}}
\newcommand{\ba}{\begin{array}}
\newcommand{\ea}{\end{array}}

\definecolor{darkblue}{rgb}{0,0,.7}

\def\1{{_{1}}}\def\2{{_{2}}}

\def\ide{1\hspace{-1mm}{\rm I}}

\def\noHe0{:\;\!\!\;\!\!:H_e(0):\;\!\!\;\!\!:}
\def\noHm0{:\;\!\!\;\!\!:H_\mu(0):\;\!\!\;\!\!:}

\def\1{{_{1}}}\def\2{{_{2}}}

\def\con{\color{black}} 

\definecolor{darkred}{rgb}{.8,0,0}

\begin{document}

%
\title{Infrared behavior of Weyl Gravity: Functional Renormalization Group approach}
%

\author{Petr Jizba}
\email{p.jizba@fjfi.cvut.cz}
\affiliation{FNSPE, Czech Technical
University in Prague, B\v{r}ehov\'{a} 7, 115 19 Praha 1, Czech Republic}
\author{Les{\l}aw Rachwa{\l}}
\email{grzerach@gmail.com}
\affiliation{FNSPE, Czech Technical
University in Prague, B\v{r}ehov\'{a} 7, 115 19 Praha 1, Czech Republic}
\author{Jaroslav K\v{n}ap}
\email{knapjaro@fjfi.cvut.cz}
\affiliation{FNSPE, Czech Technical
University in Prague, B\v{r}ehov\'{a} 7, 115 19 Praha 1, Czech Republic}
\pdfbookmark[1]{Abstract}{name1}
\begin{abstract}
Starting from an ultraviolet fixed point, we study the infrared behavior of quantum Weyl gravity in terms of a functional renormalization group (RG) flow equation. To do so, we employ two classes of Bach-flat backgrounds, namely maximally symmetric spacetimes and Ricci-flat backgrounds in the improved one-loop scheme. We show that in the absence of matter fields and with a topological term included, the effective action exhibits dynamical breaking of scale symmetry. In particular, it is shown that apart from a genuine IR fixed point that is reached at a zero-value of the running scale, the RG flow also exhibits bouncing behavior in the IR regime. 
We demonstrate that both $\beta_C$ and $\beta_E$
reach the RG 
turning point (almost) simultaneously at the same finite energy scale, irrespectively of the chosen background.
The IR fixed point itself is found to be IR-stable in the space of the considered couplings. Ensuing scaling dimensions
of both operators are also computed. Salient issues, including the connection of the observed bouncing RG flow behavior with holography and prospective implications in early Universe cosmology, are also briefly discussed.
\end{abstract}
\pacs{98.80.-k, 11.25.Hf}
\keywords{Conformal field theory, Cosmology, Functional renormalization group}

\vspace{-1mm}

\maketitle
%


\pdfbookmark[1]{I. Introduction}{name2}
\hypersetup{bookmarksdepth=-2}
\section{Introduction}
\hypersetup{bookmarksdepth}


The current constraints from Planck measurements of the CMB anisotropies indicate that the  cosmological perturbations are (nearly) scale-invariant with the value of the scalar spectral
index $n_s =0.965 \pm 0.004$ (with a 68\% confidence level)~\cite{Ade}. This is tantalizingly close to $n_s =1$, which corresponds to the exact scale-invariant fluctuations. This fact
suggests that it might be useful to describe the very early stage of the Universe in terms of some (possibly even effective) scale-invariant gravitational theories.

Theories with (classical or quantum) scale-invariance have a long and venerable history. Apart from the fact that they are instrumental in providing a dynamical origin of mass scales~\cite{Col,Ad1,Ad2,Zee,Sp,K-S,Cap}, they have a number of further desirable features; for instance, they provide an appealing framework for addressing the hierarchy problem~\cite{Shaposhnikov}, lead to naturally flat inflationary potentials~\cite{JKS}, furnish dark matter
candidates~\cite{Ayazi,Ishiwata}  or even provide a viable alternative for dark matter itself~\cite{Man1,Man2,Man3}.

In exactly scale-invariant theories, no energy scale is preferred since all are treated on equal footing.
So, in order to describe the appearance of physical energy scales (as observed at low enough energies), any phenomenologically viable scale-invariant quantum gravity theory has to exhibit the scale symmetry breaking in one way or another. For instance,  this  symmetry  can  be  explicitly  broken  by  mass  terms  or  via  dimensional  transmutation. The latter could happen either perturbatively through a Coleman--Weinberg mechanism or non-perturbatively (as it happens, for instance, by monopole condensation in quantum chromodynamics (QCD)). In these cases, the scale symmetry is typically quantum anomalous (by virtue of trace anomaly), and it is manifest only in the vicinity of non-trivial
fixed points~\cite{Tseytlin} of the renormalization group (RG) flow. Only in exceptional circumstances, such as an ${\cal N}=4$ supersymmetric Yang--Mills theory, one has an exact cancellation of this anomaly in the quantum field theory (QFT) framework.
It should perhaps be stressed that the trace anomaly is not a precursor of the spontaneous symmetry breaking (SSB) of scale-invariance, i.e., the situation where a vacuum expectation value (VEV) of some operator (order-parameter operator) supplies the needed dimensionful parameter. In fact, the symmetry that is broken in SSB is a global symmetry, not the local one (Elitzur's theorem~\cite{Elitzur}), and the trace anomaly is generally considered as harmless for global scale-invariance~\cite{Fradkin6}.
For instance, for quantum Weyl gravity (QWG) (or quantum conformal gravity),
a typical imprint of the trace anomaly in the scale-symmetric phase of QWG is the appearance of the $R^2$ (and Gauss--Bonnet $E$) term in the renormalized action even if this term is not implied by the local Weyl symmetry. Actually, the $R^2$ local term is  globally scale-invariant in spacetime with $d=4$ dimensions, but not locally, while the variation of the Gauss--Bonnet term in $d=4$ vanishes, so this term is invariant with respect to any symmetry transformation (both global and local). In addition, when the scale-invariance is spontaneously broken then the (conformal) Ward identities imply that the trace anomalies in the symmetric and broken phases are matched, though the analytical structure of the correlators is different in the two respective phases~\cite{Schwimmer}.

The inclusion of gravity in the scale-invariant framework offers far-reaching consequences. On the one hand, the breaking of the scale symmetry translates into the appearance of a pseudo-Goldstone boson (dilaton) which, due to its small mass, could potentially contribute to the early- and late-time acceleration of the Universe or to counting of the number of relativistic
degrees of freedom at Big Bang nucleosynthesis and recombination. \newline On the other hand,
the Standard Model of particle physics faces the problem of stability of the
Higgs mass against radiative corrections (the fine-tuning problem). If the full quantum theory,
including gravity, is scale-invariant, and the scale symmetry is spontaneously broken,
then the Standard-Model Higgs mass is protected from radiative corrections by an exact dilatational symmetry~\cite{Shap:2010}, cf. also~\cite{Nikolai:2007}.

Our modus operandi here will be based on the assumption that QWG can serve as a pertinent scale-invariant gravitational theory at high (pre-inflationary) energies. We will follow this premise, to address the ensuing low-energy phenomenology.
Let us first note that since the scale-invariance in the QWG is gauged (i.e., appears in the local version), the Weyl gravity is a very special theory and, in fact, unique among all higher derivative  gravity (HDG) theories. This theory possesses even bigger symmetry, namely conformal invariance, which naturally appears in the local form. That is why some authors~\cite{Tseytlin} prefer to call this theory as conformal gravity. Actually, in $d=4$, all HDG theories with four derivatives of the metric field are invariant under rigid scale
transformations. In a sense, the Weyl gravity is the simplest among HDG theories as it is determined in $d=4$ by a single (inevitably) dimensionless coupling parameter. This should be compared, e.g., with
two couplings present in a generic Stelle-type four-derivative theory~\cite{stelle1}. It should also be stressed that the Weyl gravity has a different counting of dynamical degrees
of freedom\footnote{\vspace{-2.0cm}\hypersetup{bookmarksdepth=-2}
\subsubsection*{\label{foot1}}
\hypersetup{bookmarksdepth}
For example, in $d=4$, there are $6$ degrees of freedom in contrast to $8$, which are typical for other HD gravitational theories with four derivatives.}$\hspace{-0.158cm}{}^{\ref{foot1}}$
  that cannot be obtained by any
limiting procedure from a generic HDG theory (e.g., by taking couplings in front of gauge-symmetry breaking terms to zero) due to the van~Dam--Veltman--Zakharov discontinuity~\cite{vDV} (see also~\cite{BD} and references
therein).

At the quantum level, Weyl gravity shares the same fate with all other higher derivative (HD) theories, namely the unitarity is in danger because there are perturbative states
with negative kinetic energy (ghost states~\cite{stelle1}). As a matter of fact, QWG has 6 perturbative degrees of freedom from which 2 are ghost degrees of freedom corresponding to the spin-2 massless particle.
The conventional optical theorem, in turn, implies that the $S$-matrix  is in such cases non-unitary. Obviously, ghosts are undesirable and
various approaches have been invoked to remove them (or their effects) from the observable predictions of the theory.
Diverse cures have been proposed in the literature for dealing with the ghosts issue: the Lee--Wick prescription~\cite{Lee}, fakeons~\cite{Anselmi},
the disappearance of unstable fluctuations in non-trivial backgrounds~\cite{Einhorn}, non-perturbative numerical methods~\cite{Tkach,Tomboulis,Kaku,Shapiro}, benign ghosts~\cite{Smilga2,Smilga,Narain:2016sgk},  non-local gravity~\cite{nonlrev,occurrence},
non-Hermitian quantum gravity~\cite{Man1,Man2,Man3}, etc. (see also~\cite{unita,unita2} and citations therein). One might even entertain the idea that unitarity in quantum gravity is not a fundamental concept~\cite{Hartle,Politzer,Lloyd}.
So far, none of the proposed solutions conclusively solves the problem.

It is quite possible that the unitarity problem might eventually be not as harmful as it seems. One appealing possibility for the resolution of  the unitarity issue
is Weinberg's Asymptotic Safety (AS) scenario~\cite{Weinberg}. The AS proposal/conjecture is based on the idea that quantum gravity develops a non-trivial fixed point (FP) of the
RG in the ultraviolet (UV) regime. Since the conjectured FP is non-Gaussian, then the couplings attain finite (possibly non-small) values at the end of the RG flow. Consequently, this proposal requires a non-perturbative analysis of the RG flow. Fortunately, an appropriate tool for this task --- the so-called functional
(or exact) renormalization group (FRG), has become available recently ~\cite{Reuter:1998,Wetterich:1993}. Though technically still involved,
there is a strong hope that when the non-trivial FP is found then the corresponding gravity theory will behave in a controllable way, and the FP will determine fully the non-perturbative spectrum, solve the
problem with unitarity, and tame the divergences since the RG flow stops at the FP~\cite{codello,codello_abc}. This type
of scenario (reinforced by a condition of a finite dimensionality of the critical surface on which assumed FP lies) is well-known from the QFT description of critical phenomena in condensed matter, where the
non-Gaussian FP provides a well-defined theory in the UV (or IR) regime~\cite{Niedermaier}. Moreover, when the theory sits at the FP,
then its symmetry is often enhanced and ensuing FP's are described by quantum scale-invariant or even conformal field theories (CFT).
This is one characteristic way how the exact conformal symmetry may show up on the quantum level.

Instead of debating various attitudes that can be taken toward the ghost--unitarity issue, our aim is more modest. We wish to explore, via FRG the low-energy phenomenology of the QWG and see whether it can provide a realistic cosmology and what r\^{o}le (if any) is played by ghost fields. The aim of this paper (first in a series of two) is twofold:

a) Let us assume that we start from the UV FP where the would-be quantum gravity has an exact scale-invariance. In order not to invoke any unwarranted structure, we consider only purely metric-field-based gravity without any matter field. The UV FP in question might be, for instance, one of the critical points in a series of hypothetical phase transitions that the Universe has undergone in the very early (pre-inflationary) stage of its evolution. Out of many scale-invariant HDG
candidate theories, we choose to work with the simplest one,
namely the one that has  only one coupling constant.  The latter corresponds to the quantum Weyl gravity. Precisely at the fixed point, the QWG has exact local scale-invariance. If the UV fixed point is non-Gaussian (we shall see it is not), it should be of a Banks--Zaks type~\cite{BZ} --- so that the perturbative analysis would still be applicable. Existence of such UV fixed point for QWG is only conjectured here  but there are various indications that this might be, indeed, the case~\cite{Codello,Percacci}. We start with this working hypothesis
and let the theory flow toward infrared (IR) energy scales. In the close vicinity of the UV FP, the Weyl symmetry in the renormalized Lagrangian is still preserved as only the Gauss--Bonnet term is generated in the process.
Corrections explicitly violating local scale symmetry, like the $R^2$ term, appear only
at the second or higher loop level~\cite{Tseytlin}. This is a consequence of the fact that local scale-invariance is preserved at the one-loop level, while local scale symmetry violating corrections are inevitably expected (though at higher-loop levels) due to a non-vanishing trace anomaly at the one-loop level. In this close neighborhood of the UV FP, we choose a truncation ansatz for the effective action that will be used to set up the FRG flow equation. In passing, we stress that we analyze RG flow in  the QWG  which has only dimensionless couplings; hence, most of the objections raised in the literature (cf., e.g., \cite{Donoghue:2019clr}) against the asymptotic safety program do not apply here.

b) In the next step, we will solve the ensuing RG flow equation algebraically for the two $\beta$-functions involved
and show that there exists
a non-Gaussian  IR fixed point where both $\beta$-functions simultaneously disappear. This IR FP represents a critical point after which the (global) scale-invariance is broken.
The fact that the scale-invariance gets broken is reflected via appearance of the related (composite) order-parameter field of the  Hubbard--Stratonovich (HS)
type which mediates a dynamical breakdown of the scale symmetry. In the broken phase, the order-parameter field acquires a non-trivial vacuum expectation value (VEV) via dimensional transmutation. This might in turn provide a key scalar degree of freedom needed, for instance, in various cosmologically feasible inflationary scenarios. The latter point will be explored in more detail in a subsequent paper.

The structure of the paper is as follows: to set the stage, we discuss in the next section some fundamentals of the QWG that will be needed in sections to follow. In particular, we outline the route to
quantization of Weyl gravity (WG) via functional integrals and stress some of potential problems encountered en route. We also emphasize a subtle fact that a non-dynamical spurion scalar field can be introduced in the QWG via
the HS transformation without spoiling the particle spectrum, (presumed non-perturbative) unitarity, and perturbative renormalizability. The HS field is actually an imprint of a scalar degree of freedom that would normally appear in the theory should the Weyl symmetry not decouple it
from  the  on-shell  spectrum. A second part of Section~\ref{sec1} is dedicated to the York field decomposition in Weyl gravity.
In Section~\ref{S.2.a} we employ this decomposition to construct the one-loop partition functions for the maximally symmetric spaces (MSS) and Ricci-flat backgrounds. We proceed in
Section~\ref{S.3.b} by constructing the FRG flow equation (in an Euclidean setting) for the QWG. To this end, we use the truncation prescription implied by the one-loop effective action.
We further enhance this  by including two non-perturbative effects, namely threshold phenomena and the effect of anomalous dimension of graviton field.
In Section~\ref{IV} we analyze the $\beta$-functions $\beta_C$ and $\beta_E$ that are affiliated with the Weyl tensor square and Gauss--Bonnet terms, respectively. In particular, we show that apart from a genuine IR fixed point that is reached at a zero-value of the running scale, the RG flow also exhibits bouncing behavior in the vicinity of the IR FP. We demonstrate that both $\beta_C$ and $\beta_E$
reach the RG bounce fixed point (almost) simultaneously at the same finite energy scale, irrespectively of the chosen background. The IR fixed point itself is shown to be IR-stable. Ensuing scaling dimensions for  the two operators are  also computed.
Finally, Section~\ref{SEc8} summarizes our results and discusses prospective implications for early Universe cosmology. For the reader's convenience, the paper is accompanied with
Supplemental Material (SM)~\cite{SM} that clarifies some technical and conceptual details needed in the main text.


\pdfbookmark[1]{II. Quantum Weyl Gravity}{name3}
\hypersetup{bookmarksdepth=-2}
\section{Quantum Weyl Gravity \label{sec1}}
\hypersetup{bookmarksdepth}



\pdfbookmark[2]{II. A Classical Weyl Gravity}{name4}
\hypersetup{bookmarksdepth=-2}
\subsection{Classical Weyl Gravity}
\hypersetup{bookmarksdepth}


The WG is a pure metric theory that is invariant not only under the action of
the diffeomorphism group, but also under Weyl rescaling of the metric tensor  by the
local smooth functions $\Omega(x)$: $g_{\mu\nu}(x)\rightarrow \Omega^2(x)g_{\mu\nu}(x)$.

The simplest WG action functional in four spacetime dimensions that is both diffeomorphism and
Weyl-invariant has the form~\cite{Weyl1,Bach1},
\begin{equation}
S \ = \ -\frac{1}{4\alpha^{\tiny{2}}}\int d^4x \ \! \sqrt{|g|} \ \! C_{\mu\nu\rho\sigma}
C^{\mu\nu\rho\sigma}\, ,
\label{PA1}
\end{equation}
where  $C_{\mu\nu\rho\sigma}$ is the
{\it Weyl tensor\/} which can be written as
\begin{eqnarray}
C_{\mu\nu\rho\sigma}  &=&
R_{\mu\nu\rho\sigma} \ - \ \left(g_{\mu[\rho}R_{\sigma]\nu} -
g_{\nu[\rho}R_{\sigma]\mu} \right)
\nonumber\\[1mm]
&+& \frac{1}{3}R \ \! g_{\mu[\rho}g_{\sigma]\nu}\, ,
\label{P2a}
\end{eqnarray}
with $R_{\mu\nu\rho\sigma}$ being the {Riemann curvature tensor},
$R_{\mu\rho}=R_{\mu\nu\rho}{}^\nu$  the  {Ricci tensor}, and $R= R_\mu{}^\mu$
the {scalar curvature}. Here and throughout the text, we use the time-like metric signature $(+,-,-,-)$  whenever  pseudo-Riemannian (Lorentzian) manifolds are considered. The dimensionless coupling constant $\alpha$ is conventionally chosen
so as to mimic the Yang--Mills action. On the other hand, in order to make a connection with the usual RG methodology,
it will be more convenient to consider the inverse of the coupling $\alpha^2$. 
We will denote the coupling in front of the Weyl square term in Eq.~(\ref{PA1}) as $\omega_C$
via the  identification $\omega_C \equiv 1/(4\alpha^2)$.

As for the notation for various scalar invariants (with four derivatives of the metric tensor), we accept the following conventions: for the square of
the Riemann tensor contracted naturally (preserved order of indices), that is $R_{\mu\nu\rho\sigma}R^{\mu\nu\rho\sigma}$, we use the symbol
$R^2_{\mu\nu\rho\sigma}$; the square of the Ricci tensor $R_{\mu\nu}R^{\mu\nu}$, we denote by simply $R_{\mu\nu}^2$; the square of the Ricci
curvature scalar is always $R^2$, while for the Weyl tensor square (with a natural contraction of indices) $C_{\mu\nu\rho\sigma}C^{\mu\nu\rho\sigma}$,
we employ a shorthand and schematic notation $C^2$. When the latter is treated as a local invariant (not under a volume integral, so without the possibility of
integrating by parts) in $d=4$ dimensions, one finds the following expansion of the $C^2$ invariant into standard invariants quadratic in curvature:
\begin{equation}
C^2 \ = \ R_{\mu\nu\rho\sigma}^2 \ - \ 2R_{\mu\nu}^2 \ + \ \frac{1}{3}R^2\,.
\label{weylsquare}
\end{equation}
Finally, we will also need yet another important combination of the quadratic curvature invariants, namely,
\begin{equation}
E \ = \ R_{\mu\nu\rho\sigma}^2 \ - \ 4R_{\mu\nu}^2 \ + \ R^2\, ,
\label{GBdef}
\end{equation}
which in $d=4$ is the integrand of the Euler (or Gauss--Bonnet) invariant~\cite{Percacci},
\begin{eqnarray}
\chi \ = \ \frac{1}{32\pi^2}\int d^4x \ \! \sqrt{|g|} \ \! E\, .
\end{eqnarray}
In the following, we will call the invariant $E$ in the action as a Gauss--Bonnet term.
With the help of the Chern--Gauss--Bonnet  theorem, one
can cast the Weyl action $S$ into an equivalent form (modulo topological term),
\begin{equation}
S \ = \
-\frac1{2\alpha^2}\int d^4x\ \! \sqrt{|g|} \ \!\left(R_{\mu\nu}^2 -
\frac{1}{3} R^2\right).
\label{PA2}
\end{equation}
It should be stressed that both (\ref{PA1}) and (\ref{PA2}) are Weyl-invariant only in $d=4$ dimensions.
In fact, under the (conformal or Weyl) transformation $g_{\mu \nu} \to \Omega^{2} {g}_{\mu \nu}$, the densitized square of the Weyl tensor
transforms as
\begin{eqnarray}
\sqrt{|g|} \ \! C^2 \ \to \ \Omega^{d-4} \sqrt{|g|}   C^2\, ,
\end{eqnarray}
in a general dimension $d$ of spacetime, while $\sqrt{|{g}|} \ \!E$  supplies topological invariant only in $d=4$ (the variation of this last term is a total derivative).
This is particularly important to bear in mind during the quantization where (similarly as in the Yang--Mills theories), one should  choose such a regularization method that preserves the local gauge symmetry of the underlying Lagrangian and thereby does not introduce any unwarranted symmetry breaking terms. For this reason,
one should preferentially rely on fixed-dimension renormalization schemes (as done throughout this paper) and avoid, e.g., dimensional regularization.

A variation of $S$ with respect to the metric
yields the field equation (Bach vacuum equation),
\begin{equation}
2\nabla_{\kappa} \nabla_{\lambda}C^{\mu\kappa\nu\lambda}_{\phantom{\lambda\mu\nu\kappa}}
-C^{\mu\kappa\nu\lambda}R_{\kappa\lambda} \ \equiv \ B^{\mu\nu}
\ = \  0\, ,
\label{Z42A}
\end{equation}
where $B^{\mu\nu}$  is the {\it Bach tensor\/} and $\nabla_{\alpha}$ is the usual
covariant derivative (with a Levi--Civita connection). We remind that this form of the equation of motion (EOM) is specific only to four spacetime dimensions. Moreover, the Bach tensor is always traceless ($B^\mu{}_\mu=0$) as a consequence of conformal symmetry and also is divergence-free ($\nabla_\mu B^{\mu\nu}=0$) as a consequence of diffeomorphism symmetry. One can also show that in $d=4$, one has $B^{\mu\nu}=B^{\nu\mu}$ as a consequence of being a variational derivative of the action $S$ with respect to  a symmetric metric tensor $g_{\mu\nu}$. When on a given background $B^{\mu\nu}=0$ (i.e., this configuration is a classical vacuum solution in Weyl gravity), then we say that it is Bach-flat.


\pdfbookmark[2]{II. B Quantization of Weyl Gravity}{name5}
\hypersetup{bookmarksdepth=-2}
\subsection{Quantization of Weyl Gravity}
\hypersetup{bookmarksdepth}


We formally define a quantum field theory of gravity
by a functional integral ($\hbar =  c = 1$),
\begin{equation}
Z \ = \ \sum_i \int_{\Sigma_i} {\cal D}g_{\mu\nu} \ \! e^{iS}\, .
\label{Z42Ab}
\end{equation}
Here,  ${\cal D}g_{\mu\nu}$ denotes the functional-integral measure
whose proper treatment involves  the Faddeev--Popov gauge
fixing of the gauge symmetry Diff$\times$Weyl$(\Sigma_i)$
plus the ensuing Faddeev--Popov determinant~\cite{Tseytlin}.
As for the local factors $[-\det g_{\mu \nu}(x)]^{\omega}$ in the measure, we
choose to work with the De Witt convention~\cite{Hamber}: $\omega = (d-4)(d+1)/8$. In this case, the
local factor does not contribute when the fixed-dimension renormalization in $d=4$ is employed.

The sum in (\ref{Z42Ab}) is a sum over four-topologies, that is, the sum over topologically distinct
manifolds $\Sigma_i$ (analogue to the sum over \emph{ genera}
in string theory or the sum over \emph{ homotopically} inequivalent vacua in the Yang--Mills theory), which
can potentially contain topological phase factors, e.g., the Euler--Poincar\'{e} characteristic
of $\Sigma_i$, cf. Refs.~\cite{Carlip}.
\footnote{\vspace{-2.0cm}\hypersetup{bookmarksdepth=-2}
\subsubsection*{\label{foot2}}
\hypersetup{bookmarksdepth}
The sum over four-topologies is a problematic concept since four-manifolds are generally un-classifiable
--- that is, there is no algorithm that can determine whether two arbitrary four-manifolds are homeomorphic.
On the other hand, simply connected compact topological four-manifolds are classifiable in terms of {\it Casson handles}
[as shown by M.H.~Freedman, Fields Medal (1986)]\,\cite{freedman}, which can be applied in functional integrals in Euclidean gravity.
In the Lorentzian case, one simply restricts oneself to some subset of four-manifolds. If this subset is closed under a composition of the functional
integral, then a theory thereby obtained is at least naively self-consistent.}$\hspace{-0.158cm}{}^{\ref{foot2}}$

For future convenience we note that the $R^2$-part in the Weyl action $S$ in (\ref{PA2}) can be further decomposed with the help of the Hubbard--Stratonovich (HS)
transformation~\cite{JKS,Hubbard,Stratonovich,Altland-Simons} as
\begin{eqnarray}
&&\mbox{\hspace{-11mm}}\exp(iS_{R}) \ \equiv \ \exp\left(\frac{i}{6 \alpha^2} \int d^4 x \ \! \sqrt{|g|} \ \! R^2\right)\nonumber \\[2mm]
&&\mbox{\hspace{-11mm}}= \ \int {\cal D} \phi \ \! \exp\left[-i \!\!\int d^4 x \ \! \sqrt{|g|} \ \!\left({\phi R} \ \! +  \ \! \frac{3}{8 \ \! \omega_C} \ \!\phi^2\right) \right]\!.
\label{PA2P}
\end{eqnarray}
It is not difficult to see that the essence of the HS transformation (\ref{PA2P}) is a
straightforward manipulation of a functional Gaussian integral (shifting the quadratic trinomial in the exponent).
Although an auxiliary HS field $\phi(x)$ does not have a
bare kinetic term, one might expect that due  to   radiative  corrections  it  will  develop  in
the IR regime a gradient term which will then allow us
to identify the HS boson with a genuine  propagating mode.
This scenario is, in fact, well-known from the condensed matter theory.
A quintessential example of this is obtained when
the BCS superconductivity is reduced to its low-energy
effective level. There, the HS boson coincides with the disordered
field whose dynamics is described via the celebrated Ginzburg--Landau
equation~\cite{GL,Altland-Simons}.

The $\phi$ field  can be separated into a background field $\langle{\phi}\rangle$ corresponding to a VEV of $\phi$ plus fluctuations $\delta\phi$. Since  $\langle{\phi}\rangle$ is dimensionful, it must be zero in the case when the
theory is scale-invariant. On the other hand, when the scale-invariance is broken, $\phi$ will develop a non-zero VEV. So, the HS field $\phi$ plays the role of the order-parameter field. The inner workings of this mechanism were illustrated in
Ref.~\cite{JKS}, where it was shown than on the flat background ${\phi}$ develops (in the broken phase) a non-zero VEV.
With the benefit of hindsight, we further introduce an {arbitrary} ``mixing'' hyperbolic~angle
$\vartheta~\!\!\in~\!\!(-\infty,\infty)$ and write formally
\begin{eqnarray}
S_{R} \ = \  S_{R} \cosh^2\! \vartheta  \ - \  S_{R} \sinh^2 \!\vartheta \, .
\end{eqnarray}
Applying now the HS transformation to the $S_{\rm R} \sinh^2 \!\vartheta $ part, we get
\begin{eqnarray}
S_{R} \ \! &=&\ \!
-\int \!d^4x\ \! \sqrt{|g|} \ \!
\phi R
\ \! + \ \! \frac{2\omega_C  \cosh^2\! \vartheta}{3}\!\int \!d^4x\ \! \sqrt{|g|} \ \!
 R^2
\nonumber \\[1mm]&&\!\!\!\!\!\!\!\!\hspace{0em}
+ \  \frac{3}{8 \omega_C  \sinh^2 \!\vartheta}\!\int\! d^4x\ \! \sqrt{|g|} \ \!
\phi^2\, .
\label{3rdac}\end{eqnarray}

Although the full theory described by the action $S$ is
independent of the mixing angle $\vartheta$, truncation of the
perturbation series after a finite loop order in the fluctuating
metric field will destroy this independence. The optimal result is
reached by employing the principle of minimal sensitivity~\cite{Stevenson,Kleinert-QFT}
known from the RG calculus.
There, if a perturbation theory depends on some
unphysical parameter (as  $\vartheta$ in our case), the best result is achieved if each
order has the weakest possible dependence on the parameter
$\vartheta$. Consequently, at each loop order, the value of $\vartheta$ is determined
from the vanishing of the corresponding derivative of effective action.

As discussed in~\cite{JKS}, the fluctuations of the metric $g_{\mu\nu}$
can make  $\langle{\phi}\rangle$  not only non-zero but one can also find a set of parameters in
a model's parameter space  for which $\langle{\phi}\rangle\sim M_{\rm P}^2$ where $M_{\rm P}= 2.44 \times 10^{18}$~GeV is related to the Planck energy scale. Consequently, Newton's constant $\kappa_{{_N}}$ is dynamically generated. Owing to the last term in (\ref{3rdac}), the existence of dynamical dark energy (a dynamical cosmological constant) is also an automatic consequence of the  theory. In addition,
by assuming that in the broken phase a cosmologically relevant metric is that of the Friedmann--Lema\^ i{}tre--Robertson--Walker (FLRW) type, then, modulo a topological term, the additional constraint,
\begin{eqnarray}
\int d^4x \sqrt{|g|}  \ \!  3 R_{\mu\nu}^2 = \ \int d^4x \sqrt{|g|}  \ \!  R^{2}\, ,
\end{eqnarray}
holds due to a conformal flatness of the FLRW metric~\cite{Fulling:74,exactsol}.
It was argued in \cite{JKS} that from (\ref{PA2}) and (\ref{3rdac}) one obtains in the broken phase the effective gravitational action of the form,
\begin{eqnarray}
S \ = \ - \frac{1}{2\kappa_{{_N}}^2}\int d^4x \sqrt{|g|}  \ \! (R  \ - \  \xi^2 R^2 \ - \ 2\Lambda_{\rm cc})\, ,
\label{star}
\end{eqnarray}
where both $\kappa_{{_N}}$ (Newton's constant) and  $\xi$
(Starobinsky's parameter) are dynamically generated. Note that $\xi$ has the dimension of an inverse mass and by the Planck satellite data $\xi/\kappa_{{_N}} \sim 10^{5}$ (cf. Ref.~\cite{Ade}). The action (\ref{star}) is nothing but the Starobinsky action with the cosmological constant.
We stress that the cosmological constant $\Lambda_{\rm cc}$ is entirely of a geometric origin (it descends from the QWG), and it enters with the opposite sign in comparison with the usual matter-sector induced (i.e., de Sitter) cosmological constant.

In the following sections, we will analyze in more detail the FP corresponding to the spontaneous symmetry breakdown of scale-invariance in QWG. To  reinforce our conclusions, we will use two non-trivial classes of Bach-flat backgrounds (i.e., solutions of classical Bach vacuum equation), namely MSS and Ricci-flat backgrounds.
\footnote{\vspace{-2.0cm}\hypersetup{bookmarksdepth=-2}
\subsubsection*{\label{foot3}}
\hypersetup{bookmarksdepth}
In Section \hyperref[sa]{A} of SM~\cite{SM}, we show that in $d=4$, all Einstein spaces (characterized by the condition that $R_{\mu\nu} = \Lambda g_{\mu\nu}$ with $\Lambda={\rm const}$), including both Ricci-flat and MSS manifolds as subcases, constitute an important class  of Bach-flat backgrounds.}$\hspace{-0.158cm}{}^{\ref{foot3}}$
The ensuing cosmological implications that are related to the broken phase of QWG will be discussed in the successive paper.

%
%

\pdfbookmark[2]{II. C York decomposition}{name6}
\hypersetup{bookmarksdepth=-2}
\subsection{York decomposition}
\hypersetup{bookmarksdepth}


To avoid issues related to the renormalization of non-physical sectors (i.e., Faddeev--Popov (FP) ghosts and longitudinal components of the metric field), it will be
convenient in our forthcoming reasonings to employ the York decomposition of the metric fluctuations
$h_{\mu\nu}$ defined as
\begin{eqnarray}
g_{\mu\nu}\ = \ g^{(0)}_{\mu\nu} \ + \ h_{\mu\nu}\,,
\end{eqnarray}
where we have denoted the background metric as $g^{(0)}_{\mu\nu}$. The York decomposition is
then implemented in two steps~\cite{Percacci}. In the first step (in $d=4$ spacetime dimensions),
one rewrites the metric fluctuations as
\begin{eqnarray}
h_{\mu\nu}\ = \ \bar{h}_{\mu\nu} \ + \ \frac{1}{4}g_{\mu\nu}h\, ,
\end{eqnarray}
where $h$ is a trace part of $h_{\mu\nu}$ and  $\bar{h}_{\mu\nu}$ is the corresponding  traceless part.
More specifically,
\begin{eqnarray}
g^{(0)\mu\nu}\bar{h}_{\mu\nu} &=&  \bar{h}_{\mu}{}^{\mu}\   = \  0\, , \nonumber \\[1mm]\quad
h &=&  g^{(0)\mu\nu}h_{\mu\nu} \  = \  h_{\mu}{}^{\mu}\, .
\end{eqnarray}
In our subsequent derivations, it always will be implicit that the Lorentz indices are raised or lowered
via a background metric, i.e., via $g^{(0)\mu\nu}$ or $g_{(0)\mu\nu}$, respectively.
Also, all covariant derivatives
$\nabla_\mu$ below will be understood as taken with respect to the background metric. By the symbol $\square$, we denote the covariant box operator defined as $\square\equiv\nabla^\mu\nabla_\mu$. This is the so-called Bochner Laplacian operator, it is defined with the covariant derivative $\nabla$ built on the basis of the Levi-Civita connection in metric theories. It is a standard two-derivative operator, and it arises naturally as a covariant generalization of the d'Alembertian operator in general relativity.

In the second step, one decomposes the traceless part into the transverse, traceless
tensor $\bar{h}_{\mu\nu}^\perp$  and to parts carrying the longitudinal (unphysical) degrees of freedom, namely,
\begin{eqnarray}
\bar{h}_{\mu\nu} &=& \bar{h}_{\mu\nu}^{\perp}\  + \ \nabla_{\mu}\eta_{\nu}^{\perp}\  + \ \nabla_{\nu}\eta_{\mu}^{\perp}\nonumber \\[1mm]
&&+ \  \nabla_{\mu}\nabla_{\nu}\sigma \  - \ \frac{1}{4}g_{\mu\nu}\square\sigma\, .
\label{Ydecomp}
\end{eqnarray}
These mixed-longitudinal (and traceless) parts are written in terms of an arbitrary transverse vector
field $\eta_\mu^\perp$ and a scalar (trace) degree of freedom $\sigma$. The last fields must
satisfy the usual conditions of transversality and  tracelessness, i.e.,
\begin{eqnarray}
\nabla^{\mu}\bar{h}_{\mu\nu}^{\perp}\ = \ 0,\quad\nabla^{\mu}\eta_{\mu}^{\perp} \ = \ 0,
\quad\bar{h}_{\,\,\mu}^{\perp\mu} \ = \ 0\, .
\end{eqnarray}
The true physical propagating field in QWG is the transverse and traceless field $\bar{h}_{\mu\nu}^{\perp}\equiv h^{TT}_{\mu\nu}$. Indeed, from the second variation of the
Weyl action  expanded around a generic background, it can be seen that $\bar{h}_{\mu\nu}^{\perp}$ is the only field component that propagates on a quantum level.
The vector field $\eta_\mu^\perp$ and two scalar fields $h$ and $\sigma$ completely drop out from the expansion due to diffeomorphism and conformal invariance, respectively~\cite{Irakleidou}. This is true around any background but particular examples are given in formulas below (in Eqs. (\ref{svarmss}) and (\ref{svarric})). In this way, we do not have to consider neither trace nor longitudinal degrees of freedom, nor FP ghosts in quantum dynamics of the theory.

In addition, in Section~\ref{S.3.a} we show that when the proper change of the integration measure under the functional integral (\ref{Z42Ab}) is employed,
the fixings of gauges for both the diffeomorphism and conformal symmetry are done and the ensuing Faddeev--Popov determinant is taken into account, one indeed obtains
precisely 6 propagating degrees of freedom (around flat spacetime background) --- as expected in the QWG. This counting can further be bolstered
by performing canonical Hamiltonian analysis and by counting constraints and their character. The latter leads again to 6 degrees of freedom, but in this case, the
counting holds in any spacetime background~\cite{Irakleidou}. Furthermore, we will see that the inclusion of the effect of (perturbative) zero modes will not change
this counting, though it will be key for getting the correct expression for
the partition function of the theory. In particular, in order to make the expression for the  partition function non-singular and non-vanishing, zero modes
must be handled with care.


\pdfbookmark[1]{III. FRG flow equation for QWG}{name7}
\hypersetup{bookmarksdepth=-2}
\section{FRG flow equation for QWG\label{S.3.a}}
\hypersetup{bookmarksdepth}


In order to make a comparison with existent works
on the FRG in the gravity context, we will perform our subsequent computations in
an Euclidean setting, and  our analysis will be performed exclusively in $d=4$ Euclidean space dimensions.

By performing a Wick rotation from Minkowski space to Euclidean space, the question of the resulting metric signature arises. When one does, in a standard way, only the change of the time coordinate $t\to -it_E$ (where $t_E$ is the name of the first coordinate in the Euclidean characterization of space), the resulting signature of the metric of space, is completely negative, that is $(-,-,-,-)$. It seems natural to define the corresponding GR-covariant d'Alembert operator as $\square_E=-\nabla^\mu\nabla_\mu$, where the generalization to curved Euclidean space is done by using a Bochner Laplacian. However, in all formulas that follow, we find it more convenient to use the following definition in the Euclidean signature $\square=\nabla^\mu\nabla_\mu$. We also remark that this last operator $\square$, if analyzed on the flat space background, has a negative semi-definite spectrum. We will also use a definition of the covariant Euclidean box operator (covariant Laplacian) $\Delta=\square=-\square_E$, and this last operator in the Euclidean flat space case has a spectrum which is characterized by $-k^2$, the four-dimensional Euclidean negative square of a four-momentum vector $k_\mu$ of eigenmode. From now on, the signature of the metric in Euclidean space  will be taken to be $(+,+,+,+)$.

The aim of this section is to explore both the infrared and the ultraviolet behavior of the QWG by solving
the FRG flow equation~\cite{Wetterich:1993,Reuter:1998}
for the effective average action $\Gamma_k$, which reads
\begin{eqnarray}
\partial_t \Gamma_k \ = \ \frac{1}{2} \mbox{Tr} \left[\partial_t R_k (\Gamma^{(2)} + R_k)^{-1} \right].
\label{FRG_1a}
\end{eqnarray}
 The IR-cutoff $R_k$ suppresses the contribution of modes with small eigenvalues of the covariant Laplacian  (or some other suitable differential operator)
$-\Delta \ll k^2$,  while the factor $\partial_t R_k$ removes contributions from large eigenvalues of $-\Delta \gg k^2$.  In this way, the loop integrals are both IR- and UV-finite \cite{Wetterich2}. The second variational derivative $\Gamma^{(2)}$ depends on the background metric $g_{\mu \nu}^{(0)}$, which is the argument of the running effective action
$\Gamma_k$, while $k$ is the running energy (momentum) scale or the momentum of a mode in the Fourier space. Here, we also use that $\partial_t = k \partial_k$.

Ideally, Eq.~(\ref{FRG_1a}) would require calculation of the full resummed and RG-invariant effective action.
It is, however, difficult  to proceed analytically in this way,  so we content ourselves here with
the conventional procedure, according to which one should employ some well motivated ansatz for the effective action.
In particular, in order to evaluate the RHS of Eq.~(\ref{FRG_1a}), we employ here the
Euclidean effective action in the enhanced one-loop scheme. Namely, we will consider one-loop effective action in which also the effects of the anomalous dimension
and threshold phenomena are included. Ensuing truncation will thus go beyond the usual polynomial ansatz.
On the other hand, for the LHS of Eq.~(\ref{FRG_1a}), we project the flow on the subspace of the three invariants containing precisely four derivatives of the metric (Eqs. (\ref{weylsquare}), (\ref{GBdef}), and $R^2$ invariant).
The reason why we consider effective action on the RHS being different from the effective action on the LHS is dictated by technical convenience. Namely, the RHS acts as source for the RG flow, while the LHS
contains the desired structure of the effective action that is appropriate for the extraction of the $\beta$-functions.
For more details, see Section~\ref{S.3.b}.


\pdfbookmark[2]{III. A 1-loop partition functions for MSS and Ricci-flat backgrounds}{name8}
\hypersetup{bookmarksdepth=-2}
\subsection{1-loop partition functions for MSS and Ricci-flat backgrounds\label{S.2.a}}
\hypersetup{bookmarksdepth}


For technical convenience, we choose to work with two classes of backgrounds, namely {maximally symmetric spaces} (MSS) and Ricci-flat manifolds.
As we shall see, these backgrounds will provide complementary information on $\beta$-functions of the theory that will suffice
to determine respective $\beta$-functions algebraically.
To find corresponding effective actions and $\beta$-functions, we first compute related one-loop partition functions.

We remark here that the spectrum of the covariant box operator
on non-trivial backgrounds depends on the boundary
conditions put on this operator. In the Euclidean
setup, we typically assume that our backgrounds are compact
manifolds without boundaries. This defines what
operators and which boundary conditions we speak about
below when we take functional traces of such operators.
These boundary conditions correspond to the asymptotic flatness and fall-off conditions for the fluctuations, when the space is considered in the decompactification limit
and when it is Wick-rotated back to the Minkowskian
signature.

In order to proceed with the partition function computations, it is important to set up first a notation regarding the functional determinants and ensuing functional traces.
We denote the type of space in which these determinants (and related traces) are to be evaluated
by the subscript just after the symbol ``det'', while the superscript will always denote the standard power.
The corresponding ``internal traces'' in the matrix space of fluctuations (corresponding to the field-theoretical
number of degrees of freedom (d.o.f.) in such subspaces) will be denoted by the symbol ``tr''.
We stress that the functional traces (denoted by ``Tr'') as
used, for example, in the FRG flow equation (\ref{FRG_1a}), are different because they contain also the
integrations over the background spacetime.  There are a few subspaces of fluctuations in which we would like to consider our determinants.
They mainly depend on the spin of the fluctuations and whether they are transverse,
traceless, or completely unconstrained fields. We have a description of various fields in subscripts:
\begin{itemize}
\item $0$ -- spin-0 scalar field, with 1 d.o.f., that is ${\rm tr}_{0}\hat{\ide}=1$.
\item $1\!\perp\,\equiv1T$ -- spin-1 constrained vector field $v_{\mu}^{T}\equiv v_\mu^\perp$ to be transverse $\nabla^{\mu}v_{\mu}^	{T}=0$, with 3 d.o.f., that is ${\rm tr}_{1\perp}\hat{\ide}=3$.
\item $1$ -- spin-1 unconstrained vector field $v_{\mu}$, with 4 d.o.f., that is ${\rm tr}_{1}\hat{\ide}=4$.
\item $2\!\!\perp\,\equiv2TT$ -- spin-2 fully constrained tensor rank-2 symmetric field $h_{\mu\nu}^{TT}\equiv\bar h_{\mu\nu}^\perp$, with conditions to be
 traceless and transverse: $h^{TT}{}_\mu{}^\mu=\nabla^{\mu}h_{\mu\nu}^{TT}=0$, with 5 d.o.f., that is ${\rm tr}_{2\perp}\hat{\ide}=5$.
\item $2T$ -- spin-2 partially constrained tensor rank-2 symmetric field $h_{\mu\nu}^{T}\equiv \bar h_{\mu\nu}$, with a condition to be only traceless
$h^{T}{}_\mu{}^\mu=0$, with 9 d.o.f., that is ${\rm tr}_{2T}\hat{\ide}=9$.
\item $2$ -- spin-2 fully unconstrained tensor rank-2 symmetric field $h_{\mu\nu}$, with 10 d.o.f., that is ${\rm tr}_{2}\hat{\ide}=10$.
\end{itemize}
The reader should, in particular, notice the differences in the usage of ``T'' superscript for spin-1 and spin-2 fields. The above counting of d.o.f.'s in each subspace was, of course, specific, to $d=4$ dimensions. From this moment on, we will omit the superscript and subscript $(0)$  from the background metric tensors $g^{(0)\mu\nu}$ or $g_{(0)\mu\nu}$, respectively.

Being forearmed with the above notation, we can now discuss the  one-loop partition functions of the QWG on both aforementioned classes of spacetimes.\\


\hypersetup{bookmarksdepth=-2}
\pdfbookmark[3]{III. A a) Maximally symmetric spaces}{name9}
\hypersetup{bookmarksdepth}

{\em {a) Maximally symmetric spaces }}~---~
%
\hspace{-2mm} are defined so that the Riemann curvature tensor is fully expressible through the metric tensor,
\bea
R_{\mu\nu\rho\sigma}\ = \ \frac{\Lambda}{d-1}\left(g_{\mu\rho}g_{\nu\sigma}-g_{\mu\sigma}g_{\nu\rho}\right),
\eea
where $\Lambda$ is a (real) constant parameter.
This implies that MSS are spaces of constant curvature. They are moreover conformally flat, i.e., $C_{\mu\nu\rho\sigma}=0$; hence $C^2=0$.
The examples of such spaces in the Euclidean setting are spheres and hyperboloids, while in the Minkowskian
case, we can speak of de Sitter (dS) and anti-de Sitter (AdS) spacetimes.

The second variation around the MSS background of QWG in $d=4$ in terms of $h^{TT}_{\mu\nu}$ fluctuations can be written in the form,
\begin{eqnarray}
\delta^2 S = \!\!\int\! d^4x \sqrt{g} \ \! h^{TT}{}^{\mu\nu}\!\left(\hat\square-\frac{2}{3}\Lambda \hat{\ide}\right)\!\!\left(\hat\square-\frac{4}{3}\Lambda \hat{\ide}\right)\!h^{TT}_{\mu\nu}.\,\,\label{svarmss}
\end{eqnarray}
We write a hat over all differential operators (like $\nabla$ and $\square$) from here on to emphasize that they act in a suitable matrix space of
fluctuations. After taking into account the Jacobian for a change of variables:
$h_{\mu\nu} \mapsto \{h^{TT}_{\mu\nu}, \eta_{\mu}^{\perp}, h, \sigma\}$, together with the Faddeev--Popov (FP) determinant,
and gauge fixings (for Weyl and diffeomorphic invariance),
the functional integration gives the ``one-loop partition function'',
\begin{eqnarray}
\mbox{\hspace{-3mm}}\tilde Z_{{\rm 1-loop}}^{2}
=  \frac{{\rm det}_{1T}\left(\hat \square+\Lambda \hat{\ide}\right){\rm det}_{0}\left(\hat\square+\frac{4}{3}\Lambda
\hat{\ide}\right)}{{\rm det}_{2TT}\left(\hat\square-\frac{2}{3}\Lambda \hat{\ide}\right){\rm det}_{2TT}\left(\hat\square-\frac{4}{3}\Lambda \hat{\ide}\right)}\, .
\label{16ab}
\end{eqnarray}
This result is not yet correct as it \emph{does} include a contribution from zero modes. Since the zero modes in the determinants render the partition function ill-defined (either singular or vanishing),
they have to be excluded. In Section \hyperref[sb]{B} of SM~\cite{SM},
we show that when the zero modes are properly accounted for, Eq.~(\ref{16ab}) changes into
\begin{eqnarray}
\mbox{\hspace{-5mm}}Z_{{\rm 1-loop}}^{2} &=&\frac{\det^{2}_1\left(\hat\square+\Lambda\hat{\ide}\right)\det_{1}\left(\hat\square+\frac{1}{3}
\Lambda\hat{\ide}\right)}{\det_{2T}\left(\hat\square-\frac{2}{3}
\Lambda\hat{\ide}\right)\det_{2T}\left(\hat\square-\frac{4}{3}\Lambda\hat{\ide}\right)}\nonumber \\[1mm]
&&\times \  \frac{\det_{0}\left(\hat\square+\frac{4}{3}\Lambda\hat{\ide}\right)}{\det_{0}\left(\hat\square+2\Lambda\hat{\ide}\right)}
\, .
\label{17ab}
\end{eqnarray}
At this stage, it should be noted that both (\ref{16ab}) and (\ref{17ab}) imply 6 propagating degrees of freedom. This can be seen on the level of flat space
(obtained simply by setting $\Lambda=0$) as well as on any MSS background. For this, it is enough  to take the logarithm  in the formula (\ref{17ab}) of
both sides and use properties of small internal traces in each subspace, and finally, exploit the formula for the number of degrees of freedom~\cite{Tseytlin},
\begin{equation}
N_{\rm d.o.f.} \ = \ -\frac{\log Z^2}{{\rm Tr}\log\square}\, ,
\label{DOF}
\end{equation}
and mathematical identities,
\begin{eqnarray}
\log\det\!{}_X \hat\square & = & {\rm Tr}_X \log \hat\square\, ,\nonumber \\
{\rm Tr}_X\log \hat\square & = & {\rm tr}_X \hat{\ide} \cdot {\rm Tr}\log \square\, ,
\end{eqnarray}
valid on the flat spacetime for any subspace ${X\in\{0,1T,1,2TT,2T,2\}}$.\\

\hypersetup{bookmarksdepth=-2}
\pdfbookmark[3]{III. A b) Ricci-flat manifolds}{name10}
\hypersetup{bookmarksdepth}

{\em {b) Ricci-flat manifolds}}~---~
\hspace{-2mm} are defined so that ${R_{\mu\nu}=0}$. The second variation around a Ricci-flat background in terms of $h^{TT}_{\mu\nu}$ fluctuations reads
\begin{eqnarray}
\delta^2 S = \int\! d^4x \sqrt{g} \ \! h^{TT}{}^{\mu\nu}\left(\hat\square-2\hat C\right)^{\!2}\! h^{TT}_{\mu\nu} \, .\label{svarric}
\end{eqnarray}
Analogously as in the previous MSS spacetimes, we can now employ the Jacobian of the transformation to the variables used in the York decomposition, together with the FP determinant and fixing of local symmetries to obtain
\begin{eqnarray}
Z_{{\rm 1-loop}}^{2}=
\frac{{\rm det}^3_{1}\hat\square\,{\rm det}_{0}^2\hat\square}{{\rm det}^2_{2}\left(\hat\square-2\hat{C}\right)}\, . \label{Ricflatpartition}
\end{eqnarray}
In these spacetimes, there is no correcting contribution from zero modes. Operator $(\hat\square-2\hat{C})$ in Eq.~(\ref{svarric})
represents the Laplacian and the matrix of the Weyl tensor on the
background vector bundle acting on tensor fields such as
$h^{TT}_{\mu \nu}$, which are transverse and traceless. In order to correctly account for Lorentz indices,
one should consider the square of the operator (which, in fact, descends from the second variational derivative of the action with respect to $h^{TT}_{\mu \nu}$). In particular, we have an explicit expansion according to

\bea
&&\hspace{-10mm} h^{TT}{}^{\mu\nu}\left(\hat\square-2\hat C\right)^2h^{TT}_{\mu\nu}
\nonumber \\
&&\hspace{-10mm}= \ h^{TT}_{\mu\nu}\left(\frac{g^{\mu\rho}g^{
\nu\sigma}+g^{\mu\sigma}g^{\nu\rho}}{2}\ \!\hat\square-  2C^{\mu\rho\nu\sigma}\right)^{\!2} h^{TT}_{\rho\sigma}
\label{25bb}\,,
\eea
where also the matrix square in the pair of indices is understood, so, for instance,
$(C^{\mu\rho\nu\sigma})^2 = C^\mu{}_{(\alpha}{}^\nu{}_{\beta)} \ \! C^{\alpha \rho \beta \sigma}$.
We wrote (\ref{25bb}) in its expanded form to recall that the matrix multiplication must be performed with objects explicitly symmetric in the pair of indices $(\alpha,\beta)$ because they are understood at any time to act on symmetric fluctuation fields $h^{TT}_{\alpha\beta}$.
We stress that in our expression (\ref{Ricflatpartition}), we used the determinants (and corresponding traces) in spaces of  completely unconstrained fluctuations of spin-1 and spin-2. In their report on conformal supergravity, Fradkin and Tseytlin~\cite{Tseytlin} give the following expression for the partition function:
\begin{eqnarray}
Z_{{\rm 1-loop}}^{2}=\frac{{\rm det}^3_{1}\hat\square}{{\rm det}^2_{2T}\left(\hat\square-2\hat{C}\right)}\, .
\end{eqnarray}
To find a relation between the two expressions for the one-loop partition function,
one can derive a formula,
\begin{equation}
\det\!{}_{2}\left(\hat{\square}-2\hat{C}\right)=\det\!{}_{2T}\left(\hat{\square}-2\hat{C}\right)\cdot\det\!{}_{0}\hat{\square}\, ,
\label{Fradkin_correct}
\end{equation}
which is valid on any Ricci-flat background by using the trace-free property of the Weyl tensor $\hat C$ and of traceless perturbations $h_{\mu\nu}^T$. However,
we remark that the above relation has nothing to do with zero modes.
With the formula (\ref{Fradkin_correct}),  we find a full agreement of our expression (\ref{Ricflatpartition}) with the original one due to Fradkin and Tseytlin.
Finally, one can see that our expression (\ref{Ricflatpartition}) correctly predicts  6 perturbative degrees of freedom. This can be checked by setting $\hat C=0$ in
(\ref{Ricflatpartition}) and using the counting of d.o.f. as outlined in the formula (\ref{DOF}).

\pdfbookmark[2]{III. B RG flow on MSS and Ricci-flat backgrounds}{name11}
\hypersetup{bookmarksdepth=-2}
\subsection{RG flow on MSS and Ricci-flat backgrounds}\label{S.3.b}
\hypersetup{bookmarksdepth}


We can now go back to the discussion of the FRG flow equation for the QWG. It is
clear that for the RHS of the flow equation (\ref{FRG_1a}), we will use effective actions
(and ensuing second variations) extracted from the one-loop partition functions of QWG derived in the
preceding subsection. As for the LHS of Eq.~(\ref{FRG_1a}), we will employ the
truncation ansatz to define the $\beta$-functional of the theory described by $\partial_t\Gamma_{L,k}$ (an additional subscript ``$L$'' reminds that we work with the LHS of the flow equation). For the
action $\Gamma_L$, we choose the most general possible truncation ansatz with three invariants
containing precisely four derivatives of the metric tensor (Eqs. (\ref{weylsquare}), (\ref{GBdef}), and $R^2$ invariant). We stress that the action (\ref{PA1}) of
the QWG is a subcase of this truncation. The aforementioned ansatz is motivated by the structure of possible
perturbative UV-divergences, which can be met at the one-loop level in $d=4$ in a generic HDQG theory with
four metric derivatives. On the LHS of (\ref{FRG_1a}), the coefficients of respective terms do contain
an explicit dependence on the scale $k$.  In general, the $\beta$-functional of the theory has,
within our truncation ansatz, the form,
\begin{eqnarray}
\beta_{R}R^{2} \ + \ \beta_{C}C^{2} \ + \ \beta_{{E}}{E}\, ,
\label{beta_fnl}
\end{eqnarray}
where the $\beta$-functions are defined in a standard way as RG-time $t$ derivatives of the running coupling parameters $\omega_i=\omega_i(k)$. (We have in the logarithmic RG coordinate $t=\log k/k_0$: $\beta_i=\partial_t \omega_i=k\partial_k \omega_i$.) These couplings appear in front of the corresponding quadratic in curvature terms in the truncation ansatz $\Gamma_{L,k}$.
Let us now discuss the $\beta$-functional for the above two relevant classes of Bach-flat backgrounds.

\hypersetup{bookmarksdepth=-2}
\pdfbookmark[3]{III. B a) Maximally symmetric spaces}{name12}
\hypersetup{bookmarksdepth}

{\em {a) Maximally symmetric spaces}}~---~
%
\hspace{-3mm} We recalled that they are defined so that the Riemann tensor is
\bea
R_{\mu\nu\rho\sigma}\ = \ \frac{\Lambda}{d-1}\left(g_{\mu\rho}g_{\nu\sigma}-g_{\mu\sigma}g_{\nu\rho}\right).
\label{riemmss}
\eea
Particularly, in $d=4$, this gives $R_{\mu\nu\rho\sigma}^{2}=\frac{8}{3}\Lambda^{2}$,
$R_{\mu\nu}=\Lambda g_{\mu\nu}$, so $R_{\mu\nu}^2=4\Lambda^2$ and $R=4\Lambda$, so $R^2=16\Lambda^2$. With these relations, the Gauss--Bonnet term $E = \frac{8}{3}\Lambda^{2}$ and $C^{2}=0$.
The corresponding $\beta$-functional (\ref{beta_fnl}) evaluated on this background then takes the form,
\bea
&&\mbox{\hspace{-1.5cm}}\left.(\beta_{R}R^{2}\ + \ \beta_{C}C^{2} \ + \ \beta_{E}\ \! {E})\right|_{{\rm MSS}}\nonumber \\[1mm]
&&\mbox{\hspace{1.3cm}}= \ 16\Lambda^{2}\left(\beta_{R}\ + \ \frac{1}{6}\ \! \beta_{E}\right)\! .\label{bfnlmss}
\eea
Therefore, the only combination of $\beta$-functions that can be extracted in this case is $\beta_{R}+\frac{1}{6}\beta_{E}$.

\hypersetup{bookmarksdepth=-2}
\pdfbookmark[3]{III. B b) Ricci-flat manifolds}{name13}
\hypersetup{bookmarksdepth}

{\em {b) Ricci-flat manifolds}}~---~
%
\hspace{-3mm} They are defined so that $R_{\mu\nu}=0$, and hence $R=0$. This, in turn, implies that $R_{\mu\nu\rho\sigma}^{2}=C^{2}= E$.
The $\beta$-functional thus takes the form,
\bea
&&\mbox{\hspace{-1.5cm}}\left.(\beta_{R}R^{2} \ + \ \beta_{C}C^{2} \ + \ \beta_{E}\ \!{E})\right|_{R_{\mu\nu}=0}\nonumber \\[1mm]
&&\mbox{\hspace{1.5cm}} =\ \left(\beta_{C} \ + \ \beta_{E}\right)R_{\mu\nu\rho\sigma}^{2}\, .
\label{bfnlric}
\eea
The only combination of $\beta$-functions that can be obtained in this case is $\beta_{C}+\beta_{E}$.

The above two backgrounds are indeed Bach-flat in $d~\!=~\!4$; i.e., they are vacuum solutions of the Bach equation (\ref{Z42A}).
In fact, one can make an even stronger statement, namely that the two backgrounds are
vacuum solutions of the theory,
\bea
S_{4d} \ = \ \int\!d^{4}x\sqrt{g}\left(\alpha_{R}R^{2}\ + \ \alpha_{{\rm Ric}}R_{\mu\nu}^{2}\right)\! ,
\label{eom.3a}
\eea
for arbitrary values of the parameters $\alpha_{R}$ and $\alpha_{{\rm Ric}}$.
This is because when $\Lambda={\rm const}$ in Eq. (\ref{riemmss}) (including $\Lambda =0$ case, so Ricci-flat case)
then $\nabla_{\alpha}R_{\mu\nu}=0$ (so the Ricci tensor is covariantly constant)
and also $\nabla_{\alpha}R=0$ (so the Ricci scalar is covariantly constant too). Hence, in deriving the EOM
from (\ref{eom.3a}), one can concentrate only on terms containing
curvatures and no covariant derivatives. The actual proof can be found in Section \hyperref[sa]{A} of SM~\cite{SM}.\\


\pdfbookmark[2]{III. C FRG flow: some general considerations}{name14}
\hypersetup{bookmarksdepth=-2}
\subsection{FRG flow: some general considerations \label{sec3c}}
\hypersetup{bookmarksdepth}


The quadratized action, \emph{reproducing} precisely a one-loop partition function, in QWG takes the following general form:
\begin{equation}
S_{\rm quad}  \ = \!\int\! d^4x \sqrt{g}\sum_i \phi_i K_i \phi_i\,, 
\label{squad}
\end{equation}
which shows diagonality in the space of different field fluctuations $\phi_i$. Collectively, by $\phi_i$, we denote all various fluctuation fields (possible to choose
selectively out of the set $\{h_{\mu\nu}^{TT},h_{\mu\nu}^T,h_{\mu\nu},v_{\mu}^T,v_{\mu},\sigma\}$). The kinetic operators $K_i$ are read from the kernel of
determinants of the expression for the one-loop partition functions Eqs.~(\ref{17ab}) and (\ref{Ricflatpartition}).

We note a few things here. First is that these kernels $K_i$ are differential operators containing two, four, or six covariant derivatives with respect
to the background, in each specific case. In respective cases, these differential operators are shifted by some constant vector bundle endomorphism (proportional to the
$\Lambda$ parameter and to the identity matrix $\hat\ide$ in the case of MSS) or by the Weyl tensor $\hat C$ in the matrix sense (for Ricci-flat background). Secondly, the actions (\ref{squad}), each
for the case of specific backgrounds, reproduce exactly the one-loop partition functions according to the formula,
\begin{equation}
Z_{\rm 1-loop}^2 \ = \ \det\!{}^{-1}\left(\frac{\delta^2S_{\rm quad}}{\delta \phi_i^2}\right),\label{partfn}
\end{equation}
where we also used the fact that the second variational derivative (Hessian) is diagonal in the space of different fluctuations. The key point about this formula is that here
we do not make any fixing of gauge symmetries, no FP determinant is needed, and there is no Jacobian of change of variables. The determinant of the Hessian to get the partition
function is taken plainly without any complicacy related to gauge symmetries, in general. Here, we use only physical fields as the benefit of using York decomposition. Technically, the determinant in Eq.~(\ref{partfn}) is taken in the same way as if the fluctuation fields were scalars,
without any special symmetry, which would call for a modification of the functional integral prescription. Moreover, the Hessian as the operator here is clearly non-degenerate
and without zero modes, hence taking its functional determinant does not create any problem.

Finally, we comment on the issue of factors in the partition functions (\ref{17ab}) and (\ref{Ricflatpartition}) appearing both in  numerators and
denominators. In principle, for standard scalar particles we have a contribution to the partition function at the one-loop level only in denominators. However, as we will show below,
it does not pose any problem that we have operatorial factors also in the numerators, although an interpretation in terms of standard particles is missing here (they cannot be identified neither with phantoms,
nor ghost particles). The factors in numerators can be understood as effects of the presence of  local gauge symmetries in the system being preserved by the quantization process. They can be interpreted as quantum account of constraints since they effectively decrease the number of d.o.f. of the theory. Additionally, if these factors in the numerators are considered separately, then the one-loop partition function as the generalized Gaussian integral of
the operator, is not convergent even  in the Euclidean setting.  We emphasize that in the form of the partition functions, as found in Eqs.~(\ref{17ab}) and (\ref{Ricflatpartition}),
we do not see any explicit dependence on the Weyl coupling $\omega_C$; hence, this will not show up anywhere in the action (\ref{squad}) nor in the RHS of the FRG
flow equation (\ref{FRG_1a}). Quite generally, any overall factors (and in particular their signs) of terms in the quadratized action (\ref{squad}) are irrelevant for
taking the Hessian and the ensuing functional determinants.

We factorize each of the kernel factors $K_i$ in (\ref{squad}) to monomials containing precisely two derivatives, so to monomials containing only one power of the
box operator $\hat\square$ in a corresponding representation. This we make according to the formula valid for any $i$,
\begin{equation}
K_i=\prod_j \left(\hat\square-Y_{i,j}\right)^{\pm1}\,.
\end{equation}
One can then rewrite the quadratized action (\ref{squad}) in the form,
\begin{equation}
S_{\rm quad}  = \!\int\! d^4x \sqrt{g}\sum_i \sum_j\phi_i \left(\hat\square-Y_{i,j}\right)^{\pm 1} \phi_i\,,
\label{squad2}
\end{equation}
where the shifts $Y_{i,j}$ are of the general form as explained above. The power exponents $\pm1$ should be chosen according to whether the factor is to be placed in the denominator or
in the numerator of the partition function in Eqs.~(\ref{17ab}) and (\ref{Ricflatpartition}). One can convince oneself that this form of the action reproduces again the correct form of the one-loop partition functions on
each background, but now the advantage is that all kinetic operators carry only two derivatives.

Furthermore, we now discuss the issue of the scale dependent wave-function renormalization for all fields
involved in the construction of the partition functions (so also appearing in their generating actions (\ref{squad2})).  The wave-function renormalization is obtained via the following transformation:
\begin{equation}
\phi_i\to Z_{k,i}^{1/2}\phi_i\, .
\end{equation}
The renormalization factors $Z_{k,i}$  are in the case of gauge theory with one coupling $\omega_C$ strictly related since the quantum dynamics of fluctuations is
governed by the same action term (here, the action of QWG in (\ref{PA1})). Since these fields are in the same gauge symmetry multiplet, they have the same wave-function renormalization factor~$Z_{k}^{1/2}$.

In order to take a truncation ansatz for the running effective action $\Gamma_{R,k}$ on the RHS of the FRG flow equation (\ref{FRG_1a})
that faithfully reproduces RG effects (related in particular to anomalous dimensions of quantum fields), we must take into account the
wave-function renormalization effects on fluctuations. This amounts to substitution of renormalized fields into the generating action
(\ref{squad2}) as the arguments. Therefore, we take the following RG-improved truncation
ansatz for $\Gamma_{R,k}$:

\begin{eqnarray}
\mbox{\hspace{-3mm}}\Gamma_{R,k}  &=& S_{\rm quad}\left[Z_k^{1/2}\phi_i\right]\nonumber\\[1mm]
&=& \int\! d^4x \sqrt{g}\sum_i \sum_j\phi_i Z_k\left(\hat\square-Y_{i,j}\right)^{\pm 1} \phi_i\,.
\end{eqnarray}
We construct the IR-cutoff action in a standard Wilsonian way. Namely, we use the one-loop partition function generating action (\ref{squad2}), and we modify each
quadratic in derivatives kinetic term by adding a suitably chosen IR-cutoff kernel function $R_k$. Consequently, we obtain the IR-cutoff action in the form,
\begin{equation}
\Delta S_{\rm IR } =\!\int\! d^4x \sqrt{g}\sum_i \sum_j\phi_i \left(\hat\square-Y_{i,j}+R_{k,i}\hat{\ide}\right)^{\pm 1} \phi_i\,.
\end{equation}
The role of the IR-cutoff kernel $R_{k,i}$ is to suppress the contribution to the functional integral of the field modes corresponding to eigenvalues $\lambda_n$ smaller
than the cutoff scale $k^2$ (the so-called low energy modes).
This cutoff kernel is a function of the operator, which describes the dynamics of modes. We will start with
the cutoff kernel $R_{k,i}=R_{k,i}(\hat\square)$ with the $\hat\square = \hat{\Delta}$ operator. For technical convenience, we choose as a ``cutoff
profile'', the so-called Litim cutoff function (or optimized cutoff); see Eq. (\ref{litimcutoff}) from \cite{SM}. In principle, each different subspace of fluctuation fields $\phi_i$ might have
its own cutoff function $R_{k,i}$.
However, for simplicity, we will choose them to be identical and given by one universal
function $R_k$.
Let us also remark that we do
an IR-suppression of modes (in the Wilsonian spirit)  also for factors which appear in the numerator of the partition functions. This is allowed by the
versatility of the functional RG methods and the flexibility of the flow equation (\ref{FRG_1a}). Then the regularized kinetic operator for all modes is given by
\begin{equation}
Z_{k}\left(\hat{\square}+R_{k}(\square)\hat{\ide}+Y_{i,j}\right)_{\phi_i}.
\end{equation}

Now, we can write down the Hessian operator $\Gamma^{(2)}_R$, which is used in the RHS of the FRG flow equation (\ref{FRG_1a}).
Taking second variational derivatives results in stripping the fluctuations from both sides of the quadratized action (\ref{squad2}). With this simplification in mind, we can write schematically
\begin{equation}
\Gamma^{(2)}_R \ = \ {\prod_i}{_{_{\bigotimes}}}  \prod_j Z_k\left(\hat\square-Y_{i,j}\right)^{\pm 1}_{\phi_i}\,,
\label{hessian}
\end{equation}
where, in each subspace of fluctuations $\phi_i$, we distinguish the operator by putting a mark of the subspace in the subscript after it. Similarly, we find that the regularized Hessian is the operator $\Gamma_R^{(2)}+R_k$ in (\ref{FRG_1a}) that has a schematic representation,
\begin{equation}
\Gamma^{(2)}_R+R_k\ = \ {\prod_i}{_{_{\bigotimes}}}  \prod_j Z_k\left(\hat\square+R_{k}(\square)\hat{\ide}-Y_{i,j}\right)^{\pm 1}_{\phi_i}\,.
\end{equation}
Eventually, based on the above formula and Eq.~\eqref{FRG_1a}, the FRG flow equation
takes the form,
\begin{equation}
\partial_{t}\Gamma_{L,k} =\frac{1}{2}\sum_i \sum_j \pm{\rm Tr}_{\phi_i}\left(\frac{\left(\partial_{t}R_{k}-\eta R_{k}\right)\hat{\ide}}{\hat{\square}+R_{k}\hat{\ide}
-Y_{i,j}}\right),
\label{FRG_partition}
\end{equation}
where we have defined the anomalous dimension (identical for all fluctuation fields) as $\eta = \partial_t \log Z_{k}$. Moreover, the $\pm$ signs depend on what was in the exponent on the corresponding term in the Hessian (\ref{hessian}) or in other words, whether the factor was originally in the denominator or the numerator of the partition function, respectively. The ultimate correctness of the above steps that lead to the FRG flow equation (\ref{FRG_partition}), will be verified below by a number of independent checks.

Finally, we wish to discuss our choice of the minimal consistent ansatz for the effective action appearing on the LHS of the flow equation (\ref{FRG_partition}). This
object is denoted by $\Gamma_{L,k}$, and it contains explicit dependence on the scale $k$ through the overall running couplings. As explained in the formula (\ref{beta_fnl}), the
most general truncation for the LHS may contain three terms quadratic in curvatures, each leading to terms with four derivatives of the metric. However, we will project
our resulting RG flow onto a smaller subspace without the $R^2$ term. This is motivated by the fact that at the one-loop approximation in QWG, the $\beta$-function for the
$R^2$ term is exactly zero. In this perturbative scheme, one can show that when the proper care is taken and the conformal symmetry is preserved on the
quantum level of computation of UV-divergences (and hence, in the computation of the perturbative one-loop effective action), then the ensuing $R^2$ divergences are not generated
at all~\cite{Tseytlin}. This result can be seen as a (partial) self-protection of conformal symmetry on the quantum level, since the only acceptable UV-divergences are
absorbed by the conformally covariant counterterm $C^2$ and the topological one $E$. It is expected that the $R^2$ divergences will show up at the two-loop level.
Some partial computation in this direction was presented in~\cite{Fradkin6}; however, the results are not fully conclusive, and one might still entertain some hope that the conformal
symmetry is powerful enough to prevent appearing of such non-conformal $R^2$ divergences also at two loops or even at a higher loop level. The arguments against this hope
(apart from the aforementioned partial two-loop computation in~\cite{Fradkin6}) are mostly related to the issue of conformal anomaly (CA) \cite{Capper}. However, one can try to waive them pointing to the issue of the ambiguities of CA \cite{shapiro2}.

Already at the one-loop level, despite the need for covariantly and only conformally looking counterterms, the obstacle for full quantum conformality is the presence
of CA, which is there due to non-vanishing $\beta$-functions of the theory, namely $\beta_C\neq0$  and $\beta_E\neq0$. The theory is with divergences and there are
perturbative $\beta$-functions; hence, the CA is non-vanishing. And since in the QWG, the conformal symmetry is in the local (gauged) version, then the fact that CA
is non-zero spoils the conformal symmetry on a quantum level (in particular, it destroys conformal Ward identities) basically the same way like gauge anomalies spoil
gauge symmetries in Yang-Mills theories; therefore, these anomalies have to be avoided at all cost. For consistent local conformal theory on a quantum level we would need to
have full cancellation of all UV-divergences; hence, the theory should be UV-finite and hence CA-free. Till these days, only two classes of such
theories including quantum gravitation are known. First are superconformal anomaly-free theories obtained by Fradkin and Tseytlin in the ${\cal N}=4$ conformal
supergravity models. Second are recently found perturbatively UV-finite quantum gravitational theories~\cite{superrenfin,univfin,fingauge,finconfqg,nonlrev} considered as an extension of superrenormalizable
higher derivative theories.

The situation with QWG without supersymmetry, without other matter species, and without higher-curvature operators needed to give UV-finiteness in \cite{superrenfin}
is that probably the conformal symmetry is not strong enough to constrain the quantum dynamics at higher loop orders. And starting from two loops on the $R^2$ divergences
are generated, and the conformal symmetry is completely washed out by quantum corrections. We are sure in having conformal symmetry on the classical tree-level and also
partially on the one-loop level. Since couplings in the theory are asymptotically free (more on this in Section \ref{sec4b}) and the theory is weakly coupled at high energies, then the problem of the conformal anomaly is a problem for UV-completion of the theory.
Our take on the issue of CA is that the conformal symmetry is broken at low-enough energies, and this is closely related to the dynamical breakdown of scale-invariance in the IR sector of the theory.
To explore this point more, we will mainly focus on FRG flow in the IR sector.

The fact that $\beta_R=0$ at one-loop in QWG is quite miraculous, but on more general level, this suggests that
a natural scheme should be used,
for example, in non-perturbative FRG, in which $\beta_R$ is parametrically smaller than other non-perturbative $\beta$-functions of the system. It is expected that at the perturbative two-loop level, the $\beta_R$ is expressed through higher inverse powers of the Weyl coupling $\omega_C$ of the theory. Therefore, exploiting this hierarchy of $\beta$-functions we can assume the following ansatz for $\Gamma_{L,k}$ (cf. \cite{shapiro1}):
\begin{equation}
\Gamma_{L,k} \ = \ \int\!d^{4}x\sqrt{g}\left[\omega_{C}(k)C^{2}+\omega_{E}(k)E\right],
\label{33bcd}
\end{equation}
where we have neglected the $R^2$ term and its running coupling $\omega_R=\omega_R(k)$. The choice of this ansatz for $\Gamma_{L,k}$ will allow us to  read unambiguously two $\beta$-functions, according to the Eq.~(\ref{beta_fnl}). By adopting the two backgrounds discussed above (MSS and Ricci-flat) we can read off $\beta_E$ and $\beta_C$. This will be done in Section~\ref{IV}. One notices a difference in sign in front of the $\omega_C$ coupling in Eq.~(\ref{33bcd}) compared with Eq.~(\ref{PA1}) and conventions stipulated there in Section~\ref{sec1}. This is the effect of performing a Wick--rotation to the Euclidean signature, while both types of couplings $\omega_C$ and $\alpha$ are always required to be positive.

In order to explicitly compute traces involved in (\ref{FRG_partition}), we will employ the heat kernel
technique outlined in Supplemental Material (SM). In particular, we use the formula (\ref{83aa}) from \cite{SM},
\begin{equation}
{\rm Tr}f(\Delta) \ = \ \frac{1}{(4\pi)^{d/2}}\sum_{n=0}^{+\infty}Q_{\frac{d}{2}-n}[f]B_{2n}(\Delta),
\label{34cc}
\end{equation}
where we restrict ourselves to two cases: $n=2$ and $d=4$. This choice is due to the fact that we want to project the RHS of the FRG flow equation onto the subspace spanned by the four-derivative terms present in the truncation ansatz $\Gamma_{L,k}$ (\ref{33bcd}). Moreover, let us notice that the heat kernel coefficients $B_{2n}(\Delta)$ contain exactly only terms with $2n$ derivatives of the metric tensor  (cf. Section \hyperref[sc]{C} of SM~\cite{SM}). As the operators $\Delta$,
we take in each case $\Delta=\hat{\square}+\hat Y$,  which is a two-derivative operator, possibly shifted by some endomorphism $\hat Y$ of the internal vector bundle (it acts there as a matrix multiplication, not as a differential operator).
The general IR-regulated (Euclidean) propagator of modes will have the structure,
\begin{equation}
G_{k}(z) \ = \ \frac{1}{z+R_{k}(z) + \varpi}\,,
\end{equation}
where we identify $z\equiv\hat\square$ as the main argument here and the shifts $\varpi$ (acting effectively like masses) are identified as $\varpi \ = \ Y_{i,j}$.

\vspace{1cm}


\hypersetup{bookmarksdepth=-2}
\pdfbookmark[1]{IV. Analysis of $\beta$\textunderscore C and $\beta$\textunderscore E functions}{name15}
\section{Analysis of $\beta_C$ and $\beta_E$ functions}\label{IV}
\hypersetup{bookmarksdepth}



\pdfbookmark[2]{IV. A System of two $\beta$-functions}{name16}
\hypersetup{bookmarksdepth=-2}
\subsection{System of two $\beta$-functions}
\hypersetup{bookmarksdepth}


Let  us now use  the enhanced one-loop relations for two $\beta$-functions $\beta_C$ and $\beta_E$  (implied by the ansatz (\ref{33bcd}))  in the form,
\begin{widetext}
\begin{eqnarray}
\mbox{\hspace{-10mm}}\beta_{E} &=& \frac{1}{2}(2-\eta)\left[-\frac{21}{40}\left(1-\frac{\frac{2}{3}\Lambda}{k^{2}}\right)^{-1}
\ + \ \frac{9}{40}\left(1-\frac{\frac{4}{3}\Lambda}{k^{2}}\right)^{-1} \ - \ \frac{179}{45}\left(1+\frac{\Lambda}{k^{2}}\right)^{-1}\ -\ \frac{59}{90}\left(1+\frac{\frac{1}{3}\Lambda}{k^{2}}\right)^{-1}\right.\nonumber \\[1mm]
 &+& \left. \ \frac{479}{360}\left(1+\frac{2\Lambda}{k^{2}}\right)^{-1}\ - \ \frac{269}{360}\left(1+\frac{\frac{4}{3}\Lambda}{k^{2}}\right)^{-1}\right], \label{betasys1}
\end{eqnarray}
\end{widetext}
\begin{eqnarray}
\mbox{\hspace{-0mm}}\beta_{C} \ + \ \beta_{E} \ = \frac{2-\eta}{2}\frac{137}{60}\, ,
\label{betasys2}
\end{eqnarray}
with the anomalous dimension of the graviton field,
\begin{equation}
\eta \ = \ -\frac{1}{\omega_{C}}\ \!\beta_{C}\, .
\label{73abc}
\end{equation}
Here, $\omega_{C}=\omega_{C}(k)$ represents a running coupling parameter
in front of the $C^{2}$ term in the action (\ref{PA1}) --- the so-called Weyl coupling. Explicit derivation of the results (\ref{betasys1})-(\ref{73abc}), including computation of $\eta$ up to one loop,	 can be found in Sections~\hyperref[sd]{D} and \hyperref[se]{E} of SM~\cite{SM}.

The above two $\beta$-functions $\beta_{C}$ and $\beta_{E}$ follow from the functional RG, and therefore, we can say that they are RG-improved because they include quantum effects related to both the threshold phenomena and non-trivial anomalous dimension of quantum fields. Let us recall that we have projected the full functional RG flow (\ref{FRG_1a}) onto a subspace of couplings consisting of more than just one Weyl coupling $\omega_C$. We have included also the induced effect on the running of the coupling $\omega_E$, while the effects  on $\omega_R$ were neglected due to the hierarchy in the system of $\beta$-functions, first found at the one-loop level approximation, but expected to hold also for higher loops (or even non-perturbatively). We notice that the effects of threshold phenomena show up explicitly only in the expression (\ref{betasys1}) for $\beta_E$. However, due to the relation in (\ref{betasys2}), the solution for $\beta_C$ will also contain the threshold factors. Finally, we observe that the effect of an anomalous dimension $\eta$ enters only multiplicatively in the system of $\beta$-functions (\ref{betasys1})-(\ref{betasys2}). This will have simplifying consequences when we will search for the FP's of the coupled system, both in the UV  as well as in the IR limit.

Let us now briefly discuss the reasons for the appearance of threshold phenomena in our system. As it is seen from the expressions for the one-loop partition functions Eqs.~(\ref{17ab}) and (\ref{Ricflatpartition}), the box-kinetic operator of modes is shifted in such a way as to produce IR thresholds, only in the case of MSS background. The shift by a matrix of a Weyl tensor $\hat C$ on the Ricci-flat background does not generate any threshold because the Weyl tensor is completely traceless in each pair of its indices. These shifts on MSS are analogous to massive modes in standard QFT analyzed on flat spacetime. They effectively slow down the RG flow in the IR regime because the quantum fields become heavy. As it is well-known on MSS (especially on anti-de Sitter (AdS) spacetime with a Minkowski signature) there exist bounds on the masses of ``healthy'' modes that can be considered in QFT. The dynamics of modes with mass square parameters smaller than the so-called Breitenlohner--Freedman (BF) bound put the modes in danger regarding the unitarity of the theory, basically the same way like tachyons endanger stability of flat spacetime QFT. These BF bounds depends naturally
 on the spin of modes and also on the $\Lambda$ parameter of the MSS spacetime; see Eq.~(\ref{riemmss}).
Moreover, reaching the BF bound corresponds to swapping all the modes from massive to massless; hence, some enhancement of symmetries of the theory might be expected. The BF bound is the boundary dividing healthy from unhealthy modes. In this connection, one can pose an interesting question, namely if in our expression (\ref{17ab}) for the one-loop partition function, we do not have effective masses which are below the BF bound. However, close inspection of (\ref{17ab}) reveals that we have modes in danger (with negative coefficient in front of $\Lambda$ parameter) only in the two factors belonging to the spin-2 traceless modes subsector. All other modes with scalar and vector characters are healthy. It is important to note that this discussion is purely academic here since for the issue of RG flow in the Euclidean signature the presence of modes below the BF bound will be completely inessential. Moreover, we would like to analyze the FRG flows on the energy scales ranging from UV regime ($k\to+\infty$) to IR ($k\to0$ in Euclidean), so for our purposes, this will be more than enough, and we do not have to worry here about violation of the BF bound and its physical effects.

One might be deluded by the superficial simplicity of the system of equations (\ref{betasys1})-(\ref{betasys2}) and think that by simple subtraction of (\ref{betasys1}) from (\ref{betasys2}), one gets already a full solution for $\beta_E$
and $\beta_C$. This is true only in a simple case, when the anomalous dimension $\eta$ is neglected. In general, however, one should pay attention to the dependence of $\eta$ on  $\omega_C$ and $\beta_C=\partial_t\omega_C$, which appears both in (\ref{betasys1}) and in (\ref{betasys2}). The system is still possible to be disentangled algebraically for the two $\beta$-functions. In passing we note that, as it is common with RG (global) systems, this is still a system of ordinary differential equations (ODE's) for running coupling parameters, here, respectively, for $\omega_C$ and $\omega_E$. We will not solve explicitly these ODE's for couplings here.
We will just concentrate on the FP's of this system.

By solving algebraically the above system of implicit equations, we obtain
\begin{equation}
\beta_{C} \ = \ \frac{b-X}{1+y(X-b)}, \;\;\;\; \beta_{E} \ = \ \frac{X}{1+y(X-b)}\, ,
\label{71ac}
\end{equation}
where
\begin{eqnarray}
\mbox{\hspace{-5mm}}X &=& - \ \frac{21}{40}\left(1-\frac{\frac{2}{3}\Lambda}{k^{2}}\right)^{-1} \ + \ \frac{9}{40}
\left(1-\frac{\frac{4}{3}\Lambda}{k^{2}}\right)^{-1}\nonumber \\[1mm]
&-&  \frac{179}{45}
\left(1+\frac{\Lambda}{k^{2}}\right)^{-1} \ - \ \frac{59}{90}\left(1+\frac{\frac{1}{3}\Lambda}{k^{2}}\right)^{-1}\nonumber \\[1mm]
&+ & \frac{479}{360}\left(1+\frac{2\Lambda}{k^{2}}\right)^{-1} \ - \
\frac{269}{360}\left(1+\frac{\frac{4}{3}\Lambda}{k^{2}}\right)^{-1},
\label{Xeqn}
\end{eqnarray}
with $b  =  {137}/{60}$ and $y={1}/{\omega_{C}}$. One can easily convince oneself that the origin of the common denominator $1+y(X-b)$ is entirely due to inclusion of the effect
of the anomalous dimension $\eta$. In the case when $\eta$ is neglected, the latter boils down to unity. Alternatively, this limit corresponds to taking a regime, in which the Weyl
coupling $\omega_C$ takes large values ($\omega_C\to\infty$, so $y\to0$), i.e., the regime in which the theory is very well described perturbatively (in terms of the coupling $\alpha$). When one decides to
neglect these common denominators, one is left with the simplified expressions of the form,
\begin{eqnarray}
\beta_{C} \ = \ b-X, \;\;\;\; \beta_{E} \ = X\, ,
\label{simplsys}
\end{eqnarray}
which as we will see shortly, are sufficient to cast light on the issue of existence of FP's of the FRG flow, and yet include the effect of threshold phenomena. Should we have removed the threshold phenomena from our description,
the system would acquire the form of one-loop perturbative $\beta$-functions as derived
in~\cite{Tseytlin} for QWG in a dimensional-regularization scheme, cf. Eqs.~(\ref{betae}) and (\ref{betac}) from SM~\cite{SM}. As a matter of fact, all the threshold phenomena are included in the expression called $X$.
When one takes the limit $\Lambda/k^2$ to zero,
then all threshold factors are removed, and this expression just reduces to a number $X=\beta_E^{\rm FT}={-87}/{20}$.
\begin{figure}[ht]
\vspace{-1.0cm}\hypersetup{bookmarksdepth=-2}
\subsubsection*{\label{fig.5}}
\hypersetup{bookmarksdepth}
\includegraphics[width=8.5cm]{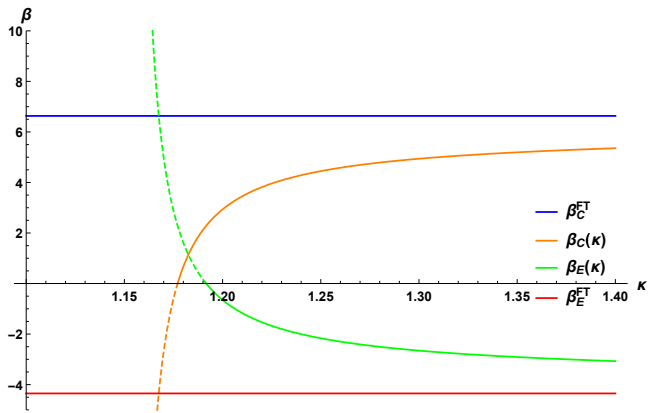}
\caption{FRG $\beta$-functions for the running couplings $\omega_C$ and $\omega_E$  in dependence on the dimensionless scale $\kappa = k/\sqrt{|\Lambda|}$. We depict the situation with $\Lambda >0$. The superscript ``{FT}'' denotes the Fradkin--Tseytlin $\beta$-functions~\cite{Tseytlin}.
Both $\beta_C(\kappa)$ and $\beta_E(\kappa)$
asymptotically approach $\beta_C^{\rm{FT}}$ and $\,\,\,\beta_E^{\rm{FT}}$, respectively, in the deep UV regime.  Running $\beta$-functions reach their
zero values in the IR region at approximately identical scales $\kappa$, cf. Eqs.~(\ref{76ab})-(\ref{77abc}). Note that at respective critical scales $\kappa_c$ both $\beta$-functions have finite values of the slope parameters $\partial_{\kappa}\beta(\kappa)|_{\kappa = \kappa_c}$.  Dashed lines are used to denote (unphysical) extensions of the running $\beta$-functions past the zero point. For simplicity's sake, we do not assume here any contribution from the anomalous dimension $\eta=0$ implying $y=0$.}
\label{fig.4}
\end{figure}


\pdfbookmark[2]{IV. B UV FP of the system of $\beta$-functions}{name17}
\hypersetup{bookmarksdepth=-2}
\subsection{UV FP of the system of $\beta$-functions}\label{sec4b}
\hypersetup{bookmarksdepth}


As already known from the seminal papers~\cite{Fradkin1, Fradkin2}, the system  of $\beta$-functions (\ref{betasys1}) and (\ref{betasys2}) reaches a trivial
Gaussian FP in the UV regime. When $\kappa = k/\sqrt{|\Lambda|} \gg 1$, the threshold effects are completely inessential and can be neglected, cf. Eqs.~(\ref{betasys1}) and (\ref{56aa}) from SM \cite{SM}.  Irrespectively of the initial values of the couplings $\omega_{C0}=\omega_C(t_0)$ and $\omega_{E0}=\omega_E(t_0)$,
the leading behaviors
of the RG running in the UV (for $t\gg1$) is $\omega_C\sim t \beta_C^{\rm FT}$ and $\omega_E\sim t \beta_E^{\rm FT}$. This means that the absolute values of the couplings
must necessarily grow in the UV regime (they decrease in the actual negative values of $\omega_E$ coupling).
In the same vein, it might be argued that in the UV regime one can also
neglect the anomalous dimension $\eta$, cf.~Eq. (\ref{73abc}).
Thus, for the UV-running, it suffices to use only the non-improved one-loop perturbation results (\ref{betae}) and (\ref{betac}) in SM \cite{SM}.

All these arguments are self-consistent and lead to the conclusion that the UV FP inevitably exists and realizes the asymptotic freedom (AF) scenario
(in much the same way as in non-Abelian gauge theories). Since our perturbation analysis is carried out in terms of the coupling $\alpha^2 \propto {1}/{\omega_C}$,
the fact that in the UV, $\omega_C\to+\infty$ bolsters even more the
correctness of our one-loop results. Actually, near the UV Gaussian FP, it is the coupling $\alpha$ that goes to zero.


\pdfbookmark[2]{IV. C IR FP's from the system of $\beta$-functions}{name18}
\hypersetup{bookmarksdepth=-2}
\subsection{IR FP's from the system of $\beta$-functions}
\label{Sec.4.ca}
\hypersetup{bookmarksdepth}


Let us now come back to the issue of IR FP's of the system. We will shortly see that the inclusion of threshold phenomena, which are present in any mass-dependent renormalization scheme, is of crucial importance in our analysis.
In fact, should we have studied only the simplified system of $\beta$-functions (\ref{betae}) and (\ref{betac}) in SM \cite{SM}, we would not find any interesting behavior of the RG flow in the IR (similarly to the case of QCD where the coupling grows stronger and gets out of the perturbative regime).
There are no any IR FP's in such a simplified scheme. To look for some less trivial behavior in the IR, we must thus include some additional non-perturbative effects.
Here, this feature is brought about by our usage of FRG and account of IR decoupling of massive modes.

In order to look for the FP's of the system in the IR, we must solve equations $\beta_C(k)=0\,$ and $\,\beta_E(k)=0$. Already here, one can see a huge simplification
because in order to find zeros, we do not need to solve the full system~(\ref{71ac}). Actually, we can completely forget the denominators and solve only Eqs.~(\ref{simplsys}), where factors $\Lambda/k^2$ are taken into account.
This signifies that the anomalous dimension $\eta$ does not influence the location of the IR FP's within the limits implied by our truncation of FRG.

Direct numerical solutions of the equations $\beta_{C}(\kappa) =  0$ and $\beta_{E}(\kappa) = 0$ reveal that they are both satisfied at (approximately) simultaneous
values of $\kappa={k}/{\sqrt{|\Lambda|}}$; see also Fig.~\hyperref[fig.5]{1}\,.
This is a smoking gun for the fixed point.  Actual numerical values for the critical energy scales $\kappa_c$,
at which the corresponding $\beta$-functions cross zero, are, respectively,
\begin{equation}
\kappa_{c,C} \ \approx \ 1.17709\quad{\rm and}\quad\kappa_{c,E}\ \approx \ 1.19163\, ,
\label{76ab}
\end{equation}
for the MSS with  $\Lambda>0$, and
\begin{equation}
\kappa_{c,C}\approx1.49722\quad{\rm and}\quad\kappa_{c,E}\approx1.52128\, ,
\label{77abc}
\end{equation}
for the MSS with $\Lambda<0$. The zeros (\ref{76ab}) and (\ref{77abc}) are automatically zeros of the system (\ref{71ac}).
One sees that the location of energy scales is almost identical (up to 2\% accuracy)
for the couplings $\omega_C$ and $\omega_E$ for both $\Lambda>0$ and $\Lambda<0$. We expect that the inclusion of higher
loop effects or extension of our truncation ansatz will make this discrepancy even smaller, such that in an exact fully non-perturbative theory,
the locations of two zeros coalesce into the one unique location of a genuine FP for both couplings.
Below we perform a general analysis of the situation near critical energy scale $\kappa_c$ for a general coupling $\omega$. The specification whether this is $\omega_C$ or $\omega_E$ will be important only for the numerical values that we quote at the end.

Now, from Fig.~\hyperref[fig.5]{1}\,, we see that the slope parameter at the zero-crossing $\partial_{\kappa}\beta(\kappa)|_{\kappa = \kappa_c} \equiv a$ is finite and positive for  $\beta_C$ and negative for $\beta_E$.
So, we can Taylor
expand the $\beta$-function $\beta(\kappa)$ around $\kappa_c$, so that
\begin{eqnarray}
\beta(\kappa) \ = \ {a} \ \!(\kappa-\kappa_c) \ +  a_2(\kappa-\kappa_c)^2 +\ \ldots \, .
\label{beta_f_1}
\end{eqnarray}
In what follows, we will concentrate only on the effects of the first term in the above Taylor expansion (with the coefficient $a$ of the first derivative), while the more refined analysis, which includes also the second derivative coefficient $a_2$ is presented in Section \hyperref[sf]{F} of SM~\cite{SM}.

It is easy to rewrite the expansion in Eq. (\ref{beta_f_1}) in terms of coupling $\omega$.
First, from (\ref{beta_f_1}), we get
\begin{eqnarray}
\mbox{\hspace{-3mm}}\omega(\kappa) - \omega_* \ = \ a\ \!(\kappa-\kappa_c) \ - \ a \kappa_c \log \frac{\kappa}{\kappa_c} \ + \ \ldots \, ,
\label{beta_f_3}
\end{eqnarray}
where $\omega_* = \omega(\kappa_c)$. Relation (\ref{beta_f_3}) can be inverted
so that we have
\begin{eqnarray}
\mbox{\hspace{-5mm}}\kappa  \ = \  - \kappa_c W\!\left(-  \exp\left[-\frac{\omega- \omega_*}{a\kappa_c} -1\right]\right)+\ \ldots\, ,
 \label{k2a}
\end{eqnarray}
where $W$ is a Lambert function. At this stage, we should recall that a Lambert function is a double-valued on the interval $(-1/e,0)$.
Since $\kappa/\kappa_c \geqslant 1$, the $W(\ldots) \leqslant -1$, and we should work with the lower branch of the Lambert function known
as $W_{-1}(\ldots)$. For small $(\omega- \omega_*)/(a\kappa_c)$ (which for both $\omega_C$ and $\omega_E$ is positive, cf. Eq.~(\ref{beta_f_3}), and actual numerical values quoted in Eqs. (\ref{numval1}) and (\ref{numval2}) below)
we can expand the RHS of (\ref{k2a}). This gives
\begin{eqnarray}
\kappa  & = &   - \kappa_c W_{-1}\!\left(-  \exp\left[-\frac{\omega- \omega_*}{a\kappa_c} -1\right]\right) \nonumber\\
&=& \ \kappa_c \ + \ \sqrt{\frac{2 \kappa_c}{a} \ \!(\omega- \omega_*)} \ + \ \frac{2}{3a}\ \!(\omega- \omega_*) \nonumber\\
&&+ \ \mathcal{O}((\omega- \omega_*)^{3/2})\, ,
\label{k2ab}
\end{eqnarray}
and so the $\beta$-function (\ref{beta_f_1}) reads
\begin{eqnarray}
\mbox{\hspace{-2mm}}\beta \ = \  \sqrt{{2a \kappa_c}\ \! (\omega- \omega_*)} \ + \ \frac{2}{3}\ \!(\omega- \omega_*) + \ldots
\label{k3ab}
\end{eqnarray}
Note that the result (\ref{k3ab}) holds true both for $\beta_C$ and $\beta_E$, since the product $a (\omega- \omega_*)\geqslant 0$ in both cases.

The fact that the $\beta$-function can develop at a finite RG scale $\kappa=\kappa_c$, a non-analytic behavior of the type (\ref{k3ab})
is well-known from holography, where it signalizes the presence of a multi-branch holographic RG flow that arises due to bounce solutions in the  bulk~\cite{Kiritsis:17,Nitti:18,Ghosh:18}.
In such cases, the corresponding $\kappa_{c}$ is merely a turning point on the way to a genuine IR fixed point. Despite that both $\beta$-functions $\beta_C$ and $\beta_E$ turn zero at $\kappa=\kappa_c$, this is not a true FP of RG flow of QFT because it happens at some finite scale $\kappa_c$. We remind that the RG flow stops when two conditions are met: all beta-functions vanish and the energy scale is $k=0$ (IR FP) or $k=+\infty$ (UV FP). Similarly, in the holographic (dual) perspective, the turning point at $\kappa=\kappa_c$ corresponds to a surface embedded in AdS-like $d=5$ geometry located at some finite radial coordinate $\rho_c\sim\kappa_c^{-1}$ . Since we know that for smaller energy scales $\kappa<\kappa_c$, we still have quantum degrees of freedom in the theory (i.e., they were not all integrated out), then this means that on the gravitational side, gravitational evolution of the dynamical $d=5$ spacetime must also continue past the bounce point with $\rho=\rho_c$ towards larger values of the AdS radial coordinate. The true IR FP shall correspond to a conformal boundary of AdS at infinite values of the radial AdS coordinate.

The gravitational bounce is an example of FLRW spacetime for which the cosmological scale factor exhibits a bounce behavior. On the bulk side, such behavior of spacetime is, of course, caused by some (exotic) matter source present \cite{hol1, hol2}. Typically the running of couplings in front of scalar operators is naturally described by dynamical bulk scalar fields with the particular mass parameter related to the scaling dimension of the operator on the boundary theory side, according to the AdS/CFT dictionary. When the $\beta$-function shows a non-analytic bouncing behavior like the one in Eq.~(\ref{k3ab}), then the similar behavior must also be exhibited by the corresponding bulk scalar field. This square-root-like singularity may mean that above the critical value $\rho_c$, the real profile of this bulk scalar field simply does not exist, or it becomes purely imaginary, which is however forbidden from the point of view of unitarity in QFT. On the other hand, the critical point can be interpreted as a joining point (or a bifurcation point depending on the direction of the flow) for two branches of solutions for the bulk scalar field. But here one sees that the gravitational spacetime easily extends beyond that critical radii surface with $\rho=\rho_c$, and one can still look for the true IR FP of the quantum system corresponding to the boundary of asymptotically AdS spacetime at $\rho\to+\infty$.

Let us note that this type of bouncing RG flow, i.e., flow that
displays one or more bounces before reaching the IR FP,  is quite
easy to encounter in a number of holography scenarios~\cite{Kiritsis:17}.
The bouncing RG flow of the above type
has been seen also in condensed-matter effective field theories~\cite{LeClair:04,Curtright:12}.

Using now the formula (\ref{k3ab}) for the expression of the $\beta$-function near
the turning FP,
we can analytically extend this behavior past the turning FP to the RG scale where $0\leqslant\kappa\leqslant \kappa_c$.
Since in this case $\kappa/\kappa_c \leqslant 1$, we should employ in (\ref{k2a}) the upper branch of the Lambert function, known as $W_0(\ldots)$.
The sole effect of this step is that $\kappa$ from (\ref{k2ab}) will be smoothly taken through the turning point with $\kappa=\kappa_c$ to $\kappa$
of the form,
\begin{widetext}
\begin{eqnarray}
\mbox{\hspace{-2mm}}\kappa  \ = \  - \kappa_c W_{0}\!\left(-  \exp\left[-\frac{\omega- \omega_*}{a\kappa_c} -1\right]\right) \ = \ \kappa_c \ - \ \sqrt{\frac{2 \kappa_c}{a} \ \!(\omega- \omega_*)} \ + \ \frac{2}{3a}\ \!(\omega- \omega_*) \ + \ \mathcal{O}((\omega- \omega_*)^{3/2})\, .
\label{k4abc}
\end{eqnarray}
\newpage
\end{widetext}
So, the key effect of the above analytical continuation is that
\begin{eqnarray}
\sqrt{\omega- \omega_*} \ \rightarrow \ - \sqrt{\omega- \omega_*}
\end{eqnarray}
(and such a flipping of a sign is present also for all terms with higher half-integer power exponents on $\omega-\omega_*$). This is a hallmark relation stemming from the square-root-like singularity of the RG flow and strictly related to holographic bounces in the bulk description.
From (\ref{k4abc}) treated as an exact relation,  we get the flow of the coupling $\omega$ in the region $0\leqslant \kappa \leqslant \kappa_c$, namely
\begin{equation}
\omega \ = \ \omega_{*} \ + \ \frac{9a}{2}\ \!\kappa_{c}^{1/3}\left[\kappa^{1/3}-\kappa_{c}^{1/3}\right]^2 .
\end{equation}
Taking this dependence  as being exact on the energy scales past the turning
FP, one can find the true FP occurring at $\kappa=0$ (i.e., in the deep IR).
This implies that the IR FP value of the coupling is
\begin{equation}
\omega_{**} \ = \ \omega(\kappa=0) \ = \ \omega_{*} \ + \ \frac{9}{2}a\kappa_{c}\, .
\label{omstar}
\end{equation}
Ensuing behavior of the $\beta$-function near $\omega_{**}$
is given by
\begin{equation}
\beta \ = \  \kappa\frac{d\omega}{d\kappa} \ = \ 3a\ \!(\kappa_{c}\kappa)^{1/3}\left[\kappa^{1/3} -\kappa_{c}^{1/3}\right].
\end{equation}

All information regarding the FP's can be extracted from the set of (critical) scaling exponents that are
defined in terms of the (negative) eigenvalues of the stability matrix at the FP, i.e.,
\begin{eqnarray}
\theta \ \in - \sigma\!\!\left.\left(\frac{\partial \beta_i}{\partial \omega_j} \right)\right|_{\omega = \omega_{**}}\, ,
\end{eqnarray}
where $\sigma$ denotes the corresponding spectrum and  $\beta_i = \{ \beta_C, \beta_E\}$, while $\omega_i = \{\omega_C, \omega_E \}$. In our case of the IR FP, the only
non-zero elements of the stability matrix are
\begin{equation}
\left.\frac{\partial\beta_C}{\partial\omega_C}\right|_{\omega=\omega_{**}} \ = \ - \sqrt{\frac{a\kappa_{c}}{2({\omega_{**}-\omega_{*}})}} \ + \ \frac{2}{3} \ = \ \frac{1}{3}\, ,
\end{equation}
and similarly,
\begin{equation}
\left.\frac{\partial\beta_E}{\partial\omega_E}\right|_{\omega=\omega_{**}} \ = \ \frac{1}{3}\, ,
\end{equation}
which implies that
\begin{equation}
\theta \ = \  \left\{ -\frac{1}{3}, -\frac{1}{3} \right\}\, .
\label{IR_stable}
\end{equation}
One refers to an IR FP as IR-stable, if all eigenvalues $\theta$ are negative, so,
from (\ref{IR_stable}), we can conclude that the FP in the IR is
completely stable in the space of all considered couplings.

Critical exponents allow for a precise definition of the (conformal) scaling dimensions of the operators
through the relation $\theta_i = d - \Delta_i$, where $\Delta_i$ is the scaling dimension
associated with a given operator. In our case, the operators $\sqrt{|g|}C^{2}$ and $\sqrt{|g|}E$ (and their related
couplings $\omega_{C}$ and $\omega_{E}$) are
classically (at tree-level) marginal with their canonical dimensions
$D_i = (\Delta_i)_{{\rm{cl}}} =4$. Due to quantum (loop) corrections, their quantum $\Delta_i \neq 4$ and the corresponding deviation from $4$, known as anomalous scaling, which is defined as $\gamma_i = \Delta_i - D_i$, equals $\gamma_i = - \theta_i = 1/3$. The latter implies that two involved operators become relevant operators near the IR FP (and also the two couplings $\omega_{C}$
and $\omega_{E}$ are IR-relevant).

For definiteness, we list below the numerical values of the products ${9}a\kappa_{c}/2$
that correspond to $\omega_{**}-\omega_{*}$ (in accordance with Eq. (\ref{omstar})). We also use a simplifying condition $y=0$. In particular,
for the case of $\Lambda>0$, we find
\begin{eqnarray}
\mbox{\hspace{-5mm}}\frac{9}{2}a_{C} \kappa_{c,C} \ \approx \ 1371.98\, , \;\;\;\;
\frac{9}{2}a_{E} \kappa_{c,E} \ \approx \ -511.201\, ,
\label{numval1}
\end{eqnarray}
while for the case of $\Lambda<0$, we have
\begin{eqnarray}
\mbox{\hspace{-5mm}}\frac{9}{2}a_{C} \kappa_{c,C} \ \approx \ 842.063\, , \;\;\;\;
\frac{9}{2}a_{E} \kappa_{c,E} \ \approx \ -490.829\, .
\label{numval2}
\end{eqnarray}

Let us recall that the value of the $\omega_{*}$ can be fixed by the initial conditions of RG flow; hence,
the same level of arbitrariness will be inherited in the IR FP values
$\omega_{**}$ of the $\omega$ couplings. It is also obvious that both couplings $\omega_{C}$ and $\omega_{E}$ at IR FP
are non-zero, and hence, the IR FP is non-Gaussian. This is, of course, very important conclusion because the Weyl gravity turns out to be a non-perturbative theory in IR along similar lines as QCD.

It should be borne in mind that the existence of the IR FP ought to be a universal property of the system independent from  particular details of the renormalization scheme used. In particular, as we emphasized above, the only crucial requirement from the renormalization scheme is its mass-dependence so the fact that the contributions from IR modes are properly secured.
In fact, it is always the case with computations done in the FRG framework that the precise locations of FP's do depend on characteristics of the renormalization process, but their existence or some other properties (related to critical exponents or the dimensionality of the critical surface) are universal, independent of gauge fixing choice, and renormalization details. Furthermore, in contrast to $\beta$-functions themselves  (that are not observable), the aforementioned properties constitute genuine observable pieces of information that can be extracted non-perturbatively from a theory, which reaches a non-trivial FP within the FRG framework.
At this point few comments are in order:

\begin{itemize}
  \item We can observe that the turning point of the RG flow  obtained above arises at the finite value of the running scale $k$, and not at $k=0$ (which is the conventional value for IR FP's). Moreover, the aforementioned value of $\kappa_c$  is background dependent (in our case, $\Lambda$-dependent). Both these points are easy to understand. The issue of the  finite value of $\kappa_c$ is related to non-analyticity near the turning FP, especially to the square-root-like singularity of the flow as seen in the Eqs. (\ref{k2ab}) and (\ref{k4abc}). Actually, one can prove that if there is a turning point of the RG flow and the non-analyticity is of the mentioned character, then this type of behavior near FP is only possible for finite non-zero $\kappa_c$. Conversely, if the FP happens at finite non-zero $\kappa_c$, then this is a turning point of the flow, and it must be continued analytically for $\kappa\leqslant \kappa_c$, if one looks for true deep IR FP. Then in such circumstances, the behavior near the turning FP can be characterized by any even order-root-like non-analyticity. Of course, the above found square root behavior is a paradigmatic example.  As for the background dependence, it should be stressed that only observable characteristics of the RG flow, such as a number of FP's or their type (scaling dimensions and related critical exponents, set of conformal primary operators,  operator product expansion coefficients, etc., or all of this as called CFT data) should be background independent. On the other hand, the actual shape of the RG flow trajectories is, in general, background dependent (see, e.g.,~\cite{Percacci,dep1,dep2,dep3,dep4,shapiro3,shapiro4}).

  \item It is evident from (\ref{simplsys}) that the IR FP obtained is entirely due to threshold phenomena, and the inclusion of the anomalous dimension $\eta$ (even with a supposedly exact non-perturbative expression for it) does not change the issue of the existence of this IR FP. Anomalous dimension is, however, responsible for the shape of the RG flow trajectory. Moreover, since threshold phenomena depend on a particular choice of the IR-cutoff kernel function $R_k(z)$, one might wonder how much such a choice influences the simultaneous running of $\beta_C$ and $\beta_E$ to FP values $\omega_{C**}$ and $\omega_{E**}$ in the IR. We accept the pragmatic  assumption that usual IR-cutoff kernels influence the observable quantities only minimally,  which is confirmed by almost all interesting examples.

\item The situation with IR FP in QWG may be compared to the decoupling of massive UV modes and the threshold phenomena, which occur for example in quantum electrodynamics (QED) due to finite-size mass of the electron -- the lightest charged particle. It is well-known that in QED, the running of the effective electric charge $e(k)$ is stopped in the IR at an energy scale around mass of the electron $k=m_e$ and with some finite value attained $e_{\rm cl}$, which we call classical (long-distance) coupling of the electron to classical electromagnetic field. For higher energy values, the corresponding $\beta$-function for the electric charge is positive. Hence, we understand that in the QED case, the $\beta$-function of the running electric charge $e(k)$ tends to zero in the IR (and its corresponding coupling to its limiting IR value $e_{\rm cl}$), actually never crossing the zero, and always keeping the positive sign. In any mass-dependent scheme, one sees that the $\beta$-function in QED attains a zero value in the IR as the effect of integrating out all modes of charged particles. And roughly below the mass of the electron $m_e$, there are no active quantum degrees of freedom, and this is the reason why the running effectively is slowed down to a full halt at $k=0$. But it would be incorrect to say that since $e(k=0)=e_{\rm cl}\neq0$, then in the IR, QED reaches a non-Gaussian FP. The stop of the RG running is due to exhaustion of all active modes and not due to some special structure of QFT of the Abelian $U(1)$ theory in the deep IR.
The situation in QWG is very different. First, the $\beta$-functions on its way from UV to IR must inevitably cross zero. This moment in the RG flow we identified with the turning FP. After the turning point when we use analytic continuation, the flow is continued towards deep IR, where the IR FP is found. However, its existence is not a virtue of only inclusion of threshold phenomena. A reason why in QWG we found an interesting IR FP can be traced back to the form of the partition function on MSS background Eq.~(\ref{17ab}) and a consistent decoupling of heavy UV modes in any mass-dependent renormalization scheme. However, it is not true that we run out of all active degrees of freedom in QWG at low energies. As we know, in the spectrum, we have only massless modes. The FP in the IR we found for non-trivial values of the couplings and this is due to special structure of QWG. Hence, this IR FP (with $\omega_{C**}$ and $\omega_{E**}$ generally not being zero) can be rightly called non-Gaussian, as opposed to the one in QED.

  \item Finally, one should remark that the existence of the IR FP is guaranteed for any value of the couplings $\omega_{E}$ and $\omega_{C}$.
  In other words, there are no initial values of $\omega_{E}$ and $\omega_{C}$ that would not run towards IR FP. This can be seen directly from the Eq.~(\ref{simplsys}) which is only $k$- but not explicitly $\omega$-dependent. Therefore, we do not find any constraint from which it could be possible to find some special values of the couplings $\omega_{E}^{*}$ and $\omega_{C}^{*}$ only for which the IR FP would occur.
\end{itemize}


\pdfbookmark[1]{V. Summary and Discussion}{name19}
\hypersetup{bookmarksdepth=-2}
\section{Summary and Discussion \label{SEc8}}
\hypersetup{bookmarksdepth}


In this paper, we show that Quantum Weyl  Gravity might provide a convenient theoretical setup for the UV-model building of phenomenologically viable quantum theory of gravity.
The present paper, the first of a series, concentrates on the existence and description of the fixed point that is responsible for the
spontaneous symmetry breakdown of the scale symmetry in the QWG. In particular, we
proceed from the hypothesis that the QWG correctly describes a physics in the vicinity of some UV fixed point.
This UV fixed point might correspond, for instance, to one of the critical
points in a series of phase transitions that the Universe has undergone in the very early
stage of its evolution. True UV-completion could be then achieved within  more fundamental theory, e.g.,  Berkowits--Witten  twistor-string theory or $\mathcal{N}=4$ conformal supergravity, which both harbor QWG in their low-energy limits (and do not have any pending unitarity issue).
The idea that the early stage of the Universe should be conformally invariant has been recently promoted by
R.~Penrose~\cite{penrose1,penrose2}, G.~'t~Hooft~\cite{hooft2}, and others~\cite{Shap:2013,Nikolai:2007,Amelino}.

In the next step, we have evolved the QWG from the presumed UV FP toward lower energies by using the functional renormalization group technique.
A novel feature of our RG analysis is that ensuing effective action ansatz goes beyond the conventional one-loop truncation through
inclusion of both the threshold phenomena and the effects of the anomalous dimension. With these, the FRG flow equation was evaluated
for two classes of Bach vacuum states, namely for the maximally symmetric spaces and Ricci-flat backgrounds.
The IR fixed point was found to be non-Gaussian and IR-stable in the space of considered couplings.
One might view this IR FP as being akin to recently studied asymptotically safe FP found in the gauge--matter--Yukawa system~\cite{Litim:2014}
 (but this time not in the UV but in the IR sector) or as a kind of gravitational analogue of
the Banks--Zaks FP known from Yang--Mills theories.
Though the two operators $\sqrt{|g|}C^{2}$ and $\sqrt{|g|}E$ are at tree-level marginal, quantum corrections
cause that both will become IR-relevant with ensuing anomalous scaling $\gamma_{C} = \gamma_{E} = 1/3$.
In addition, the logical consistency of this scheme requires the incipient UV FP to be Gaussian.

The aforesaid IR FP can be identified with the critical point at which the Weyl-invariance is spontaneously broken.
A hallmark  of  the  spontaneous  scale symmetry  breaking  is  the  existence  of  the  order-parameter field whose vacuum expectation value acquires a non-zero (dimensionful) value in the broken phase. For the case at hand, we have argued that the order-parameter field is a composite field of the Hubbard--Stratonovich type. As usual in SSB scenarios, a long-wavelength fluctuation of the latter should be identified in the broken phase with the Nambu--Goldstone mode (dilaton)~\cite{BJV,BJS}. For compatibility with an inflation-induced
large structure formation, the Weyl symmetry should be broken before (or during) inflation. So, in particular, if the presumed UV fixed point is close to the inflationary scale ($\sim 10^{15}-10^{16}$~GeV), then the asymptotic freedom in the vicinity of the UV FP  would guarantee that our RG description of the IR FP in terms of enhanced one-loop truncation is well justified. We should also stress that our RG treatment of QWG with the pre-inflationary infrared fixed point fits in a broader theoretical framework  of  the (super-)conformal inflation, which has been lately instrumental in classifying and generalizing classes of inflationary models favored by Planck data~\cite{Linde1,Linde2}.

The key observation in this context is that apart from the genuine IR FP (that is reached at zero-value of the running scale $k$), the RG flow also exhibits bouncing behavior in the vicinity of the IR FP.
In particular, both  the $\beta$-functions for $C^2$ term and Gauss--Bonnet term ($\beta$-function for the $R^2$ term is zero at our improved one-loop level) simultaneously reach the RG bounce fixed point at almost the same IR scale (up to 2\% accuracy) irrespectively of the background chosen. We noted that the observed square-root type RG bouncing can be mapped on a multi-branch (bouncing) holographic RG flow.
Although we expect that the inclusion of higher loop effects or extension of our truncation ansatz
will make the discrepancy between RG bounce FP's even smaller (in fact zero in an exact fully non-perturbative theory),
we did not present an explicit multi-loop computation confirming that this is the case, nor can we give a general proof.

There are still many questions to be understood. Here, is a partial list of them.
Our treatment is essentially based on the FRG with a particular one-loop enhanced effective action
and the Litim cutoff function (IR-cutoff kernel function $R_k$).
Though the Litim cutoff is the most conventional cutoff used in the FRG computations, one might ask how much is the structure of the IR FP
obtained (e.g., simultaneity of zeros of $\beta_{C}$ and $\beta_{E}$ and the values of $\omega_{C**}$ and $\omega_{E**}$) influenced by this
particular choice. The conventional wisdom in the FRG  posits that the structure and existence of FP's (but not the shape of the RG flow trajectories) should be
independent of the particular choice of a  cutoff function (provided it satisfies certain consistency conditions~\cite{Percacci}). This expectation has been
confirmed by a number of explicit computations in various systems~\cite{codello,Reuter_abc}.
On the other hand, any cutoff function represents an artificial term in the effective action, and every observable
becomes in one way or another cutoff dependent after the unavoidable truncations and approximations in the FRG calculations.
It might be thus interesting to make a comparison with other cutoffs on the market in order to see how robust is our prediction. To this end, one might use, for instance,  two-parameter cutoff functions of Nagy and N\'{a}ndori~\cite{Nagy:2013hka,Nandori} with parameters optimized
via principle of minimal sensitivity~\cite{MinSEn}, i.e., by requiring that the calculated observables depend least
on the cutoff kernel parameters. Another option would be to use (conformally or also gauge-) invariant cutoff kernel functions based on the proper time regularization of divergent integrals as this was suggested in~\cite{alfio}.

One can ask the question about the phenomenological implications of the considered here Quantum Weyl Gravity. This topic has been partially answered, and some applications to black hole physics (in particular to the issues of their formation and evaporation \cite{formation,bhevap} and the origin of their finite entanglement entropy \cite{entanglement}) were found themselves to be successful. Another theoretical problem is the relation between quantum conformal theories and UV-finite theories. Some works in this direction were already discussed in \cite{superrenfin,univfin,finconfqg,finads}, and the ensuing benefits of solving the issue of GR singularities were shown in \cite{spcompl0,spcompl}. Moreover, we note the existing comprehensive review on the various problems of conformal symmetry in QFT and gravity in \cite{confreview}.

It also remains to be seen to what extent our RG treatment of the QWG for the considered class of backgrounds is impeded by the presumed non-unitarity of QWG. Note, that unitarity issue was not apparently essential for our reasonings, at least not for the considered backgrounds and  given truncation ansatz. As already mentioned, the renormalizable QWG violates unitarity because it possesses
a spin-two ghost on flat background. This might well be an artifact of our ill-devised expansion around a wrong vacuum state, namely the flat spacetime. In a sense, the situation could be reminiscent of that known from the symmetry breaking model with
one real scalar field and a Higgs-like double-well potential $V(\Phi) =  \lambda(\Phi^2  - \mu^2 )^2$, with $\mu^2, \lambda >0$ and  with a tachyon in place of ghosts. We recall that the $S$-matrix unitarity means that the asymptotic ``in'' and ``out'' Fock spaces are unitarily equivalent. While the $S$-matrix is unitary for the scattering theories based upon ``true'' vacua $\Phi_0 =  \pm \mu$, this is certainly not the case when the tachyonic vacuum $\Phi_0 = 0$ is employed since incoming tachyonic Fock space states are not generally carried to the outgoing tachyonic Fock space~\cite{Jacobson}. So, an incorrectly chosen vacuum state alongside with an ensuing unstable tachyonic mode are ``culprits'' of non-unitarity.
Could a similar disappearance of unstable fluctuations in non-trivial backgrounds be in operation also in the QWG?

\section*{Acknowledgements}

It is a pleasure to acknowledge helpful conversations with
R.~Percacci, P.~Mannheim, I. Shapiro, K.~Stelle, and M.~Irakleidou.  P.J.  was  supported  by the Czech  Science  Foundation (GA\v{C}R), Grant No. 17-33812L. L.R. was supported from European Structural and Investment Fund (ESIF), EU Operational Programme Research, Development and Education No. CZ$.02.2.69/0.0/0.0/16_{-}027/0008465$. J.K. was supported by the Grant Agency of the Czech Technical University in Prague, Grant No. SGS19/183/OHK4/3T/14.

\begin{widetext}

%

\vspace{2mm}
%
\begin{center}

\hypersetup{bookmarksdepth=-2}
\pdfbookmark[1]{Supplemental Material}{name20}
\section*{\label{sm}}
\hypersetup{bookmarksdepth}

{\Large \bf Supplemental Material for ``Infrared behavior of Weyl Gravity: 
\\
\vspace{1mm} Functional Renormalization Group approach''}\\
\vspace{5mm}
Petr Jizba,${}^{1,\,*}$ \,Les\l{}aw Rachwa\l{},${}^{1,\,\dagger}$\, and\, Jaroslav K\v{n}ap${}^{1,\,\ddagger}$\\
\vspace{1mm}
{\it ${}^\mathit{1}$FNSPE, Czech Technical University in Prague, B\v{r}ehov\'{a} 7, 115 19 Praha 1, Czech Republic}
\end{center}

\vspace{6mm}

{\bf Note:} equations and citations that are related to the main text are shown in red.

\vspace{2mm}
\setcounter{equation}{0}
\setcounter{page}{1}
\thispagestyle{empty}

\pdfbookmark[2]{A. Selected exact solutions in Weyl Gravity}{name21}
\hypersetup{bookmarksdepth=-2}
\section*{A. Selected exact solutions in Weyl Gravity}
\hypersetup{bookmarksdepth}

\label{sa}

In this section we show that maximally symmetric spacetimes (MSS), Ricci-flat and Einstein spaces (ES) backgrounds used in  {\color{darkred} Sections } \hyperref[S.2.a]{\color{darkred} III A} {\color{darkred} and} \hyperref[S.3.b]{\color{darkred} III B} are all Bach-flat (in $d=4$) and, in addition, they are also solutions of classical equations of motion (EOM) associated with the action functional {\color{darkred}(\ref{eom.3a})} (also only in $d=4$). Since our derivation will be suitable for both Euclidean and Minkowskian signature, we will keep notation quite general, denoting the spacetime densities by $\sqrt{|g|}$. It is important to discuss Bach-flat spacetimes here since they being pure vacuum solutions in classical Weyl gravity (with no matter source), are good backgrounds to consider quantum perturbations around them. Moreover, the quantum partition function around such backgrounds can be taken in the Wentzel--Kramers--Brillouin (WKB) approximation form because there is no contribution from the first variation of the action {\color{darkred} (\ref{PA1})} evaluated on Bach-flat configurations and the first non-vanishing order is quadratic in quantum fluctuations.

The class of Einstein spaces is a set of such configurations of gravitational fields, that they satisfy
\begin{equation}
R_{\mu\nu}=\Lambda g_{\mu\nu}\,.\label{ESdef}
\end{equation}
We will show below that the $\Lambda$ parameter must be considered as constant. These spacetimes are vacuum background solutions of classical EOM of the Einstein's theory with a possible cosmological constant term described by the action $\int\!d^dx\sqrt{|g|}(\lambda+\omega_{\rm E-H}R$). There is a relation between $\lambda$ and $\Lambda$, but we will keep distinction between them here. Obviously, Ricci-flat spacetime are subclass for ES (for $\Lambda=0$), also MSS belong there, because from contraction of the relation {\color{darkred} (\ref{riemmss})} we get precisely (\ref{ESdef}). So far on ES there is no any restriction on the Weyl tensor and it can be non-zero (and then only in this way such an ES differs from the MSS with the same $\Lambda$). In what follows, we will assume the validity of the general relation (\ref{ESdef}), therefore, encompassing all three cases.

First, on any ES we can neglect the covariant derivatives of Ricci tensor or Ricci scalar due to
(\ref{ESdef}) and the implicit assumptions of metricity of spacetime (i.e. $\nabla_\rho g_{\mu\nu}=0$).
This simplifies a derivation of EOM a lot. We derive this fact and the constancy of $\Lambda$ by the
following argumentation. Let us consider the expression $\nabla^\mu R_{\mu\nu}-1/2\nabla_\nu R$ on general
ES. Obviously, we have it vanishing due to the second contracted Bianchi identity. On the other side,
one can use the formula (\ref{ESdef}) and the resulting $R=d\Lambda$ to write this explicitly
\begin{eqnarray}
0 &=& \nabla^\mu R_{\mu\nu} \ - \ \frac{1}{2}\nabla_\nu R \ = \ \nabla_\nu \Lambda \ + \ \Lambda \nabla^\mu g_{\mu\nu} \ - \ \frac{d}{2}\nabla_\nu \Lambda\,.
\end{eqnarray}
Now, due to metricity, we arrive at
\begin{equation}
(d-2)\nabla_\nu \Lambda \ = \ 0\,,
\label{ESLambda}
\end{equation}
which necessarily implies, that away from the special case of two dimensions $d=2$, $\nabla_\nu\Lambda=0$. This constancy immediately implies also the covariant constancy of the Ricci tensor and Ricci scalar, since in general $\nabla_\rho R_{\mu\nu}=g_{\mu\nu}\nabla_\rho\Lambda$ and $\nabla_\mu R=d\nabla_\mu \Lambda$. The reason for the exceptional case of $d=2$ dimensions is that in such number of dimensions Riemann tensor has only one algebraically independent component and therefore it as well as the Ricci tensor are completely expressed  through metric tensor and one curvature scalar, Ricci scalar $R$, that is in $d=2$ the relation $R_{\mu\nu}=\Lambda g_{\mu\nu}$  and hence $R=2\Lambda$ is valid everytime. This is general for any 2-dimensional curved manifold. But we know that there exist 2d manifolds which are with non-constant curvature, and that is why for them we must have the possibility of $\frac{1}{2}R=\Lambda=\Lambda(x)$ to be spacetime-dependent. Exactly, this case is not excluded in our proof, where for $d=2$ the Eq. (\ref{ESLambda}) is still satisfied, even if $\nabla_\nu \Lambda\neq0$.

Secondly,  we notice that when we derive the EOM from {\color{darkred} (\ref{eom.3a})} (with the purpose of analyzing vacuum solutions of the form of considered backgrounds),
we should vary only metric tensor used to raise indices and not a Ricci covariant tensor whose variation would give us only covariant derivative terms,
i.e., $\delta R_{\mu\nu}|_{{\rm no}\,\nabla}=0$. With this we can explicitly write
%
%
\bea
&&\delta\left(\sqrt{|g|}R_{\mu\nu}^{2}\right) \ = \ \sqrt{|g|}\left[\frac{1}{2}hR_{\mu\nu}^{2} \ + \ \delta\left(R_{\mu\nu}R^{\mu\nu}\right)\right]\ = \ \sqrt{|g|}\left[\frac{1}{2}hR_{\mu\nu}^{2} \ + \ \delta\left(g^{\mu\rho}g^{\nu\sigma}R_{\mu\nu}R_{\rho\sigma}\right)\right].
\eea
and so particularly, when we neglect gradient terms $\nabla_{\alpha}R_{\mu\nu}$ and $\nabla_{\alpha}R$ we get
\bea
&&\mbox{\hspace{-3mm}}\left.\sqrt{|g|}\left(\frac{1}{2}hR_{\mu\nu}^{2}\ + \ \delta\left(g^{\mu\rho}g^{\nu\sigma}R_{\mu\nu}R_{\rho\sigma}\right)\right)\right|_{{\rm no}\,\nabla}\ = \ \sqrt{|g|}\left[\frac{1}{2}hR_{\mu\nu}^{2} \ + \ \delta\left(g^{\mu\rho}g^{\nu\sigma}\right)R_{\mu\nu}R_{\rho\sigma}\right]\nonumber \\[1mm]
&&\mbox{\hspace{-3mm}}= \ \sqrt{|g|}\left(\frac{1}{2}hR_{\mu\nu}^{2}-2h^{\mu\nu}R_{\mu}{}^{\rho}R_{\nu\rho}\right)\ = \ \sqrt{|g|}h_{\alpha\beta}\left(\frac{1}{2}g^{\alpha\beta}R_{\mu\nu}^{2}\ - \ 2R^{\alpha\rho}R^{\beta}{}_{\rho}\right).
\eea
Here $h_{\alpha\beta} = \delta g_{\alpha\beta}$  and $h = g^{\alpha\beta}h_{\alpha\beta}$.
Hence for the $R_{\mu\nu}^{2}$--part of the action we get
\bea
\left.E^{\alpha\beta}\right|_{{\rm no}\,\nabla} \ = \ \left.\frac{1}{\sqrt{|g|}}\frac{\delta S}{\delta g_{\alpha\beta}}\right|_{{\rm no}\,\nabla} \ = \ \frac{1}{2}g^{\alpha\beta}R_{\mu\nu}^{2}-2R^{\alpha\rho}R^{\beta}{}_{\rho}\, ,
\eea
and this evaluated on MSS, Ricci-flat and ES backgrounds in $d=4$ gives
\begin{eqnarray}
&&\mbox{\hspace{-9mm}}\left.E^{\alpha\beta}\right|_{{\rm no}\,\nabla}
\ = \ \frac{1}{2}g^{\alpha\beta}\Lambda^{2}d\ - \ 2\Lambda^{2}g^{\alpha\beta} \ = \ \Lambda^{2}g^{\alpha\beta}\left(\frac{d}{2}\ - \ 2\right)
\ = \ \frac{d-4}{2}\Lambda^{2}g^{\alpha\beta} \ = \  0\, .
\end{eqnarray}
Similarly, we can write for the $R^{2}$-part of the action {\color{darkred} (\ref{eom.3a})}
\begin{eqnarray}
\mbox{\hspace{-5mm}}\delta\left(\sqrt{|g|}R^{2}\right)  &=& \sqrt{|g|}\left(\frac{1}{2}hR^{2}\ + \ 2R\delta R\right) \ = \  \sqrt{|g|}\left[\frac{1}{2}hR^{2} +  2R\delta\left(g^{\mu\nu}R_{\mu\nu}\right)\right]\! ,
\end{eqnarray}
and when we neglect gradient terms then
\begin{eqnarray}
&&\sqrt{|g|}\left.\left[\frac{1}{2}hR^{2}+2R\delta\left(g^{\mu\nu}R_{\mu\nu}\right)\right]\right|_{{\rm no}\,\nabla} \ = \ \sqrt{|g|}\left(\frac{1}{2}hR^{2}\ + \ 2R\delta g^{\mu\nu}R_{\mu\nu}\right)\nonumber \\[1mm]
&&= \ \sqrt{|g|}\left(\frac{1}{2}hR^{2}\ - \ 2Rh^{\mu\nu}R_{\mu\nu}\right)\ = \ \sqrt{|g|}h_{\alpha\beta}\left(\frac{1}{2}g^{\alpha\beta}R^{2}\ - \ 2RR^{\alpha\beta}\right).
\end{eqnarray}
Hence for the $R^{2}$--part of the action we obtain
\bea
\left.E^{\alpha\beta}\right|_{{\rm no}\,\nabla} \ = \ \left.\frac{1}{\sqrt{|g|}} \frac{\delta S}{\delta g_{\alpha\beta}}\right|_{{\rm no}\,\nabla} \ =  \ \frac{1}{2}g^{\alpha\beta}R^{2}\ - \ 2RR^{\alpha\beta}\, .
\eea
and this evaluated on MSS, Ricci-flat and ES backgrounds in $d=4$ gives
\begin{eqnarray}
&&\mbox{\hspace{-11mm}}\left.E^{\alpha\beta}\right|_{{\rm no}\,\nabla} \ = \ \frac{1}{2}g^{\alpha\beta}d^{2}\Lambda^{2}-2d\Lambda^{2}g^{\alpha\beta}\ = \ \frac{1}{2}g^{\alpha\beta}\Lambda^{2}\left(d^{2} -  4d\right)\ = \
\frac{d(d-4)}{2}\Lambda^{2}g^{\alpha\beta}\ = \ 0\, .
\end{eqnarray}
So, indeed, MSS, Ricci-flat and ES backgrounds are all solutions of EOM associated with the action functional {\color{darkred} (\ref{eom.3a})} and consequently they are also all Bach-flat in $d=4$ dimensions. The general ES are classical vacuum solutions of the theory {\color{darkred} (\ref{eom.3a})} in $d=4$ dimensions but the inclusion of the term quadratic in Riemann curvature tensor (of the type $\alpha_{\rm Riem}R_{\mu\nu\rho\sigma}^2$) does not change anything in this conclusion because of the usage of the Gauss--Bonnet theorem in $d=4$. In this last case we only have to redefine the couplings $\alpha_R$ and $\alpha_{\rm Ric}$ according to formulas $\alpha_R\to\alpha_R-\alpha_{\rm Riem}$ and $\alpha_{\rm Ric}\to\alpha_{\rm Ric}+4\alpha_{\rm Riem}$.

A possible solution for Riemann
tensor on a general ES can be
\bea
\mbox{\hspace{-3mm}}R_{\mu\nu\rho\sigma}\ = \ \frac{\Lambda}{d-1}\left(g_{\mu\rho}g_{\nu\sigma}\ - \ g_{\mu\sigma}g_{\nu\rho}\right)+\tilde C_{\mu\nu\rho\sigma}\,,
\eea
where $\tilde C_{\mu\nu\rho\sigma}$ is a completely trace-free arbitrary tensor playing
the role of the Weyl tensor with all its symmetries satisfied. Therefore, to completely characterize a general ES one must give one constant number $\Lambda$ and a valid form (of possibly spacetime-dependent) Weyl tensor in $d\geqslant4$.

On a general ES one can also analyze the form of the $\beta$-functional of the theory. First, we consider the curvature relations valid on any ES. One finds in $d=4$ that  $C^{2}  =  R_{\mu\nu\rho\sigma}^{2}  -  \frac{8}{3}\Lambda^{2}$ and ${E}  =  R_{\mu\nu\rho\sigma}^{2}  =  C^{2}  +  \frac{8}{3}\Lambda^{2}$, where we do not assume any particular form for the Riemann tensor beside the requirement of standard symmetries of the Riemann tensor. The ensuing $\beta$-functional reads
 \bea
 \left.(\beta_{R}R^{2}\ + \ \beta_{C}C^{2}\ + \ \beta_{E}\ \! {E})\right|_{{\rm ES}}\ &=& \ 16\beta_{R}\Lambda^{2}\ + \ \beta_{C}C^{2} \ + \ \beta_{E}\left(C^{2}\ + \ \frac{8}{3}\Lambda^{2}\right)\nonumber\\[1mm]
 &=& \ 16\Lambda^{2}\left(\beta_{R}\ + \ \frac{1}{6\ \!}\beta_{E}\right)\ + \ \left(\beta_{C} \ + \ \beta_{E}\right)C^{2}.
 \eea
It is clear that only two combinations of $\beta$-functions can be read off here, and they precisely coincide
with the two mentioned above in \hyperref[S.3.b]{{\color{darkred} Section III B}}, namely $\beta_{R}+\frac{1}{6}\beta_{E}$
and $\beta_{C}+\beta_{E}$. Using a general ES we read both these combinations of $\beta$-functions at one stroke. This was already suggested in \cite{Benedettis} for the case of higher derivative quantum gravity theories.

 We want also to comment on one more fact. Namely, if the Einstein
space is conformally flat, then it is MSS. To prove this assertion
one needs to notice only the decomposition theorem of the Weyl tensor.
In general dimension $d$ it has the following form
\bea
C_{\mu\nu\rho\sigma} \ = \ R_{\mu\nu\rho\sigma}-\frac{4}{d-2}g_{[\mu[\rho}R_{\sigma]\nu]}\ + \frac{2}{(d-1)(d-2)}g_{\mu[\rho}g_{\sigma]\nu}R\,.
\eea
If we solve this for the Riemann tensor we get obviously
\bea
R_{\mu\nu\rho\sigma} \ = \ C_{\mu\nu\rho\sigma}+\frac{4}{d-2}g_{[\mu[\rho}R_{\sigma]\nu]}\ - \ \frac{2}{(d-1)(d-2)}g_{\mu[\rho}g_{\sigma]\nu}R\,.
\eea
Now, in the case of conformal flatness we have $C_{\mu\nu\rho\sigma}=0$
and for ES $R_{\sigma\nu}=\Lambda g_{\sigma\nu}$ and $R=d\Lambda$,
so from the above formula for the Riemann we derive
\bea
R_{\mu\nu\rho\sigma} \ = \ \frac{4}{d-2}\Lambda g_{[\mu[\rho}g_{\sigma]\nu]}-\frac{2}{(d-1)(d-2)}d\Lambda g_{\mu[\rho}g_{\sigma]\nu}\ = \ \frac{2}{d-1}\Lambda g_{\mu[\rho}g_{\sigma]\nu}\,,
\eea
which is exactly the expression for the Riemann tensor of a MSS spacetime
in general dimension $d$ and with the radius square given by $\ell^{2}=\frac{d-1}{|\Lambda|}$.
Therefore we have proved that such background has to be maximally
symmetric. This confirms that indeed only the Weyl tensor quantifies
the difference how a given ES with particular value of $\Lambda$
parameter differs from a MSS with the same value of $\Lambda$. By
specifying fully the Weyl tensor $C_{\mu\nu\rho\sigma}$ and the value
of the $\Lambda$ parameter we characterize the ES completely.

Other relations worth noticing are described below.

If the space(time) is Riemann-flat, then its Riemann curvature tensor
vanishes ($R_{\mu\nu\rho\sigma}=0$), so this is a normal notion of
flatness. If the space(time) is Weyl-flat, then its Weyl tensor is
zero ($C_{\mu\nu\rho\sigma}=0$) -- the tensor of conformal curvature,
and the manifold is called conformally flat in dimensions $d>3$ (or
simply: conformal). The notions of Ricci-flatness and Bach-flatness
are natural since in the respective cases we require these corresponding
tensors to vanish. This is the origin of this nomenclature. The Ricci
and Bach tensors are special because these are the tensors whose vanishing
express vacuum gravitational equation in Einstein and Weyl gravitational
theories respectively.

We have some special relations. Every Weyl-flat space(time) is automatically
Bach-flat. In $d=4$ we have just proven that all ES are also Bach-flat.
The Riemann-flat case is a subcase to any of other flatness conditions
considered here. Ricci-flat and MSS are special cases of ES. MSS and
cosmological spacetimes (like Friedmann--Lema\^ i{}tre--Robertson--Walker (FLRW) spacetimes) are Weyl-flat. Ricci-flatness
and Weyl-flatness are quite mutually exclusive and their conjunction
implies Riemann-flatness. Weyl-flat ES must be necessarily a MSS as
proven above.


\hypersetup{bookmarksdepth=-2}
\pdfbookmark[2]{B. Contribution of zero modes needed in Section  III A}{name22}
\section*{B. Contribution of zero modes needed in \hyperref[S.2.a]{\color{darkred} Section  III A}}
\label{sb}
\hypersetup{bookmarksdepth}


Here we discuss the contribution of zero modes, which have to be excluded for the derivation of the one-loop partition function in a case of MSS background. One can envisage a method of singling out the contribution from them by analyzing partial differential equations (PDE's) for wave modes of various spins on MSS background with zero eigenvalues. Then by solving these PDE's (two-derivative with the covariant box operator) the explicit analytic solutions can be found and their counting made. Consequently, such contribution can be isolated. Physically, it is known that such zero modes on the curved manifold correspond to some isometries, which leave the physics untouched, like translations. They are also reasons for flat directions of various scalar potentials. There exists also a clear method to isolate the zero mode contribution on the level of path integral in quantum mechanics~\cite{Kleinert-books,KLMEMs,STKs,POLs,KLMGs}. However, here we will  do the counting of their contributions in a very simple field-theoretical setting on a curved MSS.

First, one can derive  for the action of the box operator ($\square$) between general arbitrary vectors
$A_\mu$ under the volume spacetime integral, the following formula
\begin{equation}
A^{\mu}\square A_{\mu}=A^{\mu}_T\square A_{\mu}^T-\phi(\square+\Lambda)\square\phi\,,
\label{zeromodevec}
\end{equation}
where we have used the York decomposition of the general vector field in a form (cf. with {\color{darkred} (\ref{Ydecomp})})
\begin{equation}
A_{\mu} \ = \ A_{\mu}^T+\nabla_{\mu}\phi\,,
\end{equation}
where $\phi$ is a scalar (longitudinal) degree of freedom. In the relation above (\ref{zeromodevec}) the volume integral is inexplicit, but this allowed us for using integration by parts under it.
Here we do not write hats over operators and matrices in the internal space since the character of these subspaces is obvious after looking at bilinears of fluctuations. The relation (\ref{zeromodevec}) is obtained after extensive commuting of covariant derivatives, doing integration by parts under the integral and exploiting the curvature relations valid on any MSS background.

We also need a relation for an action of a simple endomorphism operator $Y$ between the vector
fluctuations (again inexplicitly under volume integral). We find that
\begin{equation}
A^{\mu}YA_{\mu} \ = \ A^{\mu}_TYA_{\mu}^T-Y\phi\square\phi\,.\label{zeromodeend}
\end{equation}
Combining  (\ref{zeromodevec}) and (\ref{zeromodeend}) together we get
\begin{equation}
A^{\mu}(\square-Y)A_{\mu}=A^{\mu}_T(\square-Y)A_{\mu}^T-\phi(\square+\Lambda-Y)\square\phi\,.\label{zm1}
\end{equation}
This let us to derive the following relation between the determinants of various operators
(typically shifted box operator by some constant proportional to $\Lambda$) analyzed on MSS backgrounds
\begin{eqnarray}
\det{}'_{1}\left(\hat\square-Y\hat{\ide}\right)  \ = \ \det\!{}_{1T}\left(\hat\square-Y\hat{\ide}\right)\ \! \det\!{}_0\left(\hat\square-Y\hat{\ide}+\Lambda\hat{\ide}\right)\! ,\label{zm2}
\end{eqnarray}
where by ${\rm det}'$ we denote a determinant of the operator with the removed contribution
of zero modes. We see that from the formula (\ref{zm1}) to the one in (\ref{zm2}) we pass by
taking the determinants of the differential operators sandwiched between fluctuations and
multiplying them in total. We also neglect the overall front coefficients or signs of each terms.
The zero-mode contribution here is really in the scalar box operator $\square$ acting between scalar
fluctuations $\phi$. In general, the contributions of zero modes, which have to be neglected for the
derivation of partition function, are all determinants of operators, which explicitly depend on $Y$.
From here we derive the final expression as
\begin{equation}
\det\!{}_{1T}\left(\hat\square-Y\hat{\ide}\right)=\frac{\det\!'_1
\left(\hat\square-Y\hat{\ide}\right)}{\det_0\left(\hat\square-Y\hat{\ide}+\Lambda\hat{\ide}\right)}\,. \label{zm3}
\end{equation}
The very similar argumentation follows also in the case of spin-2 traceless fields $h_{\mu\nu}^T$, where we concentrate on computation and expansion of the following action of the operator $h_T^{\mu\nu}\square h_{\mu\nu}^T$. We use the form of the York decomposition from {\color{darkred} (\ref{Ydecomp})}. The corresponding essential formulas of the derivation in this case are listed below.
The expansion of the action of the box operator between such fluctuations is given by
\begin{eqnarray}
h_T^{\mu\nu}\square h_{\mu\nu}^T \ = \ h^{\mu\nu}_{TT}\square h_{\mu\nu}^{TT}-2\eta^{\nu}_T
\left(\square+\Lambda\right)\left(\square+\frac{5}{3}\Lambda\right)\eta_{\nu}^T
\ + \ \frac{3}{4}\sigma\square\left(\square+\frac{4}{3}\Lambda\right)\left(\square+\frac{8}{3}\Lambda\right)\sigma\,,
\end{eqnarray}
while the action of a general endomorphism $Y$ in the same setting reads
\begin{eqnarray}
h_T^{\mu\nu}Y h_{\mu\nu}^T \ = \ h^{\mu\nu}_{TT} Y h_{\mu\nu}^{TT}-2\eta^{\nu}_TY\left(\square+\Lambda\right)\eta_{\nu}^T\ + \ \frac{3}{4}\sigma Y\square\left(\square+\frac{4}{3}\Lambda\right)\!\sigma\, .
\end{eqnarray}
Combining the two formulas together we find
\begin{eqnarray}
h_T^{\mu\nu}\left(\square-Y\right) h_{\mu\nu}^T  &=&  h^{\mu\nu}_{TT}\left(\square-Y\right) h_{\mu\nu}^{TT}\ -\ 2\eta^{\nu}_T\left(\square+\Lambda\right)\left(\square+\frac{5}{3}\Lambda-Y\right)\eta_{\nu}^T\nonumber  \\ &+& \ \frac{3}{4}\sigma\square\left(\square+\frac{4}{3}\Lambda\right)\left(\square+\frac{8}{3}\Lambda-Y\right)\!\sigma\,.
\end{eqnarray}
Finally, we get for the determinants
\begin{eqnarray}
\mbox{\hspace{-8mm}}\det\!{}_{2TT}\left(\hat\square-Y\hat{\ide}\right) &=&\frac{\det'_{2T}\left(\hat\square-Y\hat{\ide}\right)}{\det_{1T}
\left(\hat\square-Y\hat{\ide}+\frac{5}{3}\Lambda\hat{\ide}\right)}\ \! \frac{1}{\det_{0}\left(\hat\square-Y\hat{\ide}+\frac{8}{3}\Lambda\hat{\ide}\right)}\, .
\label{zm4}
\end{eqnarray}

The candidate expression for the square of the partition function had the form {\color{darkred} (\ref{16ab})}, which we repeat here
\begin{equation}
 \frac{{\rm det}_{1T}\left(\hat \square+\Lambda \hat{\ide}\right){\rm det}_{0}\left(\hat\square+\frac{4}{3}\Lambda \hat{\ide}\right)}{{\rm det}_{2TT}\left(\hat\square-\frac{2}{3}\Lambda \hat{\ide}\right){\rm det}_{2TT}\left(\hat\square-\frac{4}{3}\Lambda \hat{\ide}\right)}\, .
\end{equation}
and then using the formulas (\ref{zm3}) and (\ref{zm4}) the correct expression for this quantity takes the form
\begin{eqnarray}
Z^2_{\rm 1-loop} &=&\frac{\det'^{2}_1\left(\hat\square+\Lambda\hat{\ide}\right)\det'_{1}\left(\hat\square+\frac{1}{3}
\Lambda\hat{\ide}\right)}{\det'_{2T}\left(\hat\square-\frac{2}{3}
\Lambda\hat{\ide}\right)\det'_{2T}\left(\hat\square-\frac{4}{3}\Lambda\hat{\ide}\right)}\ \! \frac{\det_{0}\left(\hat\square+\frac{4}{3}\Lambda\hat{\ide}\right)}{\det_{0}\left(\hat\square+2\Lambda\hat{\ide}\right)}\, .
\end{eqnarray}
In the main text we do not make a distinction between determinants with prime or without it since in all subsequent computations we treat all determinants of any operators on the same footing.
The correct account for zero modes in determinants is verified by the computation of UV-divergences of
the theory, which agree with the perturbative computation using, for example,  Feynman diagrams. The results of the derivation of the zero modes contribution presented here also agree with the expressions found in~\cite{Fradkin5s}.


\pdfbookmark[2]{C. Some basic features of the heat kernel techniques needed in the main text}{name23}
\hypersetup{bookmarksdepth=-2}
\section*{C. Some basic features of the heat kernel techniques needed in the main text}\label{sc}
\hypersetup{bookmarksdepth}


For the calculation of the functional traces (like in {\color{darkred} (\ref{FRG_partition})}) of the function of the covariant
d'Alembertian operator ${\rm Tr}f(\Delta)$, we can first employ the Laplace transform of the function $f(z)$, namely
\begin{equation}
f(z)\ = \ \int_{0}^{\infty}ds \ \! e^{-sz}\tilde{f}(s)\, , \;\;\;\;\; \mbox{Re}(z) >0\, .
\end{equation}
If we set $z = \Delta$ and assume that the $\Delta$-operator
has only positive definite eigenvalues then the ensuing functional trace ${\rm Tr}f(\Delta)$ reads
\begin{equation}
{\rm Tr}f(\Delta)=\!\int_{0}^{\infty}\!\!\!ds \ \!\tilde{f}(s){\rm Tr}e^{-s\Delta}=\!\int_{0}^{\infty}\!\!\!ds\ \!\tilde{f}(s){\rm Tr}K({s}).
\end{equation}
Here, ${\rm Tr}K({s})$ denotes a (functional) trace of the the heat kernel of $\Delta$, i.e. the trace of the operator $K(s) = e^{-s\Delta}$.
The terminology ``heat kernel'' stems from the fact that the operator $K(s)$
(with a flat-space Laplacian in the exponent $\Delta= - \vec\nabla^2$) is used to describe
the heat transfer in various media.  For the traced heat kernel on a general $d$-dimensional
background manifold one has an asymptotic expansion of the form~\cite{Tseytlins,Barvinskys}
\begin{equation}
{\rm Tr}K(s)\ = \ \frac{1}{(4\pi s)^{d/2}}\sum_{n=0}^{+\infty}B_{2n}(\Delta)s^{n}\, .
\end{equation}
The heat-kernel coefficients $B_{2n}$ are given
by spacetime volume integrals of the local expressions $b_{2n}$ according
to the formula
\begin{equation}
B_{2n}(\Delta)\ = \int\!d^{d}x \ \! \sqrt{g} \ \! {\rm tr}\left[b_{2n}(\Delta)\right]\, ,
\end{equation}
here ``${\rm tr}[\ldots]$'' represents trace over internal degrees of freedom.
Expansion coefficients $b_{m}$ are known as the Seeley--DeWitt coefficients.
For example, for the operator consisting of a constant shift of the
covariant d'Alembertian box operator defined by the equation
\begin{equation}
\Delta \ = \ \hat{\square} \ + \ \hat Y\, ,
\end{equation}
one has the first three $b_{2n}$ coefficients
\begin{eqnarray}
&&\mbox{\hspace{-10mm}}b_{0}(\Delta)\ = \ \hat{\ide}\, ,\nonumber \\[2mm]
&&\mbox{\hspace{-10mm}}b_{2}(\Delta)\ = \ \hat{\ide}\frac{R}{6}\ + \ \hat{Y}\, ,\nonumber \\[2mm]
&&\mbox{\hspace{-10mm}}b_{4}(\Delta)\ = \ \frac{1}{2}\hat Y^{2}-\frac{1}{6}\hat{\square}\hat Y\ + \ \frac{1}{12}\hat{\mathcal{R}}_{\mu\nu}^{2}\ + \ \frac{R}{6}\hat{Y}\ +\ \hat{\ide}\left(\frac{1}{180}R_{\mu\nu\rho\sigma}^{2}-\frac{1}{180}R_{\mu\nu}^{2}+\frac{1}{72}R^{2}-\frac{1}{30}{\square} R\right)\! ,
\end{eqnarray}
where we have defined $\hat{\mathcal{R}}_{\mu\nu}=[\hat{\nabla}_{\mu},\hat{\nabla}_{\nu}]$ and
$\hat{\square}=\hat{\nabla}^{\mu}\hat{\nabla}_{\mu}$ in accordance with definitions from \hyperref[sec3c]{{\color{darkred} Section III C}} and below in \hyperref[se]{Section E}\,.

By collecting all above results we can write
\begin{equation}
{\rm Tr}f(\Delta)\ = \ \frac{1}{(4\pi)^{d/2}}\sum_{n=0}^{+\infty}Q_{\frac{d}{2}-n}[f]\ \! B_{2n}(\Delta)\, ,
\label{83aa}
\end{equation}
where the $Q$-functionals of the function $f(z)$ are defined as
\begin{equation}
Q_{n}[f]\ = \ \int_{0}^{\infty}\!\!dt \ \! \tilde{f}(t)\ \!t^{-n}\, .
\label{Qn}
\end{equation}
By working in fixed regularization scheme with $d=4$, our $n$ in (\ref{Qn}) is always integer.
By employing the definition of the Gamma function we can rewrite Eq.~(\ref{Qn}) in more expedient form,
namely ($n > 0$):
\begin{eqnarray}
Q_{n}[f] &=&
    \frac{1}{\Gamma(n)}\!\int_{0}^{\infty}\!dz \ \! z^{n-1}f(z)\, , \nonumber \\[3mm]
Q_{-n}[f] &=&   (-1)^n f^{(n)}(0) \, , \nonumber \\[3mm]
Q_{0}[f] &=&   f(0) \, .
\end{eqnarray}

Outlined evaluation of $f(\Delta)$ via heat kernel method is used in \hyperref[sec3c]{{\color{darkred} Section III C}}  and below in \hyperref[se]{Section E}  in connection with  computations of the right hand side (RHS) of the flow equation.
For instance, in the flow equation {\color{darkred} (\ref{FRG_1a})}  we must perform an IR suppression of modes in the Wilsonian spirit. For this we have to decide about the IR-cutoff kernel function $R_k$.
We use the following (standard) form of the IR-cutoff kernel function
\begin{equation}
R_{k} \ \equiv \  R_{k}(z)\ = \ \left(k^{2}-z\right)\theta\!\left(k^{2}-z\right).
\label{litimcutoff}
\end{equation}
This is known in the literature as the optimized (or Litim) cutoff
function~\cite{Percaccis,Litim-as,Litim2s}.

The general form of the IR regularized propagators in each factor in the FRG flow equation (cf. \hyperref[S.3.b]{{\color{darkred} Section III B}}, particularly Eq. {\color{darkred} (\ref{FRG_partition})} there) is
of the form
\begin{equation}
G_{k}(z) \ = \ \frac{1}{z+R_{k}(z) +  \varpi}\, ,
\end{equation}
where in our case $\varpi$'s are shifts of the operators (looking
like mass squares of various fields). The evaluation of $Q$-functional
in this case leads to
\begin{equation}
Q_{n}\left[G_{k}(z)\partial_t R_{k}\right]\ = \
\frac{(n+1)k^{2n}}{\Gamma(n+2)}\left(1+\frac{\varpi}{k^{2}}\right)^{-1}\,,
\end{equation}
where we have, for simplicity's sake omitted a non-trivial anomalous dimension
of the operator. This is justified at the first one-loop order.
In particular, for one-loop level quantum divergent
contributions for $n=0,1$ and $2$ we have as
\begin{eqnarray}
&&Q_{0}\left[G_{k}(z)\partial_t R_{k}\right]\ = \ \left(1+\frac{\varpi}{k^{2}}\right)^{-1}
\, ,\label{Q1} \\[1mm] 
&&Q_{1}\left[G_{k}(z)\partial_t R_{k}\right]\ = \ k^{2}\left(1+\frac{\varpi}{k^{2}}\right)^{-1}\, ,\label{Q2} \\[1mm]
&&Q_{2}\left[G_{k}(z)\partial_t R_{k}\right]\ = \ \frac{1}{2}k^{4}\left(1+\frac{\varpi}{k^{2}}\right)^{-1}\, .\label{Q3}
\end{eqnarray}

The inclusion of the anomalous dimension $\eta$ is straightforward. For the $Q$-functionals, whose argument is the combination $G_{k}(z)\partial_t R_{k}$,
this amounts to multiplying all results (\ref{Q1})-(\ref{Q3}) by the common factor $(1-\eta)$. 
One sees that when $\eta=0$ this multiplication changes nothing, so the one-loop perturbative results are clearly reproduced. Due to our selected truncation
ansatz for $\Gamma_{L,k}$ in {\color{darkred} (\ref{33bcd})} we will concentrate only on $Q$-functionals, which are to be multiplied by $B_{2n}$ terms containing precisely
$4$ derivatives on the metric.
This implies that our main interest is on $Q_0$-functionals here.

\vspace{5mm}

\hypersetup{bookmarksdepth=-1}
\pdfbookmark[3]{Note on the Litim cutoff function}{name24}
\hypersetup{bookmarksdepth}

{\em Note on the Litim cutoff function}~~---~~
\hspace{-3mm}One notices that frequently we have to take functional traces with the following combination of functions of operators $f(z)=G_{k}(z)\left(\partial_tR_{k}(z)-\eta R_{k}(z)\right)$.
When we choose a specific form of the IR-cutoff kernel function~\cite{Litim-as}
\begin{equation}
R_{k} \ \equiv \  R_{k}(z)\ = \ \left(k^{2}-z\right)\theta\!\left(k^{2}-z\right)\! ,
\label{litim}
\end{equation}
($\theta(x)$ denotes a standard Heaviside step function), we can perform all the traces of functions $f$ of operators in a closed analytic form. They are all
encoded in the so-called $Q$-functionals, which we have discussed before. For our frequent
case, based on (\ref{83aa}), we have
\begin{eqnarray}
{\rm Tr}f(\Delta)|_{\rm projected} &=& Q_{0}[f]B_{4}(\Delta)\,,
\label{flowproj}
\end{eqnarray}
while for our typical $Q_0$-functional we find
\begin{eqnarray}
&&\mbox{\hspace{-9mm}}
Q_{0}\left[G_{k}(z)\left(\partial_tR_{k}-\eta R_{k}\right)\right]
\ = \ (2-\eta)\left(1+\frac{\varpi}{k^{2}}\right)^{-1}\!,
\label{38cc}
\end{eqnarray}
where we already foresee that these $Q_0$-functionals will realize in a Wilsonian spirit the IR decoupling of heavy massive modes. Therefore, they are crucial for an attempt to capture the effects of infrared threshold phenomena.

\vspace{3mm}


\pdfbookmark[2]{D. Anomalous dimension issue}{name25}
\hypersetup{bookmarksdepth=-2}
\section*{D. Anomalous dimension issue}\label{sd}
\hypersetup{bookmarksdepth}


The main object of a computation in this section is the anomalous dimension $\eta=\eta_{g^{(0)}\!,k}$
of the background graviton field $g_{\mu\nu}^{(0)}$. We derive it
in a way that parallels the derivation of the anomalous dimension in Yang-Mills (YM) gauge field theories \cite{Codellocomps}. First,
\begin{equation}
\alpha_{k} \ = \ Z_{g^{(0)},k}^{-1/2}\, ,
\end{equation}
is a gravitational coupling,
while $Z$ is the wave-function renormalization factor for background gravitational field. This relation is due to the invariance
of the covariant derivative under quantum rescaling transformations.
Next the flow of this coupling is given by and read from
\begin{equation}
\partial_t Z_{g^{(0)},k}\, ,
\end{equation}
which is obtained by evaluating the functional traces.
We have the gravitational action on the
RHS (i.e., source side) of the FRG flow equation {\color{darkred} (\ref{FRG_1a})} written as
\begin{equation}
\frac{1}{\alpha(k)^{2}}\ \! C^{2}\, ,
\end{equation}
so the identification is $\omega_{C}=\alpha^{-2}$. This suggests
that if one wants to get a $\beta$-function
looking like typical ``exact'' functional $\beta$-function of a dimensionless coupling (with characteristic non-perturbative denominators), then the anomalous dimension
of the operator $\sqrt{|g|}C^{2}$ must be taken into account.

More precisely, we derive the anomalous dimension $\eta$ by the following procedure.
First, we realize that the action important for second variations is
\begin{equation}
S \ = \ \omega_{C}\!\int\!d^{4}x\sqrt{|g|} \ \! C^{2}\, .
\label{actgrav}
\end{equation}
Let us now notice that for the YM theory, i.e., theory described by the action functional (written in a schematic way)
\begin{equation}
S_{{\rm YM}} \ = \ -\frac{1}{4g^{2}}\!\int\!d^{4}xF^{2}\, ,
\end{equation}
the anomalous dimension is defined by the relations
\begin{equation}
\eta_{\bar A} \ = \ -\partial_{t}\log Z_{\bar{A},k} \ = \ -Z_{\bar{A},k}^{-1}\partial_{t}Z_{\bar{A},k}\, .
\label{109ab}
\end{equation}
Above the wave-function renormalization factors are of the background gauge fields $\bar A$ in background field method applied to YM theories.
If we now use the one-loop results
\begin{equation}
\partial_{t}Z_{\bar{A},k} \ = \ -\frac{\partial_{t}g_{k}}{2g_{k}^{3}}\, ,
\end{equation}
and
\begin{equation}
Z_{\bar{A},k}^{-1/2} \ = \ g_{k}\, ,
\end{equation}
then we can finally rewrite (\ref{109ab}) in the form
\begin{eqnarray}
\eta_{\bar A} \ = \ 2g_{k}^{-1}\partial_{t}g_{k} \ = \ g_{k}^{-2}\partial_{t}g_{k}^{2} \ = \ -g_{k}^{-2}g_{k}^{4}\partial_{t}\left(g_{k}^{-2}\right)\ =\ 4g_{k}^{2}\partial_{t}\left(-\frac{1}{4g_{k}^{2}}\right).
\end{eqnarray}
In this way we have expressed the anomalous dimension of the background gauge field $\eta_{\bar A}$ in terms of
the (running) front coupling $g_k$ of the YM action (that is $F^2$ operator in our schematic notation) and its $t$-derivatives.
In the case of conformal gravitation described by the action (\ref{actgrav}), we get analogously
\begin{equation}
\eta \ = \ -\frac{1}{\omega_{C}}\partial_{t}\omega_{C}\,.
\label{eta-a}
\end{equation}

%


\hypersetup{bookmarksdepth=-2}
\pdfbookmark[2]{E. Relevant results from FRG flows on specific backgrounds --- used in Section IV}{name26}
\section*{E. Relevant results from FRG flows on specific backgrounds --- used in \hyperref[IV]{{\color{darkred} Section IV}}}
\label{se}
\hypersetup{bookmarksdepth}


Let us now discuss the FRG flow equation based on Eq.~{\color{darkred} (\ref{FRG_partition})} for two background spacetimes, namely MSS and Ricci-flat ones. This will substantiate our generic discussion from
\hyperref[sec3c]{{\color{darkred} Section III C}} and at the same time it will provide an important ingredient for determining the $\beta$-functions [see, Eq.~{\color{darkred} (\ref{beta_fnl})}] for QWG.\\


\hypersetup{bookmarksdepth=-1}
\pdfbookmark[3]{E. a) Maximally symmetric spaces}{name27}
\hypersetup{bookmarksdepth}

{\em {a) Maximally symmetric spaces}}~--~
\hspace{-3mm}The corresponding RG flow equation {\color{darkred} (\ref{FRG_partition})} reads
\begin{eqnarray}
2\partial_{t}\Gamma_{L,k}&=&{\rm Tr}_{2T}\left(\frac{\left(\partial_{t}R_{k}-\eta R_{k}\right)\hat{\ide}}{\hat{\square}+R_{k}\hat{\ide}-\frac{2}{3}\Lambda\hat{\ide}}\right)
\ + \ {\rm Tr}_{2T}\left(\frac{\left(\partial_{t}R_{k}-\eta R_{k}\right)\hat{\ide}}{\hat{\square}+R_{k}\hat{\ide}-\frac{4}{3}\Lambda\hat{\ide}}\right)\
- \ 2{\rm Tr}_{1}\left(\frac{\left(\partial_{t}R_{k}-\eta R_{k}\right)\hat{\ide}}{\hat{\square}+R_{k}\hat{\ide}+\Lambda\hat{\ide}}\right)
\nonumber \\[1mm]
&-&  {\rm Tr}_{1}\left(\frac{\left(\partial_{t}R_{k}-\eta R_{k}\right)\hat{\ide}}{\hat{\square}+R_{k}\hat{\ide}+\frac{1}{3}\Lambda\hat{\ide}}\right)\
+ \ {\rm Tr}_{0}\left(\frac{\left(\partial_{t}R_{k}-\eta R_{k}\right)\hat{\ide}}{\hat{\square}+R_{k}\hat{\ide}+2\Lambda\hat{\ide}}\right)\
- \ {\rm Tr}_{0}\left(\frac{\left(\partial_{t}R_{k}-\eta R_{k}\right)\hat{\ide}}{\hat{\square}+R_{k}\hat{\ide}+\frac{4}{3}\Lambda\hat{\ide}}\right)\,,
\label{32aa}
\end{eqnarray}
which we can cast, thanks to Eqs.~(\ref{83aa}), (\ref{flowproj}) and (\ref{38cc}), to the form

\begin{eqnarray}
\frac{2}{2-\eta}\partial_{t}\Gamma_{L,k} &=& \left(1-\frac{\frac{2}{3}\Lambda}{k^{2}}\right)^{-1}B_{4}
\left(\hat{\square}_{2T}-\frac{2}{3}\Lambda\hat{\ide}_{2T}\right) \ + \ \left(1-\frac{\frac{4}{3}
\Lambda}{k^{2}}\right)^{-1}B_{4}\left(\hat{\square}_{2T}-\frac{4}{3}\Lambda\hat{\ide}_{2T}\right)\nonumber \\[1mm]
&-& 2\left(1+\frac{\Lambda}{k^{2}}\right)^{-1}B_{4}\left(\hat{\square}_{1}+
\Lambda\hat{\ide}_{1}\right) \ - \
\left(1+\frac{\frac{1}{3}\Lambda}{k^{2}}\right)^{-1}B_{4}
\left(\hat{\square}_{1} \ + \ \frac{1}{3}\Lambda\hat{\ide}_{1}\right)\nonumber \\[1mm]
&+&\left(1+\frac{2\Lambda}{k^{2}}\right)^{-1}B_{4}\left(\hat{\square}_{0} +2\Lambda\hat{\ide}_{0}\right) \ - \ \left(1+\frac{\frac{4}{3}\Lambda}{k^{2}}\right)^{-1}B_{4}
\left(\hat{\square}_{0}+\frac{4}{3}\Lambda\hat{\ide}_{0}\right).
\label{39aa}
\end{eqnarray}
We express explicitly the (integrated) $B_4$ coefficients of the operator $\hat\square-Y\hat\ide$ in terms
of volume integrals of the Seeley--DeWitt (SD) coefficients $b_4$. We obviously have
\begin{equation}
B_4\left( \hat\square-Y\hat\ide\right) \ = \int\!d^4x
\sqrt{|g|} \ \! b_4\left(\hat\square-Y\hat\ide\right).
\end{equation}
For the SD coefficient $b_4$ (local and unintegrated) of the shifted covariant-box operator we can write~\cite{Tseytlins}
\begin{eqnarray}
&&\mbox{\hspace{-5mm}}b_{4}\left(\hat{\square}-Y\hat{\ide}\right) \ = \ {\rm tr}\!\left(\frac{1}{12}\hat{{\cal R}}_{\mu\nu}^{2}+\mathcal{E}\hat{\ide}+\frac{1}{2}Y^{2}\hat{\ide}-\frac{1}{6}RY\hat{\ide}\right),
\label{b4mss}
\end{eqnarray}
with
\begin{equation}
\mathcal{E} \ = \
- \ \frac{1}{360}E\ + \ \frac{1}{120}C^{2} \ + \ \frac{1}{72}R^{2},
\end{equation}
and the generalized curvature tensor $\hat{\cal R}_{\mu\nu}$ in a particular representation of fluctuation fields. The latter is
defined in a conventional way as the commutator of covariant derivatives (also acting in a specific representation), i.e.
\begin{equation}
\hat{\cal R}_{\mu\nu}=\left[\hat\nabla_\mu,\hat\nabla_\nu\right]\,.
\end{equation}
The hat over $\mathcal{R}_{\mu\nu}$ and ${\nabla}_{\mu}$ denotes that the internal indices are hidden
(representation independent form of operators).
The small traces that appear in (\ref{b4mss}) are done over these hidden matrix internal subspaces of fluctuations. Moreover, following our previous conventions,
we define the square $\hat{\cal R}_{\mu\nu}^2$ as $\hat{\cal R}_{\mu\nu}\hat{\cal R}^{\mu\nu}$, where the matrix multiplication is also
implicitly understood.

Specifically for MSS in $d=4$ one has
\begin{eqnarray}
R\ = \ 4\Lambda, \quad C^2 \ = \  0, \quad E \ = \ \frac{8}{3}\Lambda^{2} \ .
\end{eqnarray}
Consequently,
\bea\quad \mathcal{E} =  \frac{29}{135}\Lambda^{2}\eea

 and
\bea
\label{43ba}
\mbox{\hspace{-10mm}}b_{4}\left(\hat{\square}-Y\hat{\ide}\right) \ = \ {\rm tr}\left(\frac{1}{12}\hat{{\cal R}}_{\mu\nu}^{2}+\frac{29}{135}\Lambda^{2}\hat{\ide}+\frac{1}{2}Y^{2}\hat{\ide}-\frac{2}{3}\Lambda Y\hat{\ide}\right)\! .
\eea

The actual value of (\ref{43ba}) depends on the spin content of fluctuations. In particular, for scalars (spin-0) we have
\begin{equation}
\mbox{tr}\ \!\hat{{\cal R}}_{\mu\nu}^{2} \ = \ 0\, ,
\end{equation}
for vectors (spin-1) we get
\begin{equation}
\mbox{tr}\ \!\hat{{\cal R}}_{\mu\nu}^{2} \ = \ -R_{\mu\nu\rho\sigma}^{2} \ = \ -\frac{8}{3}\Lambda^{2}\, ,
\end{equation}
and finally for tensors (spin-2) with two indices we have
\begin{equation}
\mbox{tr}\ \!\hat{{\cal R}}_{\mu\nu}^{2}\ = \ -6R_{\mu\nu\rho\sigma}^{2} \ = \ -16\Lambda^{2}\, .
\end{equation}
\vspace{2mm}

\noindent The latter formula is a simple consequence of the following mathematical facts
\begin{eqnarray}
&&\hspace{-10mm}\left({\cal R}_{\mu\nu}\right)_{\alpha\beta}{}^{\gamma\delta}\ = \ 2R_{\mu\nu(\alpha}{}^{(\gamma}\delta_{\beta)}^{\delta)}\, ,\nonumber \\[2mm]
&&\hspace{-10mm}\left({\cal R}^{\mu\nu}\right)_{\alpha\beta}{}^{\gamma\delta}\left({\cal R}_{\mu\nu}\right)_{\gamma\delta}{}^{\kappa\lambda} \ = \ 2\left[R^{\mu\nu}{}_{(\alpha}{}^{\gamma}\delta_{\beta)}^{(\kappa}R_{\mu\nu\gamma}{}^{\lambda)}+R^{\mu\nu}{}_{(\alpha}{}^{(\kappa} R_{\mu\nu \beta)}{}^{\lambda)}\right], \nonumber  \\[2mm]
&&\hspace{-10mm}{\rm tr}\ \!\hat{{\cal R}}_{\mu\nu}^{2}={\rm tr}\left[\left({\cal R}^{\mu\nu}\right)_{\alpha\beta}{}^{\gamma\delta}
\left({\cal R}_{\mu\nu}\right)_{\gamma\delta}{}^{\kappa\lambda}\right] \ = \ -6R_{\mu\nu\rho\sigma}^{2},
\label{61aa}
\end{eqnarray}
valid for symmetric spin-2 fluctuations on which these tensors act.  In particular, the latter action can explicitly be written as
\begin{eqnarray}
\left( \hat{\cal R}_{\mu\nu}h\right)_{\alpha\beta} \ = \
\left({\cal R}_{\mu\nu}\right)_{\alpha\beta}{}^{\gamma\delta}h_{\gamma\delta}\, .
\end{eqnarray}
In the notation employed in Eqs.~(\ref{61aa}) we used standard idempotent symmetrization brackets between two indices.

By employing the fact that ${\rm tr}\hat{\ide}_{2T}=9$ we obtain
\begin{eqnarray}
&&b_{4}\left(\hat{\square}_{2T}-\frac{2}{3}\Lambda\hat{\ide}_{2T}\right) \ = \
-\frac{7}{5}\Lambda^{2}\, , \label{48aa}\\[2mm]
&&b_{4}\left(\hat{\square}_{2T}-\frac{4}{3}\Lambda\hat{\ide}_{2T}\right) \ = \
\frac{3}{5}\Lambda^{2}\, .
\end{eqnarray}
Similarly because ${\rm tr}\hat{\ide}_{1}=4$ we have
\begin{eqnarray}
&&b_{4}\left(\hat{\square}_{1}+\Lambda\hat{\ide}_{1}\right) \ = \
\frac{716}{135}\Lambda^{2}\, , \\[2mm]
&&
b_{4}\left(\hat{\square}_{1}+\frac{1}{3}\Lambda\hat{\ide}_{1}\right)
\ = \
\frac{236}{135}\Lambda^{2}\, .
\label{49aa}
\end{eqnarray}
Finally, we use ${\rm tr}\hat{\ide}_{0}=1$ to get
\begin{eqnarray}
&&b_{4}\left(\hat{\square}_{0}+2\Lambda\hat{\ide}_{0}\right) \ = \
\frac{479}{135}\Lambda^{2}\, , \\[2mm]
&&
b_{4}\left(\hat{\square}_{0}+\frac{4}{3}\Lambda\hat{\ide}_{0}\right)\ = \ 
\frac{269}{135}\Lambda^{2}\, .
\label{50aa}
\end{eqnarray}
Inserting now the results (\ref{48aa})-(\ref{50aa}) to (\ref{39aa}) and integrating over the volume of space, the RG flow equation on the MSS boils down to

\begin{eqnarray}
\partial_{t}\Gamma_{L,k}&=&\frac{1}{2}(2-\eta)\!\int\!d^{4}x\sqrt{g}
\left[\left(1-\frac{\frac{2}{3}\Lambda}{k^{2}}\right)^{-1}\left(-\frac{7}{5}\Lambda^{2}\right)+
\left(1-\frac{\frac{4}{3}\Lambda}{k^{2}}\right)^{-1}\left(\frac{3}{5}\Lambda^{2}\right)\right. \nonumber \\[2mm]
&-&\left.2\left(1+\frac{\Lambda}{k^{2}}\right)^{-1}\frac{716}{135}\Lambda^{2}-\left(1+\frac{\frac{1}{3}
\Lambda}{k^{2}}\right)^{-1}\frac{236}{135}\Lambda^{2}+\left(1+\frac{2\Lambda}{k^{2}}\right)^{-1}\frac{479}{135}
\Lambda^{2}-\left(1+\frac{\frac{4}{3}\Lambda}{k^{2}}\right)^{-1}\frac{269}{135}\Lambda^{2}\right].
\label{51abc}
\end{eqnarray}
This should  be compared with the $\beta$-functional of the theory on the
LHS of (\ref{39aa}), and consistently with {\color{darkred}(\ref{bfnlmss})} and {\color{darkred} (\ref{33bcd})}, namely (in $d=4$)
\begin{eqnarray}
\partial_{t}\Gamma_{L,k} &=& \left.\int\!d^{4}x\sqrt{g} \ \!\left(\beta_{C}C^{2}+\beta_{E}E\right)\right|_{\rm MSS} \ = \ \int\!d^{4}x\sqrt{g}\ \! \left(\frac{8}{3}\Lambda^{2}\beta_{E}\right),
\label{52ab}
\end{eqnarray}
in agreement with \hyperref[S.3.b]{{\color{darkred} Section~III~B}}.
Hence, on MSS we can write for $\beta_{E}$, cf. (\ref{51abc}) and (\ref{52ab}), that
\begin{eqnarray}
\beta_{E} &=& \frac{1}{2}(2-\eta)\left[-\frac{21}{40}\left(1-\frac{\frac{2}{3}\Lambda}{k^{2}}\right)^{-1} \ + \
\frac{9}{40}\left(1-\frac{\frac{4}{3}\Lambda}{k^{2}}\right)^{-1}
\ - \ \frac{179}{45}\left(1+\frac{\Lambda}{k^{2}}\right)^{-1}\right.   \nonumber \\[2mm]
&-& \left.\frac{59}{90}
\left(1+\frac{\frac{1}{3}\Lambda}{k^{2}}\right)^{-1}
\ + \ \frac{479}{360}
\left(1+\frac{2\Lambda}{k^{2}}\right)^{-1} \ - \ \frac{269}{360}\left(1+\frac{\frac{4}{3}\Lambda}{k^{2}}
\right)^{-1}\right].
\label{56aa}
\end{eqnarray}
The factors of the type $(1\pm a{\Lambda}/{k^{2}})^{-1}$, where $a={\rm const}$
describe the so-called threshold phenomena (i.e., decoupling of the flow in IR due to IR masses of modes)~\cite{Percaccis}.

A simple consistency check of (\ref{56aa}) can be done by considering (\ref{51abc}) without threshold phenomena (i.e., ${{\Lambda}/{k^{2}} \to0}$)
and with  no anomalous dimension  (i.e., ${\eta = 0}$). In this case the coefficients on the RHS of (\ref{51abc}) sum up to
\begin{equation}
-\frac{7}{5}+\frac{3}{5}-2\times\frac{716}{135}-\frac{236}{135}+\frac{479}{135}-\frac{269}{135}\ = \
-\frac{58}{5}\, ,
\end{equation}
which gives the one-loop perturbative coefficient $b_{4}(H)$ (cf. \hyperref[sec3c]{{\color{darkred} Section~III~C}}).
This number is related to perturbative logarithmic UV-divergence coming at one-loop with the $E$ term
in the effective action. By employing (\ref{52ab}) and simplifying by the factor $\mbox{$8/3$}$  we obtain the one-loop perturbative $\beta$-function of the $E$ term (Gauss--Bonnet term), in the form
\begin{equation}
\beta_E^{\rm FT} \ = \ -\frac{87}{20}\, .
\label{betae}
\end{equation}
This coincides with the value computed by Fradkin and Tseytlin in Ref.~\cite{Tseytlins}.\\


\hypersetup{bookmarksdepth=-1}
\pdfbookmark[3]{E. b) Ricci-flat manifolds}{name28}
\hypersetup{bookmarksdepth}

{\em {b) Ricci-flat manifolds}}~---~
\hspace{-3mm}Here the one-loop partition function is given by {\color{darkred} (\ref{Ricflatpartition})}. The ensuing RG flow equation
can be therefore written as

\begin{equation}
\partial_{t}\Gamma_{L,k}=2{\rm Tr}_{2}\left(\frac{\left(\partial_{t}R_{k}-\eta R_{k}\right)\hat{\ide}}{\hat{\square}-2\hat{C}+R_{k}\hat{\ide}}\right)-3{\rm Tr}_{1}\left(\frac{\left(\partial_{t}R_{k}-\eta R_{k}\right)\hat{\ide}}{\hat{\square}+R_{k}\hat{\ide}}\right)-2{\rm Tr}_{0}\left(\frac{\left(\partial_{t}R_{k}-\eta R_{k}\right)\hat{\ide}}{\hat{\square}+R_{k}\hat{\ide}}\right).
\end{equation}

Again as in {\color{darkred} (\ref{33bcd})}, the minimal consistent ansatz for the LHS effective action is
\begin{equation}
\Gamma_{L,k} \ = \ \int\!d^{4}x\sqrt{g}\left(\omega_{C}(k)C^{2}+\omega_{E}(k)E\right).
\end{equation}
At this stage one can exactly follow steps (\ref{83aa})-(\ref{38cc}).
Hence, the RG flow equation on Ricci-flat backgrounds reads
\begin{eqnarray}
&&\mbox{\hspace{-9mm}}\frac{2}{2-\eta}\partial_{t}\Gamma_{L,k} \ = \ 2B_{4}\left(\hat{\square}_{2}-2\hat{C}\right)-3B_{4}
\left(\hat{\square}_{1}\right)-2B_{4}\left(\hat{\square}_{0}\right).
\end{eqnarray}
By employing the formula (5.16) from Ref.~\cite{Tseytlins} we obtain
\begin{eqnarray}
&&\mbox{\hspace{-9mm}}b_{4}\left(\hat{\square}\right)={\rm tr}\left(\frac{1}{12}\hat{{\cal R}}_{\mu\nu}^{2}+\mathcal{E}\hat{\ide}\right),\nonumber \\[2mm]
&&\mbox{\hspace{-9mm}}b_{4}\left(\hat{\square}-Y\hat{\ide}\right) \ = \ {\rm tr}\left(\frac{1}{12}\hat{{\cal R}}_{\mu\nu}^{2}+\mathcal{E}\hat{\ide}+\frac{1}{2}Y^{2}\hat{\ide}-\frac{1}{6}RY\hat{\ide}\right).
\end{eqnarray}
On Ricci-flat backgrounds one has $E=C^2$ and $R=0$, hence
\begin{eqnarray}
\mbox{\hspace{-5mm}}\mathcal{E}\ = \ \frac{1}{180}E+\frac{1}{120}\left(C^{2}-E\right)+\frac{1}{72}R^{2}
\ = \ \frac{1}{180}C^{2}\, .
\end{eqnarray}
Consequently,
\begin{eqnarray}
&&\mbox{\hspace{-9mm}}b_{4}\left(\hat{\square}\right)\ = \ {\rm tr}\left(\frac{1}{12}\hat{{\cal R}}_{\mu\nu}^{2}+\frac{1}{180}C^{2}\hat{\ide}\right), \nonumber \\[2mm]
&&\mbox{\hspace{-9mm}}b_{4}\left(\hat{\square}-Y\hat{\ide}\right)\ = \ {\rm tr}\left(\frac{1}{12}\hat{{\cal R}}_{\mu\nu}^{2}+\frac{1}{180}C^2\hat{\ide}+\frac{1}{2}Y^{2}\hat{\ide}\right).
\label{62ch}
\end{eqnarray}
Clearly, the actual value of (\ref{62ch}) depends on the spin content --- i.e., on the character of fluctuations. For scalars (spin-0) we have
\begin{equation}
{\rm tr}\ \!\hat{{\cal R}}_{\mu\nu}^{2} \ = \ 0\, ,
\label{63cd}
\end{equation}
for vectors (spin-1) we can write
\begin{equation}
{\rm tr}\ \!\hat{{\cal R}}_{\mu\nu}^{2} \ = \ -R_{\mu\nu\rho\sigma}^2 \ = \ -C^{2}\, .
\end{equation}
Finally, for symmetric tensors (spin-2) with two indices (rank-2) we obtain
\begin{eqnarray}
&&{\rm tr}\ \! \hat{{\cal R}}_{\mu\nu}^{2} \ = \ -6R_{\mu\nu\rho\sigma}^{2}\ = \ -6C^{2}\, .
\end{eqnarray}
An important, but somewhat less intuitive relation holds for the trace of the matrix square of the Weyl tensor (${\rm tr}\ \!\hat{C}^{2}$), namely
\begin{eqnarray}
&&{\rm tr}\ \!\hat{C}^{2}\ =
\frac{1}{2}\delta_{\alpha}^\kappa\delta_{\beta}^{\lambda}C^{\alpha\gamma\beta\delta}C_{\gamma\kappa\delta\lambda}+\frac{1}{2}
\delta_{\alpha}^\kappa\delta_{\beta}^{\lambda}C^{\alpha\delta\beta\gamma}C_{\gamma\kappa\delta\lambda}\ = \ \frac{1}{2}C^{\alpha\gamma\beta\delta}C_{\alpha\gamma\beta\delta}+\frac{1}{2}C^{\alpha\delta\beta\gamma}
C_{\gamma\alpha\delta\beta}
\nonumber\\
&&=\frac{1}{2}C^{{\alpha\beta\gamma\delta}}C_{\alpha\beta\gamma\delta}+\frac{1}{4}
C^{\alpha\beta\gamma\delta}C_{\alpha\beta\gamma\delta}\ = \ \frac{3}{4}C^{\alpha\beta\gamma\delta}C_{\alpha\beta\gamma\delta} \ = \ \frac{3}{4}C^2\,,
\label{65cd}
\end{eqnarray}
where we have also used cyclic property of the Weyl tensors (first Bianchi identity). We recall that the action of the matrix Weyl tensor $\hat C$ on symmetric fluctuations is explicitly given by $\left(\hat{C}h\right)_{\alpha\beta}=C_{\alpha}{}^\gamma{}_\beta{}^\delta h_{\gamma\delta}$.

By inserting (\ref{63cd})-(\ref{65cd}) to (\ref{62ch}) we obtain the following simple relations:
\begin{eqnarray}
&&\mbox{\hspace{-2mm}}b_{4}\left(\hat{\square}_{2}-2\hat{C}\right) \ = \ \frac{1}{12}\times\left(-6C^{2}\right)+\frac{1}{180}\times10C^{2}+2{\rm tr}\ \!\hat{C}^{2} \ = \ \frac{19}{18}C^{2}\, ,\nonumber \\[2mm]
&&\mbox{\hspace{-2mm}}b_{4}\left(\hat{\square}_{1}\right) \ = \ -\frac{1}{12}C^{2}+\frac{1}{45}C^{2} \ = \ -\frac{11}{180}C^{2}\, ,\nonumber \\[2mm]
&&\mbox{\hspace{-2mm}}b_{4}\left(\hat{\square}_{0}\right) \ = \ \frac{1}{180}C^{2}\, .
\end{eqnarray}
In writing the previous traces we have also employed that ${\rm tr}\hat{\ide}_{2}=10$,
${\rm tr}\hat{\ide}_{1}=4$ and ${\rm tr}\hat{\ide}_{0}=1$. Consequently,  the RG flow equation on Ricci-flat backgrounds reads

\begin{equation}
\partial_{t}\Gamma_{L,k} \ =\frac{1}{2}  (2-\eta)\!\int\!d^{4}x\sqrt{g}\left[2\times\frac{19}{18}C^{2}-3\times
\left(-\frac{11}{180}\right)C^{2}-2\times\left(\frac{1}{180}\right)C^{2}\right]\ =  \frac{2-\eta}{2} \!\int\!d^{4}x\sqrt{g}\left(\frac{137}{60}C^{2}\right).
\label{67aa}
\end{equation}
At this stage we should realize that a $\beta$-functional of the theory on the
LHS of the FRG flow equation {\color{darkred} (\ref{FRG_partition})}, when evaluated on Ricci-flat manifolds in $d=4$ gives
\begin{eqnarray}
\partial_{t}\Gamma_{L,k} &= &\left. \int\!d^{4}x\sqrt{g}\left(\beta_{C}C^{2}+\beta_{E}E\right)\right|_{R_{\mu\nu}=0}\ = \ \int\!d^{4}x\sqrt{g}\left(\beta_{C}+\beta_{E}\right)C^{2}\, .
\end{eqnarray}
This is done consistently with the Eq.~{\color{darkred} (\ref{bfnlric})}.
With this the RG flow equation (\ref{67aa}) boils down to a relation
\begin{equation}
\left(\beta_{C}+\beta_{E}\right)C^{2} \ =\frac{1}{2} (2-\eta)\left(\frac{137}{60}C^{2}\right).
\label{69aa}
\end{equation}
Similarly, as in the MSS case, we can perform a consistency check with the known one-loop result of Ref.~\cite{Tseytlins}
by considering  no anomalous dimension, i.e. $\eta=0$ (in this case we do not have any effect due to threshold phenomena). Within such a framework  Eq.~(\ref{67aa})
reduces to
\begin{equation}
\beta_{C}+\beta_{E} \  = \ \frac{137}{60}\, .
\end{equation}
From this and Eq.~(\ref{betae}) we can also obtain a perturbative one-loop expression for the $\beta$-function $\beta_C$ of the Weyl coupling in QWG
\begin{equation}
\beta_{C}^{\rm FT}=\frac{199}{30}\,.
\label{betac}
\end{equation}
This is again  in  a full agreement with the result of Fradkin and Tseytlin from Ref.~\cite{Tseytlins}.\\


\hypersetup{bookmarksdepth=-2}
\pdfbookmark[2]{F. Extension of results of Section IV C to higher order}{name29}
\section*{F. Extension of results of \hyperref[Sec.4.ca]{{\color{darkred}  Section IV C}} to higher order}
\label{sf}
\hypersetup{bookmarksdepth}


In this section we discuss the existence and characteristics of the IR FP, when the second term in Eq. {\color{darkred} (\ref{beta_f_1})} containing the second derivative $\beta''(\kappa)$, is properly taken into account. We
expand the $\beta$-function $\beta(\kappa)$ around $\kappa_c$ to the second order, so that
\begin{eqnarray}
\beta(\kappa) \ = \ {a} \ \!(\kappa-\kappa_c) \ + a_2(\kappa-\kappa_c)^2\,.
\label{nbeta_f_1}
\end{eqnarray}
 It is slightly more complicated to rewrite this expansion in terms of coupling $\omega$, but with the help of series operations like composition and inverting, we can achieve it without too much effort. Generally, we must remember that we will work to the accuracy given by the second or higher than the leading one order expansion in $(\kappa-\kappa_c)$ or $(\omega-\omega_*)$ and hence all forthcoming expressions have this series character. This means that for example, we cannot integrate exactly, like this was done in Eq. {\color{darkred} (\ref{beta_f_3})} and leave the result as the logarithm -- we must expand to the second after the leading order in $(\kappa-\kappa_c)$.

We state our definitions of the parameters $a$ and $a_2$
\bea
a=\left.\frac{d\beta(\kappa)}{d\kappa}\right|_{\kappa=\kappa_{c}},\quad a_2=\left.\frac{1}{2}\frac{d^{2}\beta(\kappa)}{d\kappa^{2}}\right|_{\kappa=\kappa_{c}}\,.
\eea

Next, integrating from (\ref{nbeta_f_1}) we get
\begin{eqnarray}
\mbox{\hspace{-3mm}}\omega(\kappa) - \omega_* \ =\frac{a}{2\kappa_c} (\kappa-\kappa_c)^2+\frac{a_2\kappa_c-a}{3\kappa_c^2} (\kappa-\kappa_c)^3\, ,
\label{nbeta_f_3}
\end{eqnarray}
where $\omega_* = \omega(\kappa_c)$. Relation (\ref{nbeta_f_3}) can be inverted, but for this we cannot use anymore compact formulas with Lambert function. However, inverting the series is possible and we get

\begin{eqnarray}
\mbox{\hspace{-2mm}}\kappa = \kappa_c + \sqrt{\frac{2 \kappa_c}{a} (\omega- \omega_*)} + \frac{2(a-a_2\kappa_c)}{3a^2} (\omega- \omega_*) \,,
\label{nk2ab}
\end{eqnarray}
\con and so the consistent (to this accuracy) series for the $\beta$-function (\ref{nbeta_f_1}) reads
\begin{eqnarray}
\mbox{\hspace{-6mm}}\beta =  \sqrt{{2 a\kappa_c}\ \! (\omega- \omega_*)} \ + \ \frac{2a + 4 a_2 \kappa_c}{3 a}\ \!(\omega- \omega_*) .
\label{nk3ab}
\end{eqnarray}
Note that the result (\ref{nk3ab}) holds true both for $\beta_C$ and $\beta_E$, since the product $a (\omega- \omega_*)\geqslant 0$ in both cases. Now, one can see that due to the dependence on $a_2$ in the second term in Eq. (\ref{nk3ab}) it is not enough to this order of accuracy to keep only non-zero the  $a$ coefficient. Consistency requires that if we write first two terms in Eq. (\ref{nk3ab}) we must also start with two leading terms in Eq. (\ref{nbeta_f_1}). Hence, in full generality both $a$ and $a_2$ coefficients must be taken into account. Of course, one finds that the limit $a_2\to0$ reproduces precisely all the results mentioned in the main text. The analysis presented in \hyperref[Sec.4.ca]{{\color{darkred} Section IV C}} may be viewed as a first approximation, correct in a case when we can neglect the impact of the second derivative at the crossing point $a_2$. Strictly we can apply it only in the regime when $a_2 \kappa_c\ll a$. However, the characteristics of the turning FP at $\kappa=\kappa_c$ is unchanged and also its holographic interpretation remains fully valid.

The important lesson that we have taken in \hyperref[Sec.4.ca]{{\color{darkred} Section IV C}} was how in that case analytically continue the dependence of $\beta=\beta(\omega)$ past the turning FP at $\kappa=\kappa_c$. Abstracting from the usage of Lambert function we remind that the crucial thing was the flip of the sign on the terms containing square roots, according to the rule

\begin{eqnarray}
\sqrt{\omega- \omega_*} \ \rightarrow \ - \sqrt{\omega- \omega_*}\, .
\end{eqnarray}

The sole effect of this step is that $\beta(\omega)$ from (\ref{nk3ab}) will be smoothly taken through the turning point at $\kappa_c$ to $\kappa<\kappa_c$
of the form

\begin{eqnarray}
\mbox{\hspace{-6mm}}\beta =  -\sqrt{{2 a\kappa_c}\ \! (\omega- \omega_*)} \ + \ \frac{2a + 4 a_2 \kappa_c}{3 a}\ \!(\omega- \omega_*) .
\end{eqnarray}
For consistency we also flip the sign of the second derivative coefficient $a_2\to-a_2$ since this one always multiplies terms with integer powers of $(\omega-\omega_*)$, while the first derivative coefficient $a$ we keep unchanged.

So, the key effect of the above analytical continuation is that the expression for $\beta(\omega)$ on a different branch reads

\begin{eqnarray}
\mbox{\hspace{-6mm}}\beta =  -\sqrt{{2 a\kappa_c}\ \! (\omega- \omega_*)} \ + \ \frac{2a - 4 a_2 \kappa_c}{3 a}\ \!(\omega- \omega_*) .
\label{nk4abc}
\end{eqnarray}
For the form
of the $\beta$-function in Eq. (\ref{nk4abc}) we keep only first two terms as written in Eq. (\ref{nbeta_f_1})
since we have worked in the one-loop scheme keeping $\omega$
very large. Expanding the behavior near the turning FP we can continue
the RG flow towards lower values of energies (lower than $\kappa=\kappa_{c}$) consistently with this order of approximation.

From (\ref{nk4abc}) we get the flow of the coupling $\omega$ in the region $0\leqslant \kappa \leqslant \kappa_c$, namely
\begin{eqnarray}
\hspace{-7mm}\omega \ = \ \omega_{*} \ + \frac{9 a^3
   \kappa_c^{\frac{a+4
   a_2 \kappa_c}{3 a}}
   \kappa^{\frac{2 a-4 a_2
   \kappa_c}{3 a}}}{2
   (a-2a_2 \kappa_c)^2} -  \frac{9 a^3 \left(2
   \kappa_c^{\frac{2 a+2a_2
   \kappa_c}{3 a}}
   \kappa^{\frac{a-2 a_2 \kappa_c}{3
   a}}-\kappa_c\right)}{2
   (a-2a_2 \kappa_c)^2}\, ,\label{omsolved}
\end{eqnarray}
provided that  $(a-2a_2 \kappa_c)/a>0$.  
Taking this dependence  as being exact on the scales past the turning
FP one can find the true IR FP occurring at $\kappa=0$.
This implies that the IR FP value of the coupling is
\begin{equation}
\omega_{**} \ = \ \omega(\kappa=0) \ = \ \omega_{*} \ + \ \frac{9 a^3 \kappa_c}{2 (a-2 a_2 \kappa_c)^2}\, .\label{omstar2}
\end{equation}
Ensuing behavior of the $\beta$-function near $\omega_{**}$
is given by
\begin{eqnarray}
\beta \ = \   \kappa\frac{d\omega}{d\kappa}\ = \frac{3 a^2 \kappa_c^{\frac{a+4 a_2 \kappa_c}{3 a}} \kappa^{\frac{a-2 a_2
   \kappa_c}{3 a}} }{a-2a_2 \kappa_c}  \ \! \left(\kappa^{\frac{a-2a_2 \kappa_c}{3
   a}}-\kappa_c^{\frac{a-2a_2 \kappa_c}{3 a}}\right)\, .
   \label{betasolved}
\end{eqnarray}
It is again confirmed that when $(a-2a_2 \kappa_c)/a>0$, then the limit $\kappa\to0$ makes $\beta$-function vanish.

One can also study the stability matrix in our more generalized case considered here. The only
non-zero elements of the stability matrix are
\bea
\left.\frac{\partial\beta_C}{\partial\omega_C}\right|_{\omega=\omega_{**}} \ =\frac{2 (a-2a_2 \kappa_c)}{3 a}-\frac{\sqrt{a \kappa_c}}{
   \sqrt{2(\omega_{**}-\omega_*)}}=\frac{a-2a_2 \kappa_c}{3 a} = \frac{1}{3}\left(1-\frac{2a_2 \kappa_c}{a}\right)\,,
\eea
similarly,
\begin{equation}
\left.\frac{\partial\beta_E}{\partial\omega_E}\right|_{\omega=\omega_{**}} \ = \ \frac{1}{3}\left(1-\frac{2 a_2 \kappa_c}{a}\right)\, ,
\end{equation}
which implies that the set of critical exponents is
\begin{equation}
\theta \ = \  \left\{ -\frac{1}{3}\left(1-\frac{2 a_2 \kappa_c}{a}\right), -\frac{1}{3}\left(1-\frac{2 a_2 \kappa_c}{a}\right) \right\}\, ,
\label{nIR_stable}
\end{equation}
where the parameters $a$, $a_2$ and $\kappa_c$ are for $\omega_C$ and $\omega_E$ couplings respectively.

We shall comment on the size of the anomalous dimensions $\theta$ obtained here and compare them with the values quoted in the main text ($\theta=1/3$). The fact that anomalous dimensions after the inclusion of the terms with the second derivative $a_2$ differ quite significantly (order of magnitude or more) is a confirmation of the lack of perturbativity near the IR FP. However, the perturbative computation to a higher order shows that the FP at $k=0$ is still IR-stable. Since this FP is non-Gaussian, the values of couplings are not necessarily small and we cannot trust anymore the perturbation calculus. This is confirmed because with the next order the anomalous dimension grow fast. On one side this proves that we are in a consistent situation but on a different side this calls for a usage of a different calculational techniques since we cannot rely on the standard perturbation approaches and techniques. We show that indeed the FP is in the non-perturbative regime and to accurately describe its critical exponents we need to use some non-perturbative tools as well. Unfortunately, we could not use FRG to do this for the last part of the RG flow, but we hope to manage this in the future.

\vspace{8mm}
\noindent${}^*$\, Electronic address: \href{mailto: p.jizba@fjfi.cvut.cz}{p.jizba@fjfi.cvut.cz}\\
${}^\dagger$\, Electronic address: \href{mailto: grzerach@gmail.com}{grzerach@gmail.com}\\
${}^\ddagger$\, Electronic address: \href{mailto: knapjaro@fjfi.cvut.cz}{knapjaro@fjfi.cvut.cz}\\

\hypersetup{bookmarksdepth=-2}
\pdfbookmark[2]{References}{name30}

\hypersetup{bookmarksdepth}
\vspace{30cm}
\end{widetext}


\begin{thebibliography}{200}

\bibitem{Ade}
N. Aghanim at al. (Planck Collaboration), Planck 2018 results. VI. Cosmological parameters, arXiv:1807.06209.
\bibitem{Col} S. Coleman and E. Weinberg, Phys. Rev. D {\bf 7}, 1888 (1973).
\bibitem{Ad1} S.L. Adler, Phys. Rev. Lett. {\bf 44}, 1567 (1980).
\bibitem{Ad2} S.L. Adler, Rev. Mod. Phys. {\bf 54}, 729 (1982).
\bibitem{Zee} A. Zee, Ann. Phys. (N.Y.) {\bf 151}, 431 (1983).
\bibitem{Sp} B.L. Spokoiny, Phys. Lett. B 147, 39 (1984).
\bibitem{K-S} H. Kleinert and H.-J. Schmidt, Gen. Relativ. Gravit. {\bf 34}, 1295 (2002).
\bibitem{Cap} S. Capozziello and V. Faraoni, {\it Beyond Einstein Gravity; A Survey of
Gravitational Theories for Cosmology and Astrophysics} (Springer,
London, 2011).


\bibitem{Shaposhnikov} M.~Shaposhnikov and A.~Shkerin, J. High Energy Phys. {\bf 10}, 024 (2018).


\bibitem{JKS} P.~Jizba, H.~Kleinert, and F.~Scardigli, Eur. Phys. J. C {\bf 75}, 245 (2015).
\bibitem{Ayazi} S.Y.~Ayazi and A.~Mohamadnejad, Eur. Phys. J. C  {\bf 79}, 140 (2019).
\bibitem{Ishiwata} K.~Ishiwata, Phys. Lett. B {\bf 710}, 134 (2012).

\bibitem{Man1} P.D. Mannheim, Phys. Rev. D {\bf 85}, 124008 (2012).
\bibitem{Man2} P.D. Mannheim, Astrophys. J. {\bf 391}, 429 (1992).

\bibitem{Man3}   J.~G.~O'Brien and P.~D.~Mannheim,
  Mon.\ Not.\ Roy.\ Astron.\ Soc.\  {\bf 421}, 1273 (2012).
  
\bibitem{Tseytlin} E.S.~Fradkin and A.A.~Tseytlin, Phys.~Rep.~{\bf 119},  233 (1985).
\bibitem{Elitzur} S.~Elitzur, Phys. Rev. D. {\bf 12}, 3978 (1975).

\bibitem{Fradkin6}
  E.S.~Fradkin and A.A.~Tseytlin,
  Phys.\ Lett.\  {\bf 134B}, 187 (1984).

\bibitem{Schwimmer}
 A.~Schwimmer and S.~Theisen,
  Nucl.\ Phys.\ B {\bf 847}, 590 (2011)
  [arXiv:1011.0696 [hep-th]].


\bibitem{Shap:2010}  M.~Shaposhnikov and Ch.~Wetterich,	Phys.~Lett.~{\bf 683B}, 196 (2010).
\vspace{2cm}
\bibitem{Nikolai:2007} K.A.~Meissner and H.~Nicolai, Phys. Lett. {\bf 648B}, 312 (2007).
\bibitem{stelle1} K.S. Stelle, Phys. Rev. D {\bf 16}, 953 (1977).

\bibitem{vDV} H.~van Dam and M.~Veltman, Nucl. Phys. {\bf B 22}, 397 (1970);   V.~I.~Zakharov,
  JETP Lett.\  {\bf 12}, no. 9, 312 (1970)
  [Pisma Zh.\ Eksp.\ Teor.\ Fiz.\  {\bf 12}, 447 (1970)], \newline 
  \href{http://www.jetpletters.ac.ru/ps/1734/article_26353.shtml}{http://www.jetpletters.ac.ru/ps/1734/}  \\
  \href{http://www.jetpletters.ac.ru/ps/1734/article_26353.shtml}{article\_26353.shtml}\,.

\bibitem{BD} D.G.~Boulware and S.~Deser, Phys. Rev. D {\bf6}, 3368 (1972).


\bibitem{Lee}
  T.~D.~Lee and G.~C.~Wick,
  Nucl.\ Phys. {\bf B 9}, 209 (1969);
  Phys.\ Rev.\ D {\bf 2}, 1033 (1970).

\bibitem{Anselmi} D.~Anselmi,  J. High Energy Phys. {\bf 1802}, 141 (2018);   Class.\ Q.\ Grav.\  {\bf 36}, 065010 (2019); J. High Energy Phys. {\bf 1811}, 021 (2018).

\bibitem{Einhorn} M.B.~Einhorn and D.R.T.~Jones, Phys. Rev. D {\bf 96}, 124025 (2017).

\bibitem{Tkach} V.I.~Tkach,
  Mod.\ Phys.\ Lett.\ A {\bf 27}, 1250131 (2012).


\bibitem{Tomboulis}
  E.~Tomboulis,
  Phys.\ Lett.\  {\bf 70B}, 361 (1977).
\bibitem{Kaku}
  M.~Kaku,
  Phys.\ Rev.\ D {\bf 27}, 2819 (1983).


\bibitem{Shapiro}
  G.~Cusin, F.~de O.Salles and I.~L.~Shapiro,
  Phys.\ Rev.\ D {\bf 93}, no. 4, 044039 (2016).


\bibitem{Smilga2} A.V.~Smilga,
   	Nucl. Phys. {\bf B 706}, 598 (2005).

\bibitem{Smilga} A.V.~Smilga,
  J.\ Phys.\ A {\bf 47}, no. 5, 052001 (2014).

\bibitem{Narain:2016sgk}
  G.~Narain,
  Eur.\ Phys.\ J.\ C {\bf 77}, 683 (2017).


\bibitem{nonlrev}
L.~Modesto and L.~Rachwal,
Int.\ J.\ Mod.\ Phys.\ D {\bf 26}, 1730020 (2017).


\bibitem{occurrence}  A.S.~Koshelev, L.~Modesto, L.~Rachwal, and A.A.~Starobinsky,
J. High Energy Phys. {\bf 1611}, 067 (2016).



\bibitem{unita} F.~Briscese and L.~Modesto,  Phys. Rev. D {\bf 99}, 104043 (2019)
[arXiv:1803.08827 [gr-qc]].

\bibitem{unita2} M.~Christodoulou and L.~Modesto, JETP Lett. {\bf 109}, 286 (2019)
[arXiv:1803.08843 [hep-th]].


\vspace{2cm}
\bibitem{Hartle} J.B.~Hartle, Phys. Rev. D {\bf 49}, 6543 (1994).

\bibitem{Politzer} H.D.~Politzer, Phys. Rev. D {\bf 46}, 4470 (1992).

\bibitem{Lloyd} S.~Lloyd, L.~Maccone, R.~Garcia-Patron, V.~Giovannetti, and Y.~Shikano, Phys. Rev. D {\bf 84}, 025007 (2011).

\bibitem{Weinberg} S.~Weinberg, Ultraviolet divergences in quantum theories of gravitation, in "General Relativity: An Einstein Centenary Survey",
edited by S. W. Hawking and W. Israel (Cambridge University Press, Cambridge, England, 1979), Chapter 16,  pp.790--831.

\bibitem{Reuter:1998}
  M.~Reuter,
  Phys.\ Rev.\ D {\bf 57}, 971 (1998);   
  Nucl.\ Phys. {\bf B 427}, 291 (1994);   Nucl.\ Phys. {\bf B 417}, 181 (1994).   


\bibitem{Wetterich:1993} C.~Wetterich, Phys. Lett. B {\bf 301}, 90 (1993).

\bibitem{codello}
  A.~Codello, R.~Percacci, and C.~Rahmede,
  Ann. Phys.\ (Amsterdam)  {\bf 324}, 414 (2009).

\bibitem{codello_abc}
  D.~Benedetti, P.~F.~Machado, and F.~Saueressig,
  Nucl.\ Phys.\  {\bf B 824}, 168 (2010); S. Weinberg, arXiv:0903.0568 [hep-th].


\bibitem{Niedermaier}
  M.~Niedermaier and M.~Reuter,
  Living Rev.\ Relativity  {\bf 9}, 5 (2006).

%
\bibitem{BZ} T.~Banks and A.~Zaks, Nucl. Phys.  {\bf B196}, 189 (1982).


\bibitem{Percacci} R.~Percacci, {\em An Introduction to Covariant Quantum Gravity and Asymptotic Safety} (World Scientific, New York, 2017).


\bibitem{Codello}  A.~Codello and  R.~Percacci, Phys. Rev. Let. {\bf 97}, 221301 (2006).


\bibitem{Donoghue:2019clr}
  J.~F.~Donoghue,
  arXiv:1911.02967 [hep-th].


\bibitem{SM}
See Supplemental Material to this article at
\href{http://link.aps.org/supplemental/10.1103/PhysRevD.101.044050}{http://link.aps.org/supplemental/10.1103/}\\
\href{http://link.aps.org/supplemental/10.1103/PhysRevD.101.044050}{PhysRevD.101.044050} and \hyperref[sm]{here} for finer
technical details, which include Refs. 
\cite{Tseytlin,Percacci,Benedetti,Kleinert-book,KLMEM,STK,POL,KLMG,Fradkin5,Barvinsky,Litim-a,Litim2,Codellocomp}.

\bibitem{Weyl1} H.~Weyl, Math. Zeitschr. {\bf 2},  384 (1918).

\bibitem{Bach1} R.~Bach, Mathematische Zeitschrift {\bf 9},  110 (1921).

\bibitem{Hamber} H.W.~Hamber, {\it Quantum Gravitation, The Feynman Path Integral Approach}  (Springer, Berlin, 2009).

\bibitem{Carlip} S.~Carlip, Class.~Quantum~Grav.~{\bf 15}, 2629 (1998).


\bibitem{freedman} M. H. Freedman, 
J. Diff. Geom. {\bf 17} (3), 357-453  (1982).

\bibitem{Hubbard}  K.~Hubbard, Phys. Rev. Lett. {\bf 3}, 77 (1959).

\bibitem{Stratonovich} R. L. Stratonovich, Sov. Phys. Dokl. 2, 416 (1957).


\bibitem{Altland-Simons} A.~Altland and B.~Simons, {\it Condensed Matter Field Theory} (Cambridge University Press, Cambridge, England, 2013).

\bibitem{GL}   V.~L.~Ginzburg and L.~D.~Landau,
  Zh.\ Eksp.\ Teor.\ Fiz.\  {\bf 20}, 1064 (1950).


\bibitem{Stevenson} P.M.~Stevenson, Phys. Rev. D {\bf 23}, 2916 (1981).

\bibitem{Kleinert-QFT} H.~Kleinert and V.~Schulte-Frohlinde, {\it Critical Properties of $\phi^4$-Theories} (World Scientific, Singapore, 2001).

\bibitem{Fulling:74}
S.~Fulling, L.~Parker, and B.~Hu,
Phys.\ Rev.\ D {\bf 10},  3905 (1974).


\bibitem{exactsol} Y.~D.~Li, L.~Modesto and L.~Rachwal,
  J. High Energy Phys. {\bf 1512}, 173 (2015)
  [arXiv:1506.08619 [hep-th]].


\bibitem{Irakleidou}
  M.~Irakleidou and I.~Lovrekovic,
  Phys.\ Rev.\ D {\bf 93}, 104043 (2016).

\bibitem{Wetterich2}
  C.~Wetterich,
  Phys.\ Rev.\ D {\bf 98}, no. 2, 026028 (2018).

\bibitem{Capper}
  D.~M.~Capper and M.~J.~Duff,
  Phys.\ Lett.\ A {\bf 53}, 361 (1975).



\bibitem{shapiro2}
M. Asorey, E.V. Gorbar, and I.L. Shapiro, 
 Class. Quant. Grav. {\bf 21},  163 (2003).

 \bibitem{superrenfin}
L. Modesto and L. Rachwal,
  Nucl.\ Phys. {\bf B 889}, 228 (2014).

\bibitem{univfin}
L. Modesto and L. Rachwal,
  Nucl.\ Phys. {\bf B 900}, 147 (2015).


\bibitem{fingauge} L. Modesto, M. Piva, and L. Rachwal,
Phys.\ Rev.\ D {\bf 94}, 025021 (2016).



\bibitem{finconfqg} L.~Modesto and L.~Rachwal,
arXiv:1605.04173 [hep-th].


\bibitem{shapiro1}
G. de Berredo-Peixoto and I. L. Shapiro, 
Phys. Rev. D {\bf 70},  044024 (2004).

\bibitem{Fradkin1}
  E.S.~Fradkin and A.A.~Tseytlin,
  Phys.\ Lett.\  {\bf 104B}, 377 (1981).

\bibitem{Fradkin2}
  E.S.~Fradkin and A.A.~Tseytlin,
  Nucl.\ Phys. {\bf B 201}, 469 (1982).

\bibitem{Kiritsis:17} E.~Kiritsis, F.~Nitti, and L.S.~Pimenta, Fortschr. Phys. {\bf 65}, 1600120 (2017).

\bibitem{Nitti:18}
F.~Nitti, L.S.~Pimenta, and D.A.~Steer, J. High Energ. Phys. {\bf 7}, 22 (2018).

\bibitem{Ghosh:18}
J.K.~Ghosh, E.~Kiritsis, F.~Nitti  {\em et al.}, J. High Energ. Phys. {\bf 05}, 34 (2018).


\bibitem{hol1} R.~Percacci and L.~Rachwal, 
 Fortsch.\ Phys.\  {\bf 62}, 887 (2014).

\bibitem{hol2} D.F.~Litim, R.~Percacci, and L.~Rachwal, 
 Phys.\ Lett.\ B {\bf 710}, 472 (2012);  J.\ Phys.\ Conf.\ Ser.\  {\bf 343}, 012098 (2012).



\bibitem{LeClair:04} A.~LeClair, J.M.~Roman, and G. Sierra,  Nucl. Phys. {\bf B 700}, 407 (2004).

\bibitem{Curtright:12} T.L.~Curtright, X.~Jin, and C.K.~Zachos, Phys.
Rev. Lett. {\bf 108}, 131601 (2012).



\bibitem{dep1}
  G.~Narain and R.~Percacci,
  Acta Phys.\ Polon.\ B {\bf 40}, 3439 (2009)
  [arXiv:0910.5390 [hep-th]], \newline\href{https://www.actaphys.uj.edu.pl/R/40/12/3439/pdf}{https://www.actaphys.uj.edu.pl/R/40/12/3439/pdf}\,.

\bibitem{dep2}
  N.~Ohta, R.~Percacci, and A.~D.~Pereira,
  Eur.\ Phys.\ J.\ C {\bf 77}, 611 (2017).



\bibitem{dep3}
  I.~L.~Shapiro and A.~G.~Zheksenaev,
  Phys.\ Lett.\ B {\bf 324}, 286 (1994).


\bibitem{shapiro3}
G. de Berredo-Peixoto, A. Penna-Firme, and I. L. Shapiro, 
Mod. Phys. Lett. A {\bf 15}, 2335 (2000).


\bibitem{shapiro4}
J. D. Goncalves, T. de Paula Netto, and I. L. Shapiro,
Phys. Rev. D {\bf 97},  026015 (2018).


\bibitem{dep4}
  V.~F.~Barra, P.~M.~Lavrov, E.~A.~Dos Reis, T.~de Paula Netto, and I.~L.~Shapiro,
  arXiv:1910.06068 [hep-th].


\bibitem{penrose1}
R.~Penrose,  {\it Cycles of Time: An Extraordinary New View of the Universe} (The Random House, London, 2010).

\bibitem{penrose2}
R.~Penrose, Found. Phys. {\bf 44}, 557 (2014).

\bibitem{hooft2} G.~'t Hooft, arXiv:1410.6675 [gr-qc].

\bibitem{Shap:2013} R.~Armillis,  A.~Monin,  and  M.~Shaposhnikov, J. High Energy Phys. {\bf 10}, 030 (2013).

\bibitem{Amelino} G.~Amelino-Camelia, M.~Arzano, G.~Gubitosi, and J.~Magueijo,  Int. J. Mod. Phys. D {\bf 24}, 1543002 (2015).

\bibitem{Litim:2014}  D.F.~Litim and F.~Sannino,
  J. High Energy Phys. {\bf 1412}, 178 (2014).

\bibitem{BJV} M.~Blasone, P.~Jizba, and G.~Vitiello, {\em Quantum Field Theory and its Macroscopic Manifestations}
(World Scientific \& ICP, London, 2010).

\bibitem{BJS} M.~Blasone, P.~Jizba, and L.~Smaldone, Phys. Rev.  D {\bf 100}, 045027 (2019).

\bibitem{Linde1} S.~Ferrara, R.~Kallosh, A.~Linde, A.~Marrani, and A.~Van Proeyen, Phys. Rev. D {\bf 82}, 045003 (2010); Phys. Rev. D {\bf 83}, 025008 (2011).

\bibitem{Linde2} I.~Bars, P.~Steinhardt, and N.~Turok, Phys. Rev. D {\bf 89}, 043515 (2014).


\bibitem{Reuter_abc}
M.~Reuter and F.~Saueressig, {\em Quantum Gravity and the Functional Renormalization Group: The Road Towards Asymptotic Safety} (Cambridge University Press, Cambridge, England, 2018).

\bibitem{Nagy:2013hka}
  S.~Nagy, B.~Fazekas, L.~Juhasz, and K.~Sailer,
  Phys.\ Rev.\ D {\bf 88}, 116010 (2013).

\bibitem{Nandori} I.~N\'{a}ndori, J. High Energy Phys. {\bf 04}, 150 (2013).

\bibitem{MinSEn} P.M.~Stevenson, Phys. Rev. D {\bf 23}, 2916 (1981).

\bibitem{alfio}  A.~Bonanno and G.~Lacagnina,
  Nucl.\ Phys. {\bf B 693}, 36 (2004);
 A.~Bonanno, S.~Lippoldt, R.~Percacci, and G.~P.~Vacca,
 arXiv:1912.08135 [hep-th].

\bibitem{bhevap} C.~Bambi, L.~Modesto, S.~Porey, and L.~Rachwal,
J. Cosmol. Astropart. Phys. {\bf 1709}, 033 (2017).


\bibitem{formation}  C.~Bambi, L.~Modesto, S.~Porey, and L.~Rachwal,
Eur.\ Phys.\ J.\ C {\bf 78}, 116 (2018).

\bibitem{entanglement} S.~Giaccari, L.~Modesto, L.~Rachwal, and Y.~Zhu, 
  Eur.\ Phys.\ J.\ C {\bf 78}, 459 (2018). 

\bibitem{finads} A.S.~Koshelev, K.~Sravan Kumar, L.~Modesto, and L.~Rachwal,
Phys.\ Rev.\ D {\bf 98}, 046007 (2018).

\bibitem{spcompl0} L.~Modesto and L.~Rachwal,
J.\ Phys.\ Conf.\ Ser.\  {\bf 942}, 012015 (2017).


\bibitem{spcompl} C.~Bambi, L.~Modesto, and L.~Rachwal,
J. Cosmol. Astropart. Phys. {\bf 1705}, 003 (2017).


\bibitem{confreview} L. Rachwal
\textit{Conformal Symmetry in Field Theory and in Quantum Gravity} - Review, Universe {\bf 4}, 125  (2018);
arxiv:1808.10457v3 [hep-th].



\bibitem{Jacobson} T.~Jacobson, N.C.~Tsamis, and R.P.~Woodard, Phys. Rev. D {\bf 38}, 1823 (1988).


\bibitem{Benedetti}
  D.~Benedetti, P.~F.~Machado, and F.~Saueressig,
  Mod.\ Phys.\ Lett.\ A {\bf 24}, 2233 (2009).

\vspace{50cm}

\bibitem{Kleinert-book} H.~Kleinert, {\it Path Integrals in Quantum Mechanics, Statistics, Polymer
Physics, and Financial Markets} (World Scientific, Singapore, 2006).

\bibitem{KLMEM} H.~Kleinert, Phys. Lett. A~{\bf 114}, 263 (1986); Mod. Phys. Lett. A~{\bf 03}, 531 (1988);
Phys. Lett. B {\bf 189}, 187 (1987).

\bibitem{STK} H.~Kleinert, Phys. Lett. B {\bf 174}, 335 (1986); Phys. Rev. Lett. {\bf 58}, 1915 (1987).


\bibitem{POL} A.M.~Polyakov, Phys. Lett. B {\bf 59}, 79 (1975).

\bibitem{KLMG} H.~Kleinert, Phys. Lett. B~{\bf 196}, 355 (1987).


\bibitem{Fradkin5}
  E.~S.~Fradkin and A.~A.~Tseytlin,
  Nucl.\ Phys. {\bf B 234}, 472 (1984).
  

\bibitem{Barvinsky} A. O.~Barvinsky and G. A.~Vilkovisky, Phys. Rep. {\bf 119}, 1 (1985).

\bibitem{Litim-a} 
  D.~F.~Litim,
  PoS QG {\bf -Ph}, 024 (2007)

\bibitem{Litim2}D.~F.~Litim,
  Phys.\ Rev.\ D {\bf 64}, 105007 (2001);  Phys.\ Lett.\ B {\bf 486}, 92 (2000).

\bibitem{Codellocomp}
  A.~Codello, R.~Percacci, L.~Rachwal, and A.~Tonero,
  Eur.\ Phys.\ J.\ C {\bf 76}, no. 4, 226 (2016).

\vspace{50cm}


\end{thebibliography}

\begin{thebibliography}{99}


\vspace{5mm}
\bibitem{Benedettis}
  D.~Benedetti, P.~F.~Machado, and F.~Saueressig,
  Mod.\ Phys.\ Lett.\ A {\bf 24}, 2233 (2009)
\bibitem{Kleinert-books} H.~Kleinert, {\it Path Integrals in Quantum Mechanics, Statistics, Polymer
Physics, and Financial Markets} (World Scientific, Singapore, 2006).



\bibitem{KLMEMs} H.~Kleinert, Phys. Lett. A~{\bf 114}, 263 (1986); Mod. Phys. Lett. A~{\bf 03}, 531 (1988);
Phys. Lett. B {\bf 189}, 187 (1987).

\bibitem{STKs} H.~Kleinert, Phys. Lett. B {\bf 174}, 335 (1986); Phys. Rev. Lett. {\bf 58}, 1915 (1987).

\bibitem{POLs} A.M.~Polyakov, Phys. Lett. B {\bf 59}, 79 (1975).

\bibitem{KLMGs} H.~Kleinert, Phys. Lett. B~{\bf 196}, 355 (1987).

\bibitem{Fradkin5s}
  E.~S.~Fradkin and A.~A.~Tseytlin,
  Nucl.\ Phys.\ B {\bf 234}, 472 (1984).

\bibitem{Tseytlins} E.S.~Fradkin and A.A.~Tseytlin, Phys.~Rep.~{\bf 119},  233 (1985).

\bibitem{Barvinskys} A.O.~Barvinsky and G.A.~Vilkovisky, Phys. Rep. {\bf 119}, 1 (1985).


\bibitem{Percaccis} R.~Percacci, {\em An Introduction to Covariant Quantum Gravity and Asymptotic Safety} (World Scientific, New York, 2017).

\bibitem{Litim-as} D. F.~Litim, PoS QG {\bf -Ph} 024 (2008).

\bibitem{Litim2s}D.~F.~Litim,
  Phys.\ Rev.\ D {\bf 64}, 105007 (2001)
  [hep-th/0103195];  Phys.\ Lett.\ B {\bf 486}, 92 (2000).

\bibitem{Codellocomps}
  A.~Codello, R.~Percacci, L.~Rachwal, and A.~Tonero,
  Eur.\ Phys.\ J.\ C {\bf 76}, no. 4, 226 (2016)
  [arXiv:1505.03119 [hep-th]].


\end{thebibliography}
\end{document}